\documentclass{article}
\usepackage{graphicx} 
\usepackage{jcappub}
\usepackage[dvipsnames]{xcolor}
\usepackage[normalem]{ulem}
\usepackage[noabbrev]{cleveref}
\usepackage{csquotes}
\usepackage{placeins}
\usepackage{enumitem}

\usepackage{pifont}

\usepackage{orcidlink}

\title{
Smooth sailing or ragged climb? --- robustness of power spectrum de-wiggling for ShapeFit parameter compression}

\abstract{The baryonic features in the galaxy power spectrum offer tight, time-resolved constraints on the expansion history of the Universe but complicate the measurement of the broadband shape of the power spectrum, which also contains precious cosmological information. In the context of ShapeFit, the broadband information is compressed into a single parameter, the slope of the power spectrum at the pivot scale, $m$, designed to be sensitive to matter-radiation equality and the baryonic suppression. To calculate this parameter, two steps are necessary: 1) smoothing the power spectrum to remove the baryonic oscillations and 2) calculating the derivative of the power spectrum ratio at the pivot scale. In this work we compare thirteen methods designed to separate the broadband and oscillating components and examine their performance. The systematic uncertainty between different de-wiggling procedures is at most $2$\%, depending on the scale. For the obtained slope, we show that the de-wiggling procedures impart large (50\%) differences, but as long as the theory and data pipelines are set up consistently, this is of no concern for cosmological inference given the precision of existing and on-going surveys. However, it still motivates the search for more robust ways of extracting the slope. We show that post-processing the power spectrum ratio before taking the derivative makes the slope values far more robust. We further investigate eleven ways of extracting the slope and highlight the two most successful ones. We derive a systematic uncertainty on the slope $m$ of $\sigma_{m,\mathrm{syst}} = 0.023 |m| + 0.001$ by studying the behavior of the slopes in different cosmologies within and beyond $\Lambda$CDM and the impact in cosmological inference. In cosmologies with a feature in the matter-power spectrum, such as in the early dark energy cosmologies, this systematic uncertainty estimate does not necessarily hold, and further investigation is required.}
\date{Apr 2025}

\author[a,b,c]{Katayoon Ghaemi~\orcidlink{0009-0007-3975-4079},}
\emailAdd{ghaemi@cppm.in2p3.fr}
\author[d,e]{Nils Sch\"oneberg~\orcidlink{0000-0002-7873-0404},}

\emailAdd{nils.science@gmail.com}
\author[c,f]{Licia Verde~\orcidlink{0000-0003-2601-8770}}

\affiliation[a]{Aix-Marseille Université, CNRS/IN2P3, CPPM, Marseille, France}
\affiliation[b]{Dipartimento di Fisica e Astronomia “Galileo
Galilei” Università di Padova, I-35131 Padova, Italy}
\affiliation[c]{ICCUB Institut de Ci\`encies del Cosmos, Universitat de Barcelona, Mart\'i i Franqu\`es 1, E08028 Barcelona, Spain}
\affiliation[d]{University observatory, Faculty of Physics, Ludwig-Maximilians-Universität, Scheinerstr. 1, 81677 Munich, Germany}
\affiliation[e]{Excellence Cluster ORIGINS, Boltzmannstr. 2, 85748 Garching, Germany}
\affiliation[f]{ICREA, Instituci\'o Catalana de Recerca i Estudi Avan\c{c}at, Pg. Lluis Companys 23, Barcelona, 08010, Spain}

\begin{document}

\maketitle

\section{Introduction}\label{sec:introduction}

The power spectrum of galaxies has proven to be a treasure trove of cosmological information. It is well known that the potential wells caused by the clustering of baryons and cold dark matter (CDM) are fundamental for seeding the formation of galaxies and galaxy clusters. Therefore, the late-time galaxy power spectrum captures a lot of information about the energy densities in the early universe imprinted in current day galaxy positions, for example through the baryonic acoustic oscillations (BAO) and its corresponding baryonic suppression of growth, as well as the overall broadband shape related to the transition from radiation-dominated to matter-dominated growth and the spectral index of the primordial fluctuation power spectrum.

\begin{figure}
    \centering
    \includegraphics[width=0.5\linewidth]{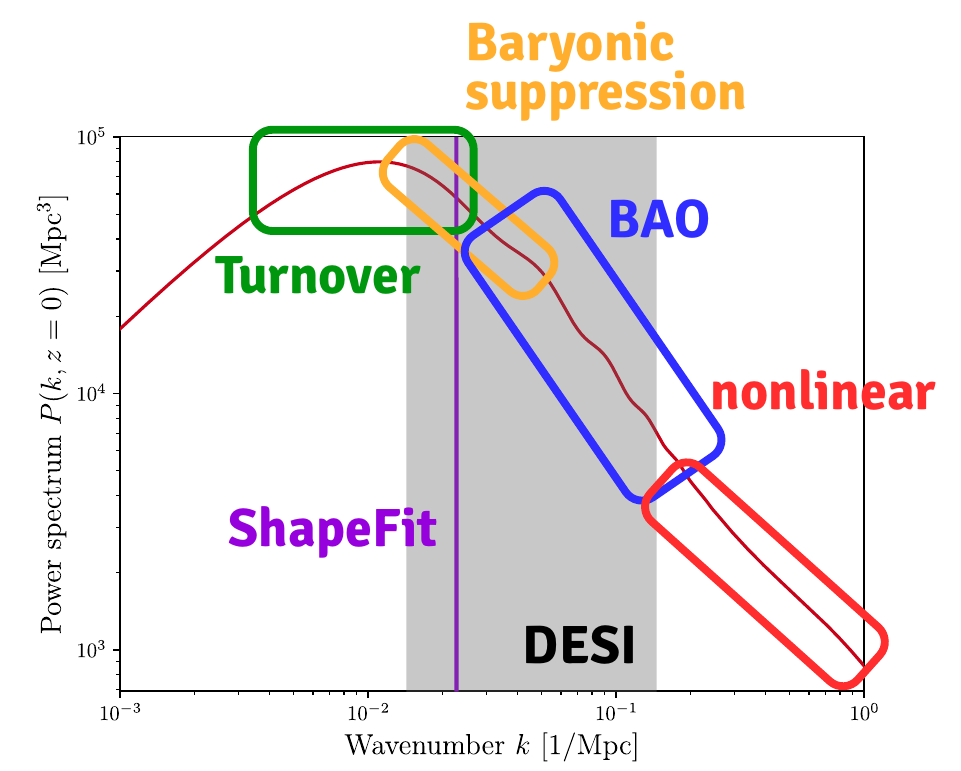}
    \caption{A schematic representation of the relevant information imprinted in the matter power spectrum. The scales relevant to DESI correspond to $0.02h/\mathrm{Mpc}$-$0.2h/\mathrm{Mpc}$ \cite{DESI:2024hhd}, while we use the ShapeFit pivot wavenumber of $0.03h/\mathrm{Mpc}$ \cite{brieden2021shapefit}.}
    \label{fig:power_spectrum}
\end{figure}

Since its first detection in the early Two degree of Field Galaxy Redshift Survey (2dFGS) and the Sloan Digital Sky Survey (SDSS) Luminous Red Galaxies (LRG) samples \cite{Miller:2001cf,2dFGRS:2005yhx,SDSS:2005xqv}, the BAO signature
has been a staple of cosmological analyses from galaxy surveys, 
, culminating in the SDSS Baryon Oscillation Spectroscopic Survey (BOSS) and its extension (eBOSS) as well as the Dark energy Spectroscopic Instrument (DESI) BAO analyses \cite{eBOSS:2020yzd,DESI:2024mwx}, giving the most precise constraints on the Universe's expansion history from BAO to date. However, the BAO oscillations (\enquote*{wiggles}) are not the only information imprinted in the power spectrum: its broadband shape encodes additional valuable information through various effects. We schematically show the most notable features of the power spectrum in \cref{fig:power_spectrum}. 

The turnover results from the transition between radiation-dominated and matter-dominated growth. It occurs at scales which are slightly too large for current survey footprints, and can thus be measured only with high statistical uncertainty \cite{Bahr-Kalus:2023ebd}. While the non-linear scales are in principle accessible to many current galaxy surveys, the limited understanding of non-linear effects restricts their use: the maximum investigated wavenumbers  often extend only mildly into the non-linear regime, limited by the range of validity of perturbative or effective field theory approaches.

The grey band in \cref{fig:power_spectrum} shows the investigated wavenumber range (for DESI) and the red rounded square shows the non-linear enhancement of structure formation in the fully non-linear regime. Even at the only mildly non-linear scales currently included in most analyses, the mode-coupling introduced by non-linearities induces a suppression of the BAO feature (e.g., 
\cite{Crocce:2007dt,Sanchez:2008iw,Padmanabhan:2009yr,Sherwin:2012nh,Prada:2014bra}). 
Most state-of-the-art analyses use the effective theory of large scale structure (EFTofLSS) to model the non-linear power spectrum of tracers
(see \cite{Porto:2016pyg,Ivanov:2022mrd} for a review on EFTofLSS). To compute this suppression, for EFTofLSS modeling it is useful and customary to decompose the power spectrum into the oscillatory (\enquote*{wiggly}) component, and the broadband (\enquote*{de-wiggled}) component.

In addition to the above clearly visible features, there are two more subtle features imprinted in the power spectrum. The logarithmic curvature of the power spectrum is related to the time of growth between the horizon entry of a mode during radiation domination and the eventual start of matter domination, leading to a weak dependence on the scale of matter-radiation equality. The  baryonic suppression  on the other hand, is a clear feature after the turnover of the power spectrum and results from the lack of growth of baryonic overdensities due to their involvement in the acoustic oscillations.
The baryonic suppression and the logarithmic curvature are measurable on large scales e.g., \cite{Philcox:2020xbv,brieden2021model,Brieden:2022lsd} (where non-linear effects are irrelevant) and directly relate to the ratio of baryon to cold dark matter in the early Universe.

In order to use these two features to constrain cosmological parameters, there are two main approaches: fitting the full shape of the power spectrum (as in e.g., \cite{Simon:2022csv,DAmico:2019fhj,Troster:2019ean,Ivanov:2019hqk,DESI:2024hhd}) or compressing the information into a single variable, such as with the ShapeFit approach developed and employed in Refs.~\cite{brieden2021shapefit,brieden2021model,brieden2022model}. While the former can in principle extract all information from the power spectrum, a cosmological model needs to be assumed \textit{a priori}, and therefore the data analysis needs to be completely re-done for each cosmological model of interest. 
Instead, the latter approach is in principle independent of the cosmological model: the analysis can be performed once and be re-interpreted for different models. In addition, it is somewhat more interpretable due to the localization of the parameter constraints to specific features in the power spectrum.

Both approaches rely to some degree on the decomposition of the power spectrum into a wiggly and de-wiggled component (either through their use of EFTofLSS for the full modeling of the power spectrum, or for the computation of the ShapeFit parameter, see \cref{ssec:theory_shapefit} below).
Therefore, the de-wiggling of the power spectrum is an important part of any modern LSS analysis pipeline. It is not surprising then that a vast number of de-wiggling methods have been proposed in the literature. Full shape analyses based on EFTofLSS are largely insensitive to the choice of de-wiggling algorithm once marginalization over nuisance parameters capturing the IR resummation is performed \cite{vlah2016perturbation, MoradinezhadDizgah:2020whw, Euclid:2023tog}. A thorough comparison of all proposed methods and a corresponding assignment of systematic uncertainties for the de-wiggled power spectrum in the context of ShapeFit --in particular for the corresponding extraction of the shape parameter--  for a variety of cosmological models has, to our knowledge, not been performed before.

The paper is structured as follows. In \cref{sec:dewiggling} we  describe in detail the different de-wiggling methods proposed in the literature and estimate a systematic uncertainty in the de-wiggled power spectrum. The reader not too keen on technical details of the algorithms can skip sections \ref{ssec:analytical} to \ref{ssec:peakremoval} at first sitting: the findings are summarized in \cref{ssec:dewiggle_summary}. In \cref{sec:ShapeFit} we then focus on different methods of extracting the shape information using the ShapeFit formalism, and estimate a systematic uncertainty on the ShapeFit slope parameter~$m$. Once again the reader not to keen on technical details can skip the first part of section \ref{ssec:derivatives} and jump directly to the comparison in \cref{ssec:comparison}. In \cref{sec:results} we then test the systematic uncertainty for different cosmologies within and beyond the $\Lambda$CDM model.

We conclude in \cref{sec:conclusion}.

\section{Broadband/BAO 
decomposition}\label{sec:dewiggling}

The separation (or decomposition) of the power spectrum (BAO) wiggles from the broadband shape has been an important part of cosmological analyses since 
about two decades ago, when it was realized that nonlinear structure formation affects the baryonic oscillation feature in a non-trivial way \cite{Seo:2005ys, Eisenstein:2006nj}. The first methods were based on either semi-analytical fitting methods, such as the Eisenstein-Hu transfer function \cite{Eisenstein:1997jh,eisenstein1998baryonic}, or simple polynomial fits as in \cite{2010MNRAS.404...60R}. As interest in the analysis of the baryonic features in the power spectrum grew, the approaches diversified. Today there is a large collection of different methodologies for decomposing the power spectrum into baryonic oscillations (the \enquote*{wiggly} part) and the broadband shape (the \enquote*{no-wiggle} part). We discuss a large, nearly exhaustive collection of methods in \cref{sec:dewiggling} below, focusing on the latest implementations of any given method.

\begin{figure}
    \centering
    \includegraphics[width=0.49\linewidth]{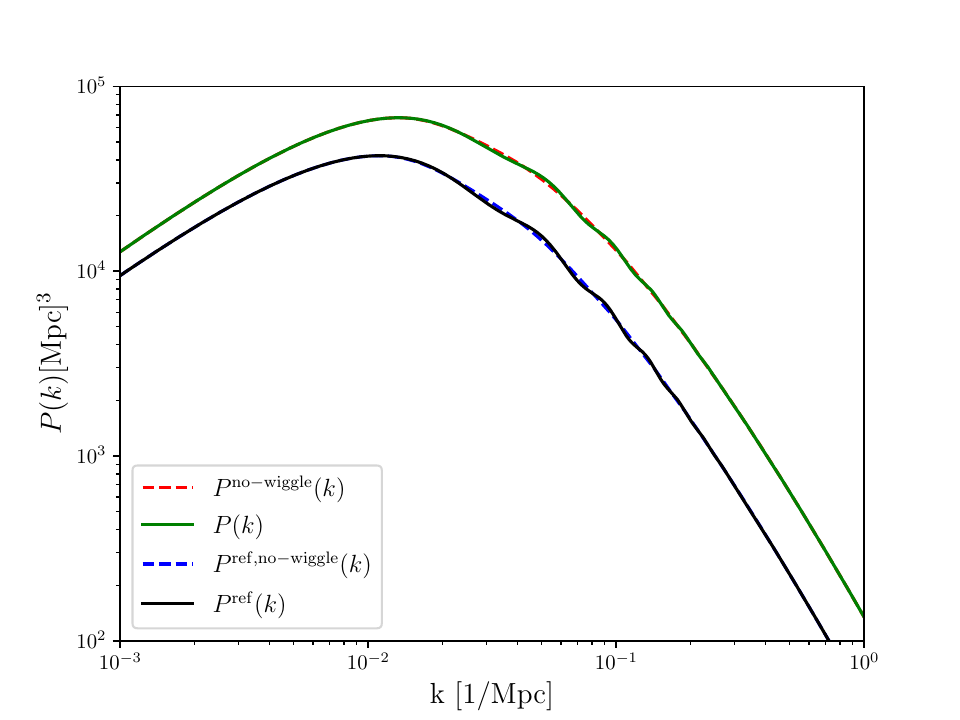}
    \includegraphics[width=0.49\linewidth]{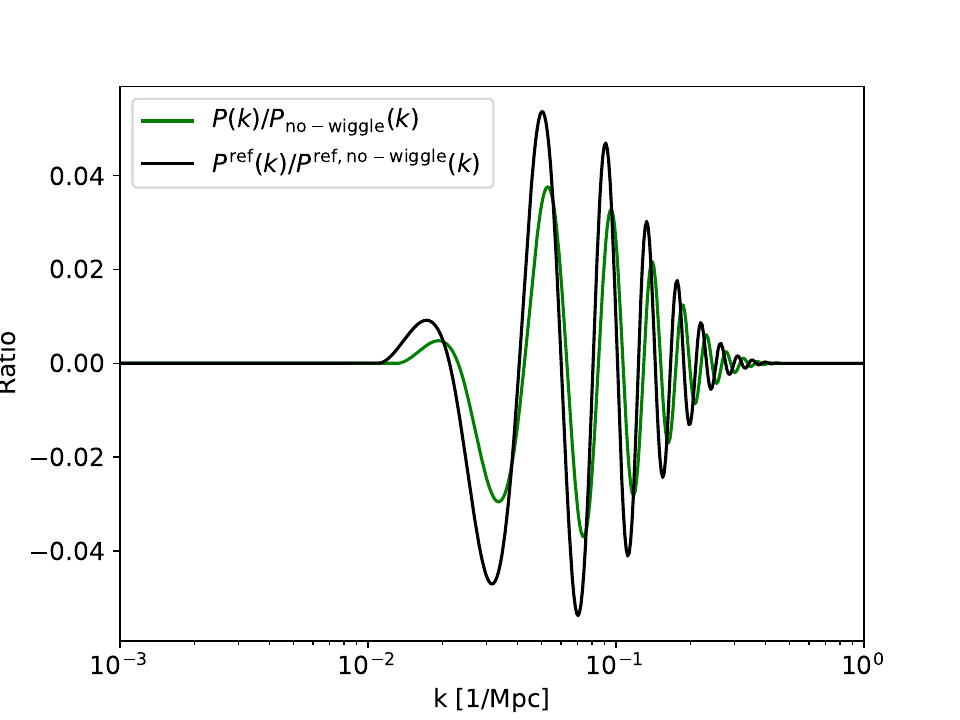}

    \caption{\textbf{Left:} Power spectra for the fiducial (black) and showcase (green) cosmology
    (see table \ref{tab:cosmologies} for the corresponding cosmological parameters values), and their corresponding de-wiggled power spectra (blue and red dashed lines, respectively). The example de-wiggling algorithm used here is the \enquote{Cubic Inflections} algorithm of \cref{ssec:inflections}. \textbf{Right:} Ratio of the power spectrum to the no-wiggle power spectrum, for both cosmologies, highlighting the baryonic acoustic oscillations.}
    \label{fig:dewiggle}
\end{figure}

These \textit{de-wiggling} methods have gained traction in particular for their use for accurately predicting the nonlinear power spectrum in halo models \cite{Smith:2002dz,Takahashi:2012em, bird2012massive, Mead:2020vgs, Mead:2015yca} and effective field theories of large-scale structure growth (EFTofLSS) \cite{Porto:2016pyg,Ivanov:2022mrd,Simon:2022csv,DAmico:2019fhj,Troster:2019ean,Ivanov:2019hqk,DESI:2024hhd}.
These algorithms typically attempt to isolate the BAO wiggle from the power spectrum, allowing one to compute a smoothed version of the power spectrum: they remove the oscillations themselves but not the overall baryonic suppression. We show an example of such a decomposition in \cref{fig:dewiggle}, both for a fiducial/reference cosmology and a somewhat arbitrary showcase cosmology (see \cref{tab:cosmologies} for the parameters defining these cosmologies).

Recently, a new compressed parameter scheme \enquote{ShapeFit} has also been introduced \cite{brieden2022model,brieden2021model,brieden2021shapefit}, which is based on determining the broadband slope at a wavenumber of interest. This approach also requires efficient de-wiggling schemes, but robustly extracting the baryonic suppression requires special care and we focus on such specific application in \cref{sec:ShapeFit}.

\begin{figure}
    \centering
    \includegraphics[width=0.9\linewidth]{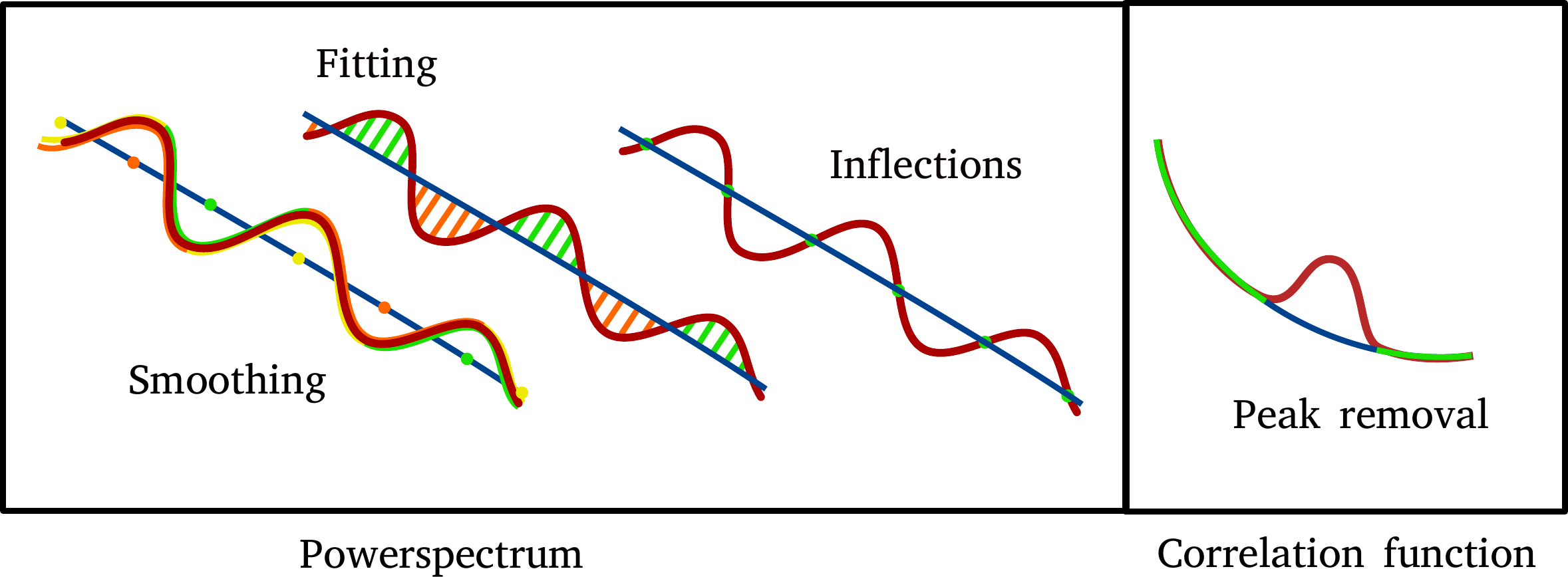}
    \caption{Schematic overview of de-wiggling methods. The red line corresponds to the combined broadband spectrum with additional wiggles (as a power spectrum on the left, and as a correlation function on the right) while blue corresponds to the dewiggled broadband.
    The green/yellow/orange lines and points show how the corresponding de-wiggled broadband function is obtained. For the \enquote{Smoothing} method, the yellow/orange/green lines show sufficiently long smoothing windows at certain locations across the function, whose averages are correspondingly shown as yellow/orange/green points, through which a smooth function is drawn. For the \enquote{Fitting} method, the green/orange lines show the mean squared deviation that has to be minimized in any approach that fits a smooth function to the oscillations. For the \enquote{Inflections} method, the green point show the inflections of the oscillations (after subtracting a rough broadband trend) through which the blue function is fitted. Finally, for the \enquote{Peak removal} method, the correlation function is fitted where there is no BAO peak by a smooth function, resulting in the blue curve directly.}
    \label{fig:schematic}
\end{figure}

De-wiggling methods common in the literature can roughly be divided into four main numerical approaches. We schematically display these four approaches in \cref{fig:schematic} and list them below:
\begin{enumerate}
    \item \textbf{Smoothing:} Just numerically smoothing the oscillations can give viable results if the smoothing kernel width is approximately a full oscillation wavelength. At that point the oscillation integrated over the smoothing kernel approximately cancels out, leaving just the broadband shape. This method is discussed in \cref{ssec:smoothing}.
    \item \textbf{Fitting:} Based on fitting a smooth function through the oscillations. The BAO oscillations are assumed to be symmetric around the brodband, meaning that the amplitude of the oscillation above and below the broadband should balance out. When using a least-square fitting method, the BAO oscillations are forced to be symmetric and this indirectly determines the broadband.
    This method is discussed in \cref{ssec:fitting}.
    \item \textbf{Inflections:} Based on constructing a smooth function passing through the inflections of the oscillations. The inflection points usually coincide with the zero-point of the oscillations, and those correspond to points where the wiggly power spectrum crosses the smooth broadband spectrum. This method is discussed in \cref{ssec:inflections}.
    \item \textbf{Peak removal:} Instead of working with the power spectrum, 
    the correlation function is used, for which the BAO oscillation feature is known to be well localized. After removing this local feature (for example by just fitting a smooth function through the surrounding scales) the resulting correlation function can be transformed back into a power spectrum without the BAO wiggles. This method is discussed in \cref{ssec:peakremoval}.
\end{enumerate}

We discuss each dewiggling method individually in the corresponding section of appendix~\ref{app:smoothing} where the reader will find all the necessary details for appreciating the performance, advantages and disadvantages of each method and all the information for their implementation.  We summarize their performances
in \cref{ssec:dewiggle_summary} and \cref{tab:dewiggles}. 

We use the \texttt{class} code \cite{class} for generating the linear power spectra, and use our custom python code to implement the different de-wiggling methods as well as to convert the power spectrum to a correlation function. The corresponding code is made publicly accessible from \texttt{https://github.com/katayoongh/ShapeFitSmoothing}. For early dark energy cosmologies, we use instead the \texttt{AxiCLASS} code from \cite{AxiClass1,AxiClass2}.

\subsection{On using analytical formulae}\label{ssec:analytical}

An analytic fitting formula for the shape of the transfer function (in the absence of massive neutrinos) is provided in \cite{eisenstein1998baryonic}, both for the cases when the BAO wiggles are present and when they are neglected. We refer to it as \textit{EH98}. This function has also been extended in the context of massive neutrinos in~\cite{Eisenstein:1997jh}. The advantage of this approach is that this model is analytical. However, the fitting formula was designed for galaxy surveys from the 2000s and is not suitable for the precision of present-day galaxy surveys. Additionally, in practice the model cannot be used beyond the most commonly discussed cosmological models -- for many cases no generalization exists. This is a generic problem for any kind of analytical fitting formula -- its usage will be restricted only to those models for which it has been explicitly constructed.

While we do not further consider the 
use of the analytical formulae to model the de-wiggled power spectrum, we recognize that the EH98 formula can still be helpful when smoothing as it typically captures the overall shape and order-of-magnitude of the true power spectrum, see below. Moreover, the general form can still be helpful in a semi-analytic approach whereby the individual (otherwise analytically determined) coefficients of the formula are numerically fit instead, see \cref{ssec:fitting}.

\subsection{Overview and summary}\label{ssec:dewiggle_summary}

\Cref{tab:dewiggles} presents a summary of the performance of the different methods presented in detail in \cref{app:smoothing}, where the algorithms are also named.
In particular, we observe that a few methods are not very well suited for the oscillation/broadband decomposition in the context of a consistent numerical study of the BAO, for example due to a strong dependence on hyperparameters or unsatisfactory removal of the wiggles. Therefore, we single out a \enquote*{golden sample} of six of the most promising de-wiggling methods, including the \textbf{Simple Gaussian}, \textbf{B-spline fit} (with averaging over multiple splines), \textbf{EH fit}, \textbf{Cubic Inflections}, \textbf{EH inflections (version 2)}, and \textbf{Fast sine transform} methods.

The golden sample outlined above is selected on the basis of being useful for applications concerned with the full BAO range and for recovering the broadband shape in the full wavenumber range which is typically of interest for galaxy surveys. However, for specific applications focused on specific scales or features, another selection might be more optimal. For this reason we keep considering the entire collection of thirteen methods until \cref{ssec:comparison} at which point we focus on these six most robust ones.

\begin{figure}[t]
    \centering
    \includegraphics[width=0.49\linewidth]{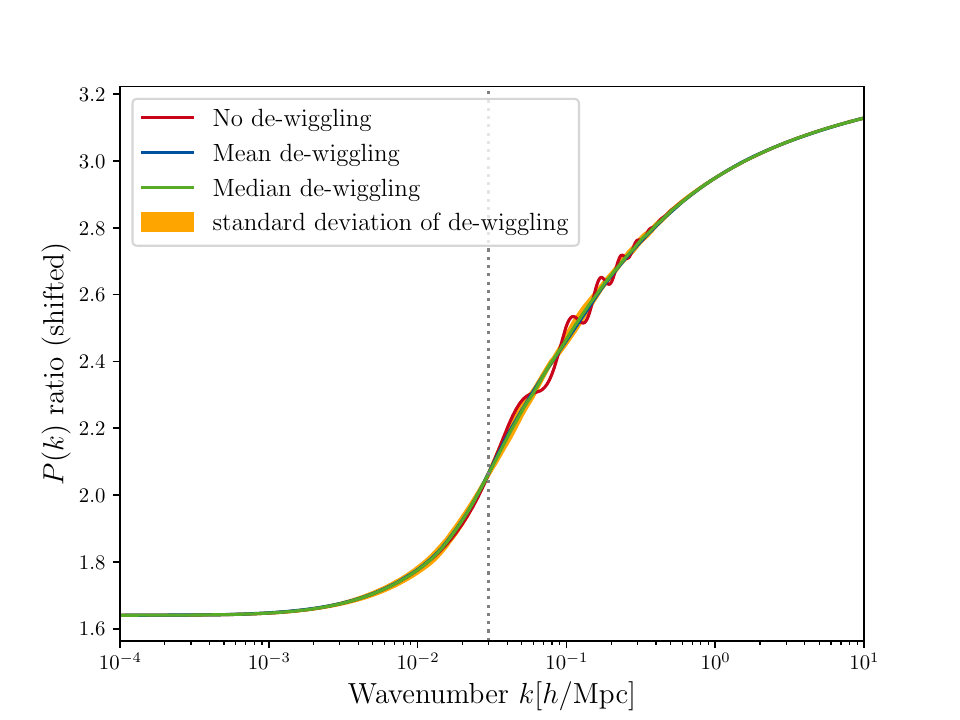}
    \includegraphics[width=0.49\linewidth]{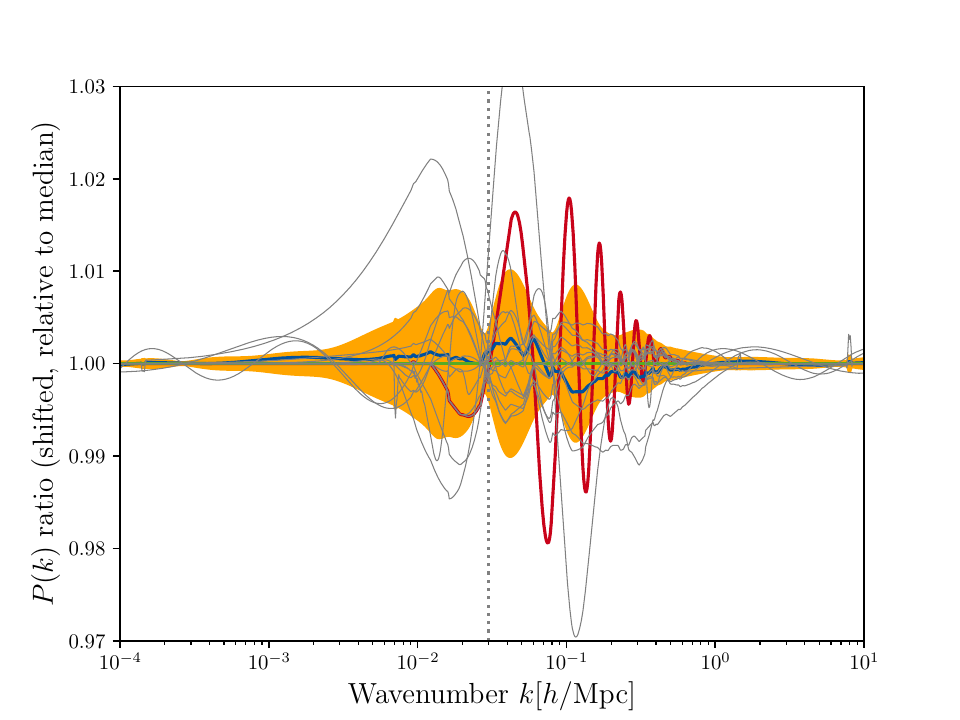}
    \caption{Ratio of the (shifted) power spectra for the computation of \cref{eq:mdef}, using the fiducial and showcase cosmologies of \cref{tab:cosmologies}. Left: Absolute ratio, showing the ratio without de-wiggling, and for all the de-wiggling methods the mean, median, and the 10-90\% quantile range. Right: Relative ratio, normalized by the median. The dashed  vertical grey line represents the pivot wavenumber $k_p$ at which the ShapeFit slope (see \cref{sec:ShapeFit}) is  evaluated.}
    \label{fig:dewiggle_comparison}
\end{figure}

All thirteen de-wiggling methods are compared in \cref{fig:dewiggle_comparison}, where we show the mean and median of the different methods as well as their standard deviation as an orange band around the median. Compared to this median, we find deviations up to around 2\%, strongly dependent on the scale, and particularly large at either side of $k_p= 0.03h/\mathrm{Mpc}$. This can be understood  as follows. The first peak of the BAO coincides with the turnover of the power spectrum due to the closeness of the baryon drag and matter-radiation equality times. Therefore what is considered the first BAO peak and what is considered the turnover of the power spectrum is not uniquely defined (see  \cref{app:illdefined} for an explicit toy example). This ambiguity is what causes the marked differences among the different de-wiggling methods exactly around this scale.

This issue persists virtually unaltered also in the \enquote{gold sample}: these methods also do not treat the first BAO/power spectrum turnover decomposition  consistently. We will return to this in \cref{ssec:mtrue}, as it has important consequences for the interpretation of ShapeFit results.

\begin{table}[t]
    \centering
    \begin{tabular}{c|c l| c}
        Method name & Used for  & & Reason\\
        $[$Citation$]$ & gold sample & & for rejection\\ \hline
        Analytical \cite{Eisenstein:1997jh,eisenstein1998baryonic}& No & {\color{red}\ding{55}}& Inflexible for non-trivial cosmologies\\
        Simple Gaussian \cite{vlah2016perturbation}& Yes & {\color{ForestGreen} \ding{51}}  & --- \\
        Polynomial fit \cite[app.~A]{hinton2016extraction}& No & {\color{red}\ding{55}} & Runge-phenomenon at relevant $k$ \\
        Cubic Spline fit \cite{2010MNRAS.404...60R} & No &  {\color{red}\ding{55}} & Clear distortions even in BAO range \\
        Cubic Spline fit \cite[app.~B]{blas2016time} & No & {\color{red}\ding{55}} & Strong hyperparameter dependence\\
        B-spline fit \cite[app.~A]{vlah2016perturbation} & Yes &  {\color{ForestGreen} \ding{51}} & --- \\
        Univariate spline fit [this work]& No & {\color{red}\ding{55}} & Strong hyperparameter dependence\\
        EH fit \cite[app.~B]{blas2016time} & Yes (only $\Lambda$CDM) & {\color{orange}\ding{51}} & Fails for non-trivial cosmologies\\
        Cubic Inflections \cite{audren2013conservative, brinckmann2019montepython}& Yes & {\color{ForestGreen} \ding{51}} & --- \\
        EH inflections \cite{brieden2021shapefit}& No & {\color{red}\ding{55}} & Updated version exists \\
        EH inflections (version 2) \cite{brieden2021shapefit} & Yes & {\color{ForestGreen} \ding{51}} & --- \\
        Hankel transform \cite[Sec.~2.2.1]{BOSS:2013sqq} & No & {\color{red}\ding{55}} & Divergence at small $k$ \\
        Fast sine transform \cite[Sec.~4.2]{chudaykin2020nonlinear} & Yes & {\color{ForestGreen} \ding{51}}& ---
    \end{tabular}
   \caption{Summary of the methods of de-wiggling and corresponding selection for the gold sample or reason for not including it. The \enquote{EH fit} method is included in the gold sample, but cannot be applied to all cosmologies, see \cref{sec:ShapeFit}.}
    \label{tab:dewiggles}
\end{table}

\section{ShapeFit} \label{sec:ShapeFit}

ShapeFit, first introduced by \cite{brieden2021shapefit}, is a new approach to analyze the power spectrum that is quickly gaining popularity in the cosmology community (see also \cite{brieden2021shapefit, brieden2021model, brieden2022model} for more details on the method, which involves computing a derivative of the large-scale de-wiggled power spectrum). This framework bridges the standard approaches of BAO and redshift space distortions (RSD) with the full-modeling approach. While the BAO and RSD are easily interpretable and can be used to isolate the features of the power spectrum from which the constraints are coming from, the full modeling approach extracts the most information from the power spectrum. ShapeFit can constrain parameters almost as tightly as the full-modeling approach while retaining the model-agnostic and interpretable nature of the standard BAO/RSD approaches. These properties make this approach a promising tool for current and future galaxy surveys. When combined with the BAO+BBN data, ShapeFit can provide an additional constraint on the Hubble constant $H_0$ by isolating information from matter-radiation equality; for a more detailed discussion, see \cite{schoneberg2022bao+}. ShapeFit relies on the broadband/BAO decomposition (as studied in \cref{sec:dewiggling}) and involves the evaluation of a slope: a derivative of the numerically-evaluated broadband component of the power spectrum. The evaluation of the derivative requires some care.

In practice, the differences between different de-wiggling methods as well as different ways of obtaining the derivative will combine to impart a systematic uncertainty on the  value of the ShapeFit slope parameter $m$, which is \textit{a priori} not straightforward to evaluate. In addition, systematic biases in $m$ do not necessarily propagate to a systematic shift in the recovered cosmology.
Therefore, we proceed here as follows. First, in \cref{ssec:theory_shapefit} we introduce the theoretical concepts at the heart of ShapeFit and in \cref{ssec:derivatives} we discuss in detail  different  implementations of numerical derivatives. Second, in \cref{sec:mfromdata} we  stress the importance for cosmological inference of matching the way $m$ is obtained from the data.   In \cref{ssec:mtrue} we then discuss how to mitigate the underlying ambiguity in the definition of $m$  and argue
that it is possible to find a way of calculating the slope $m$ that is highly consistent between different ways of numerically implementing the required derivative. Finally, in \cref{sec:results} we recommend a procedure to obtain a robust and consistent $m$ value from a given theory power spectrum and associate to it a systematic error budget.

\subsection{Theory}\label{ssec:theory_shapefit}

The ShapeFit parameter $m$ is effectively defined to be the slope of the underlying transfer function\footnote{The transfer function in this sense is just (the square root of) the ratio of the linear power spectrum and the primordial power spectrum.} for a given power spectrum at a pivot wavenumber $k_p$\,, compared to that of an arbitrarily chosen reference cosmology (ref). This ratio is mainly sensitive to the baryonic suppression and the equality scale, see \cref{sec:introduction} and \cite{brieden2021shapefit}.

Operationally, the slope $m$ can be computed by first computing the overall slope $\mu$ of the ratio of the linear power spectra and subsequently subtracting the slope of the ratio of the primordial power spectra $n$. We can write
\begin{equation}\label{eq:mdef}
\begin{aligned}
    \mu &= \frac{\partial \ln \left(\frac{P^\mathrm{no-wiggle}_\mathrm{lin}(k/s)}{P^\mathrm{ref,no-wiggle}_\mathrm{lin}(k)}\right)}{\partial \ln k} \raisebox{0.3cm}{$\left.\rule{0cm}{0.7cm}\right|_{k=k_p}$} \equiv \left.\frac{\partial \ln R(k,s)}{\partial \ln k} \right|_{k=k_p} \\
    n &= \frac{\partial \ln \left(\frac{P_\mathrm{prim}(k/s)}{P^\mathrm{ref}_\mathrm{prim}(k)}\right)}{\partial \ln k} \raisebox{0.3cm}{$\left.\rule{0cm}{0.7cm}\right|_{k=k_p}$}\\
    m &= \mu - n\\
\end{aligned}
\end{equation}
Here $s$ is the rescaling of the sound horizon scale, $s = r_d/r_d^\mathrm{ref}$. Note that for the almost scale-invariant power spectrum adopted in $\Lambda$CDM ($P_\mathrm{prim} \propto A_s k^{n_s-1}$), we simply have $n= n_s - n_s^\mathrm{fid}$\,. While the parameter that can be best measured in current data is $\mu=m+n$, future surveys might be able to distinguish between effects from $m$ and $n$. Given that currently large-scale-structure analyses are often used in combination with a prior on the slope $n$ (containing primordial information), we focus on the early-universe information contained in $m$.
However, we caution that this aspect of constraining primarily $\mu=m+n$ has to be taken into account when interpreting the constraints on $m$ for a fixed $n$, which are commonly shown in the literature (for example in \cite{Brieden:2022lsd,DESI:2024jis}, see also \cite{Jiang:2025ylr}). In what follows we refer to the ratio of the no-wiggle power spectra that appears in \cref{eq:mdef} (first line) as $R$.

While the computation of a slope at a point might sound like a trivial task, there are several aspects to be considered. First, since the scale $k_p$ is chosen to lie in the linear regime (avoiding non-linear corrections) where the baryon suppression begins, it unavoidably is impacted by the baryon acoustic oscillations, see also \cref{fig:schematic}. Therefore, it is important to define the slope in such a way that is insensitive to the precise nature and phase of the BAO for it to be a robust model-independent measure. This has been recognized already in the original ShapeFit paper \cite{brieden2021shapefit}, leading to the definition presented in \cref{eq:mdef}, where instead of a full power spectrum~$P_\mathrm{lin}(k)$ one uses a de-wiggled power spectrum~$P^\mathrm{no-wiggle}_\mathrm{lin}(k)$. However, the ambiguity of the separation of the first BAO peak and the power spectrum turnover, yielding the large variance among different de-wiggling algorithms, also affects $m$. This problem motivates looking beyond the point-wise definition of a derivative and to investigate methods that are less sensitive to small local fluctuations of the shape of the $R$ function.

It is important to note that, despite the value of $m$ being sensitive to the choice of the algorithm,  the methodology adopted in the original ShapeFit implementation \cite{brieden2021model}  which then is adopted for applications to state of the art data \cite{DESI:2024hhd, Lai:2024bpl}, has been shown to yield unbiased cosmological inference at very high precision \cite{PTchallenge}. To understand why this is the case, to quantify potential biases on $m$ for the next generation of surveys,  and to offer a transparent way to connect the value of $m$ to a theory model, it is important to take a deep dive into derivatives methods and their performance.

The reader less interested in numerical implementation details of such methods may skip to \cref{ssec:comparison}.

\subsection{Derivatives}\label{ssec:derivatives}

While the naive gradient method that computes the numerical finite difference between subsequent samples may intuitively appear to be the most accurate representation of the derivative, in practice the different de-wiggling methods do not agree very well on the local derivative, yielding a potentially large theoretical uncertainty.

In order to be robust to the differences between de-wiggling methods (see \cref{fig:dewiggle_comparison}), it is possible to define a more non-local version of the derivative. There are many such approaches, and we present them below, showing that they result in modest differences for the extracted slope.  
As such, we are trading sensitivity to the exact de-wiggling method with non-locality of the derivative computation.

\begin{figure}[h]
    \centering
    \includegraphics[width=0.49\linewidth]{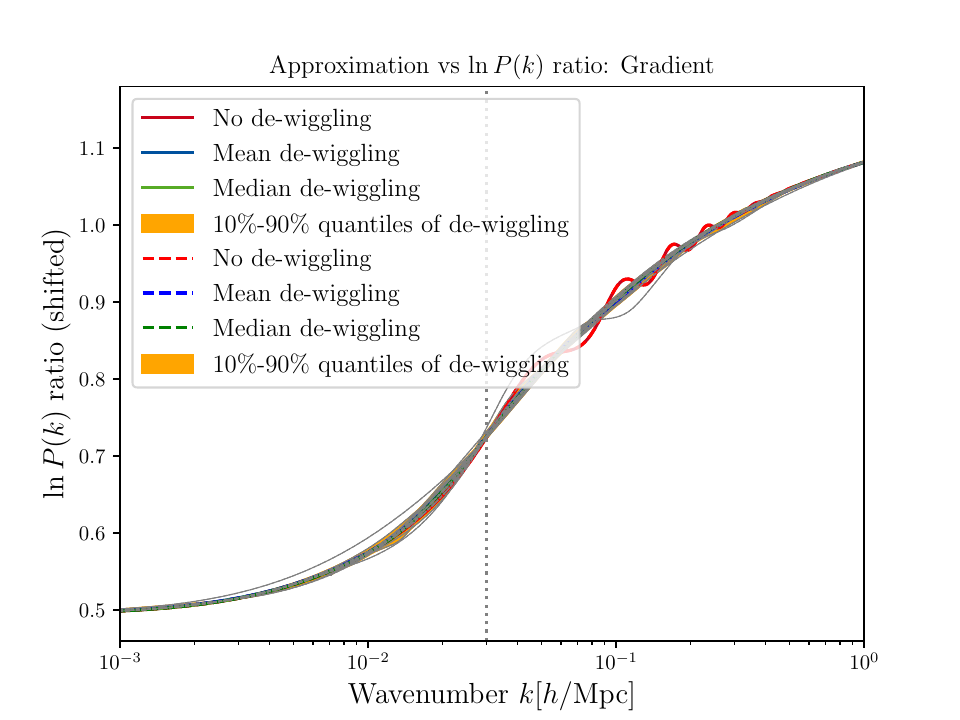} 
    \includegraphics[width=0.49\linewidth]{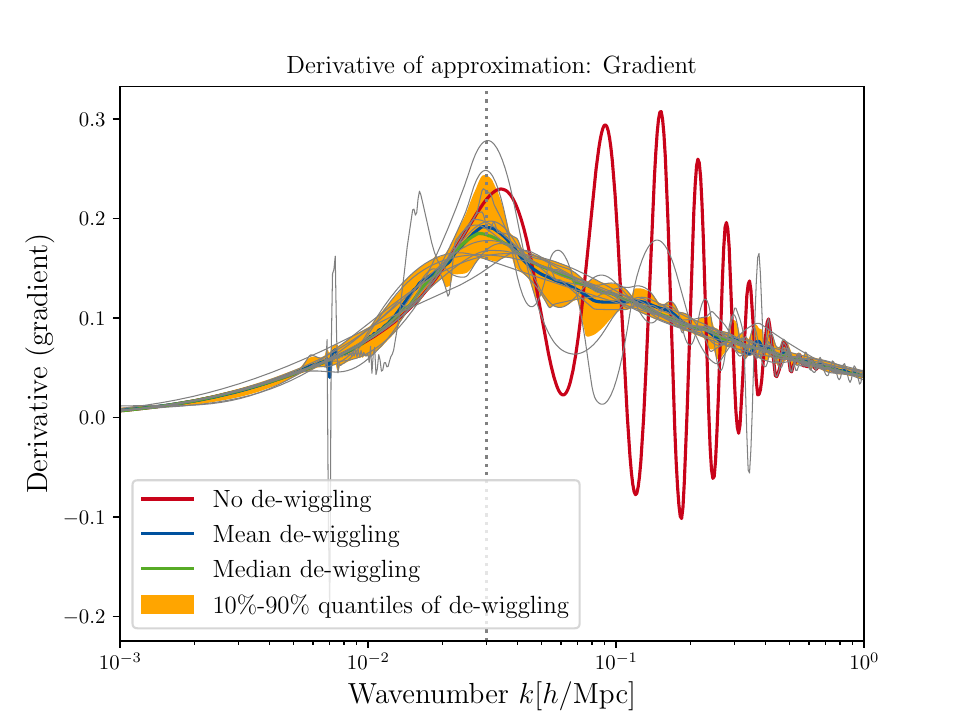} 
    \caption{Gradient method for the derivative. Left: Approximation of the logarithm of the ratio used for \cref{eq:mdef} (approximation in solid, true function in dashed lines). 
    Right: Derivative of the functional approximation (whose value at the pivot scale $k_p$ is taken as $m$).}
    \label{fig:gradient_deriv}
\end{figure}

This consideration can also be seen as a sort of
bias-variance trade-off for the final systematic error on~$m$. Say that for a given way of computing the derivative, the variance is simply given by the spread of the $m$ values between different de-wiggling methods (below we use all the 13 presented above). Then we can say that the bias instead arises from how strongly the function is approximated in order to compute the derivative. For example, a derivative method that always assigns the slope $1$ to any input will have a zero variance but a large bias. The gradient method has no bias, but a large variance, see \cref{fig:gradient_deriv}. 

Each derivative method is based on approximating the true function by a sufficiently smooth function in a given interval (see below for examples). The bias can be indirectly estimated from how different the smooth function approximation is from the true function ($R$) at/around the pivot point. Below we report as bias the typical difference between $R$ and its smooth function approximation in a range of scales around $k_p$ (around $k \in [0.02 h/\mathrm{Mpc},0.05 h/\mathrm{Mpc}]$), but we caution that it may not be directly interpreted as the bias for $m$ (see \cref{ssec:mtrue}). In tests and figures we compute the ratio between the fiducial/reference cosmology and the showcase cosmology, as listed in \cref{tab:cosmologies}.

\paragraph{Gradient}

This method is simply using the numerical second-order central finite difference formula for subsequent samples to compute a numerical approximation of the derivative. For the gradient method the functional approximation is simply the linear interpolation between two subsequent samples. Therefore, the approximation bias at any sampling location is by definition is zero (and sub-permille in a neighborhood around it, at least with sufficient sampling points), while the variance between different de-wiggling methods remains rather large. In particular, in this case, we find $m = 0.193 \pm 0.035$ (mean $\pm$ standard deviation among the 13 dewiggling methods), with a maximal spread of $\Delta m = \max(m)-\min(m) = 0.13$.

\paragraph{Spline Derivative}

In this method, a univariate spline (see \cref{app:ssec:univariate_spline}) of degree 5 is fitted to the power spectrum ratio, and its first derivative is evaluated at the pivot point. The fit can be performed globally, across all wavenumbers $k$ (global) or locally, only in a given $k$ range (local).
While the local spline derivative is more sensitive to local fluctuations, the global one might under-estimate real differences in the shape when they cannot be represented with a fifth-order polynomial.\footnote{We use a smoothing $\mathfrak{s}=3$ for the global version (in order to reproduce the original implementation) and $\mathfrak{s}=0.3$ for the local version (decreased to give a good fit still).} The global spline derivative is the one originally implemented in the context of ShapeFit for \cite{brieden2021shapefit}.

\begin{figure}[h]
    \centering
    
    \includegraphics[width=0.49\linewidth]{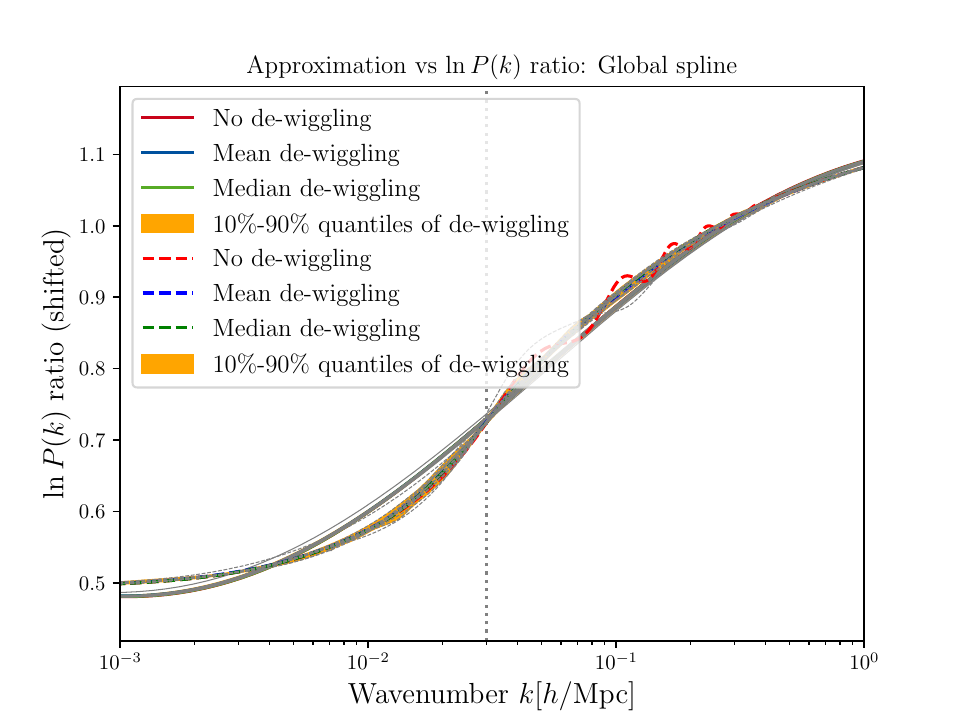} \includegraphics[width=0.49\linewidth]{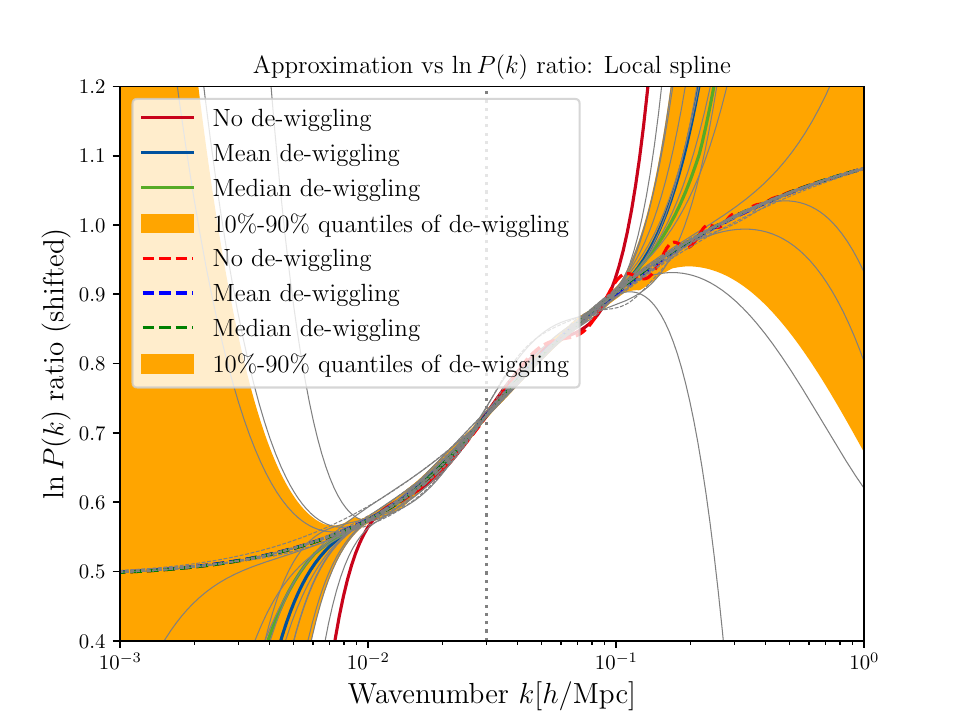}\\
    \includegraphics[width=0.49\linewidth]{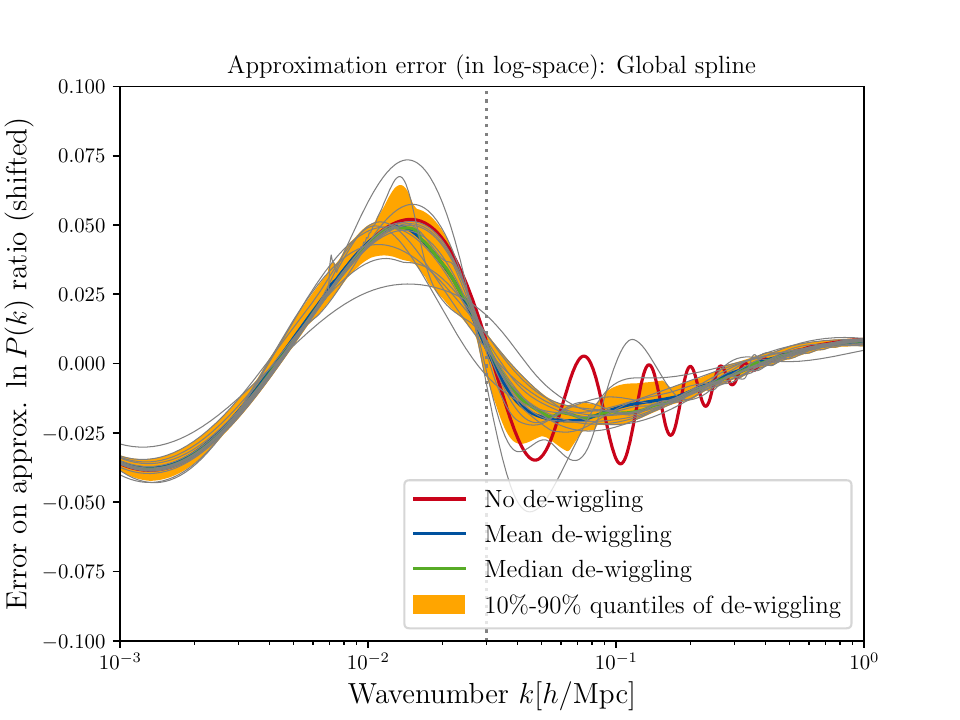} \includegraphics[width=0.49\linewidth]{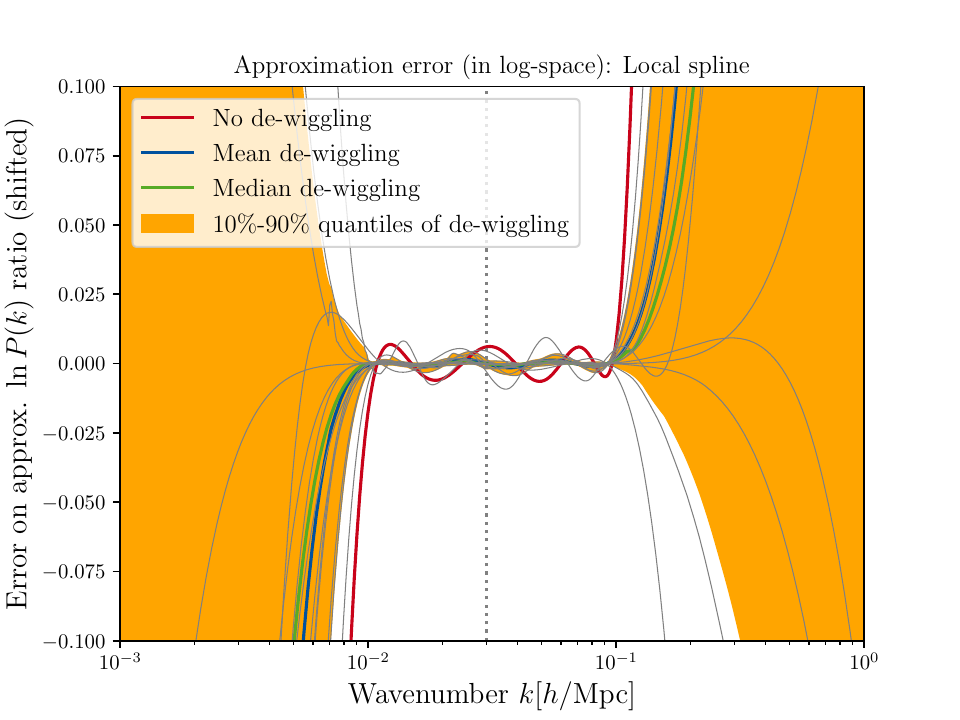}\\
    \includegraphics[width=0.49\linewidth]{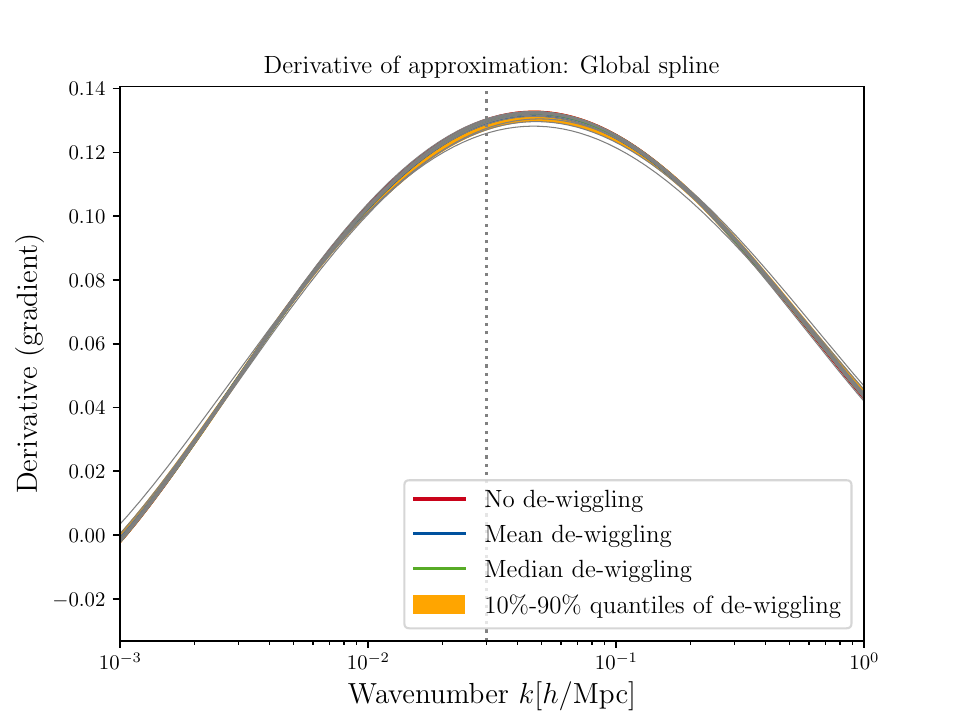} \includegraphics[width=0.49\linewidth]{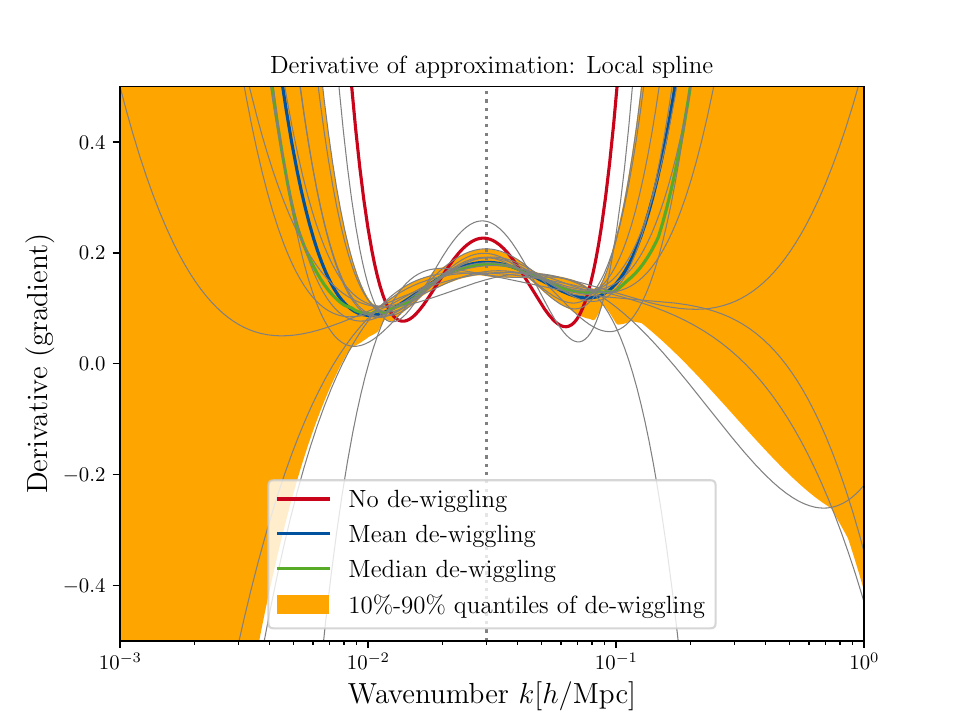} \\
    \caption{Top: Approximation of the logarithm of the ratio used for \cref{eq:mdef} (approximation in solid, true function in dashed lines). Middle: Relative error of that approximation. Bottom: Derivative of the functional approximation (whose value at the pivot scale $k_p$ is taken as $m$). Left: Global spline. Right: Local spline (between $k_\mathrm{min}=0.01 h/\mathrm{Mpc}$ and $k_\mathrm{max}=0.1 h/\mathrm{Mpc}$).}
    \label{fig:spline_derivative}
\end{figure}

The results are displayed in \cref{fig:spline_derivative} for both the local and global versions. We find  for the global spline $m = 0.1293 \pm 0.0013$ and a maximal deviation of $\Delta m = 0.0046$ and for the  local spline method $m = 0.1856 \pm 0.028$ and $\Delta m = 0.11$.
In this case, we can see that the functional approximation through a spline does introduce some bias, especially for the global method: the middle panels of \cref{fig:spline_derivative} show that the approximation error is around $7\%$ for the global spline method and only around $<1\%$ for the local one. We can immediately observe the trade off between bias and variance: the global spline method has smaller variance but larger bias.

\begin{figure}[h]
    \centering
    
    \includegraphics[width=0.49\linewidth]{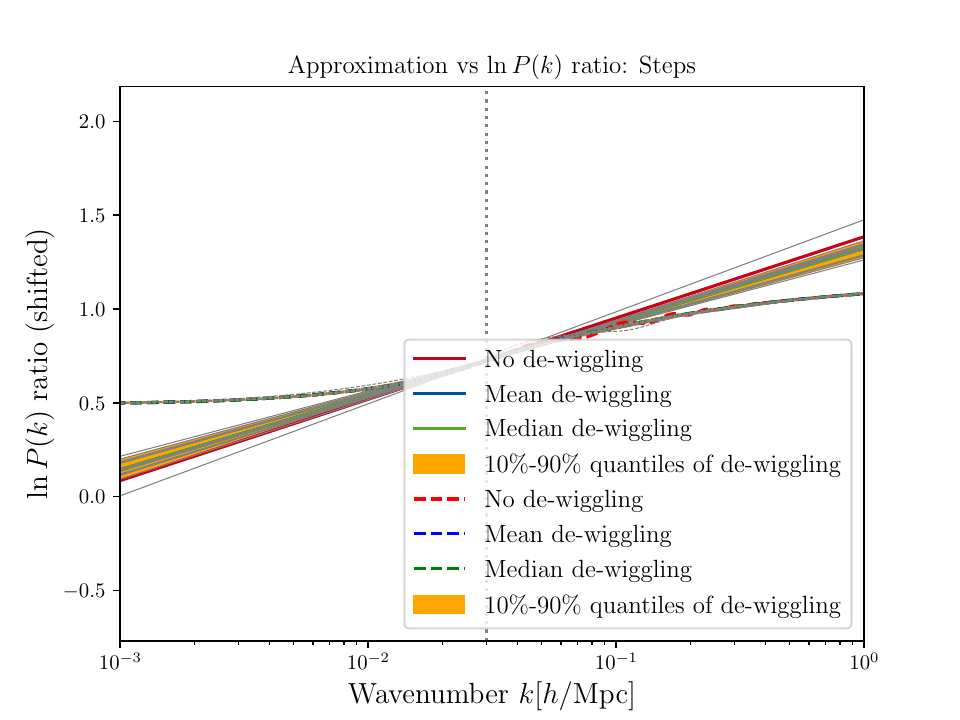} 
    \includegraphics[width=0.49\linewidth]{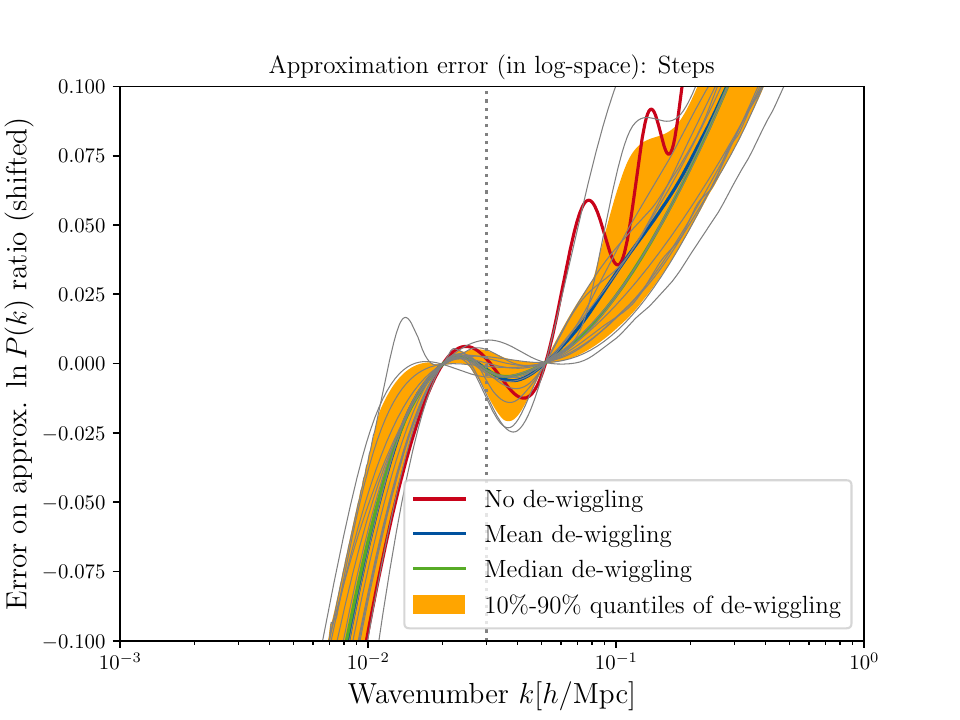} 
    \includegraphics[width=0.49\linewidth]{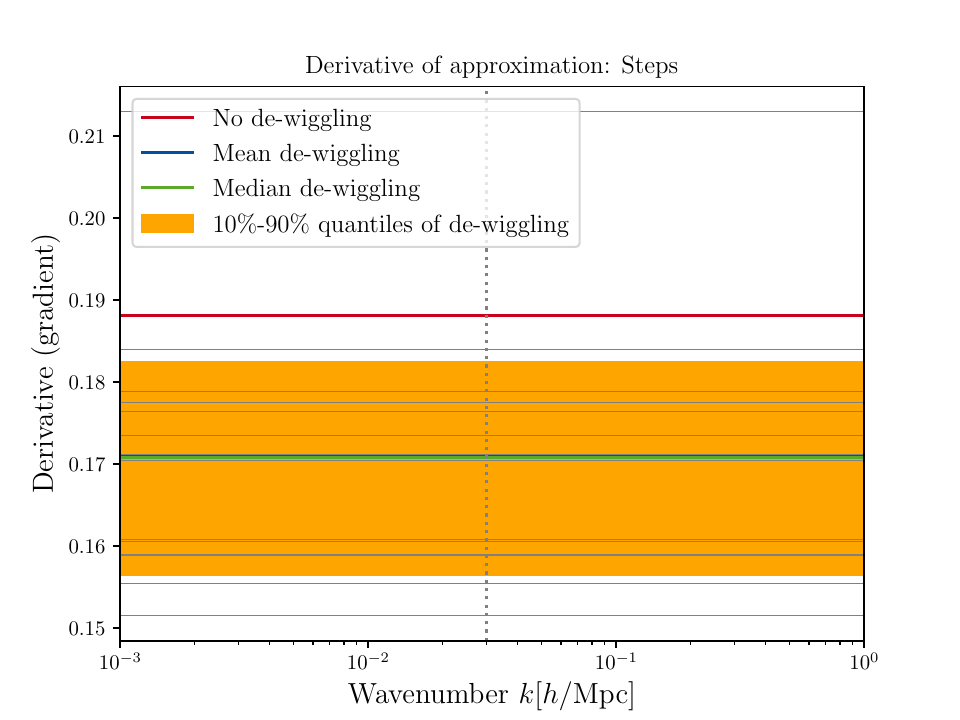} 
    \caption{Top left: Approximation of the logarithm of the ratio used for \cref{eq:mdef} (approximation in solid, true function in dashed lines). Top right: Relative error of that approximation. Bottom: Derivative of the functional approximation (whose value at the pivot scale $k_p$ is taken as $m$). Steps method with $\Delta \ln k = 0.6$.}
    \label{fig:steps_derivative}
\end{figure}
\paragraph{Linear derivative (Steps)}

This method simply selects two $k$ values $k_1$, $k_2$ that are not subsequent samples and uses these to compute the slope using the usual finite difference approximation:
\begin{equation}
    \frac{\ln R(k_2) - \ln R(k_1)}{\ln k_2 - \ln k_1}
\end{equation}
where $R$ is the ratio of the (shifted) power spectra required for \cref{eq:mdef}. In \cref{fig:steps_derivative} we show the corresponding results. Very similarly to the local spline method, this method trades off variance with bias. In order to reach a $1-2\%$ approximation error, the steps have to be chosen relatively closely to the pivot wavenumber (in this case $\Delta \ln k = \ln k_2 - \ln k_1 = 0.6$). The resulting spread is $m=0.172 \pm 0.015$ with $\Delta m = 0.061$. Comparing to the local spline method, we observe about half the variance while still keeping the bias at less than $2\%$. We show different step sizes in \cref{fig:derivative_comparison,ssec:mtrue} below.

\begin{figure}[h]
    \centering
    
    \includegraphics[width=0.49\linewidth]{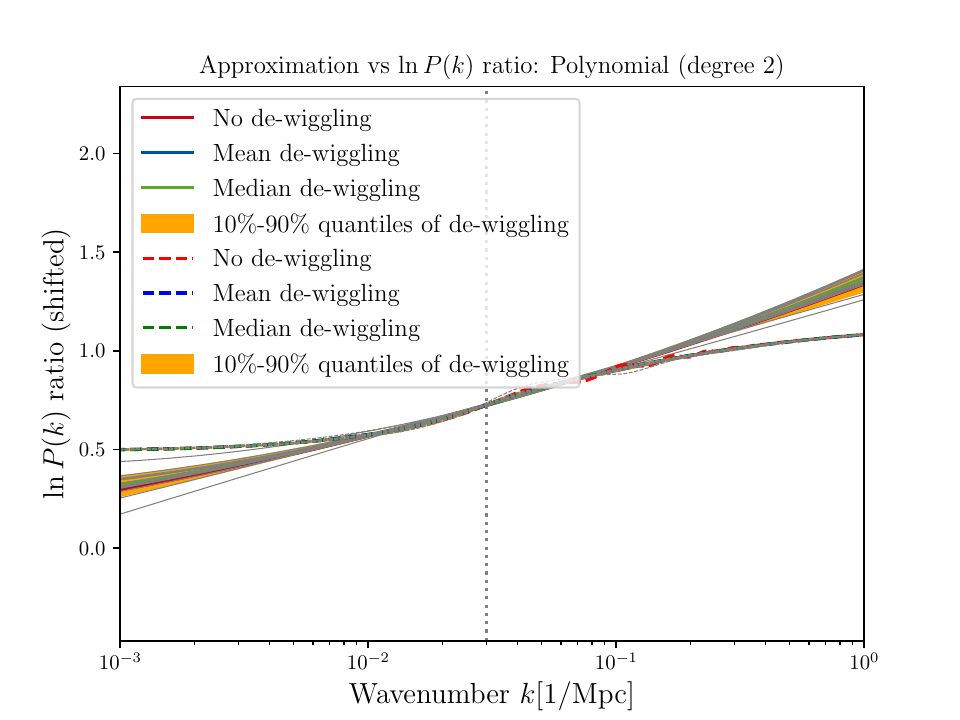} 
    \includegraphics[width=0.49\linewidth]{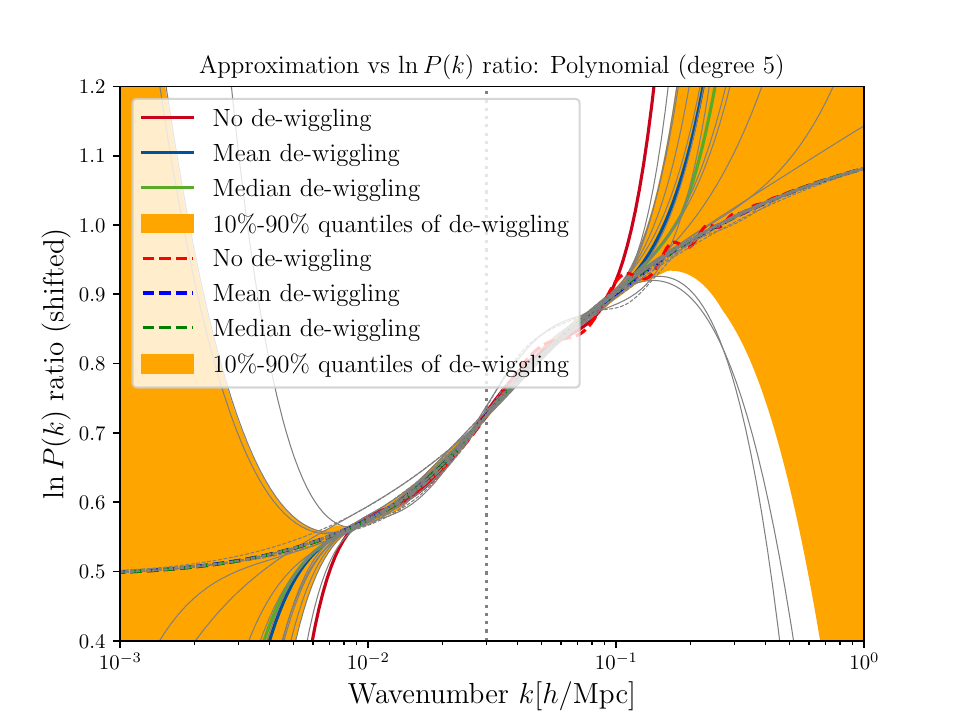} 
    \includegraphics[width=0.49\linewidth]{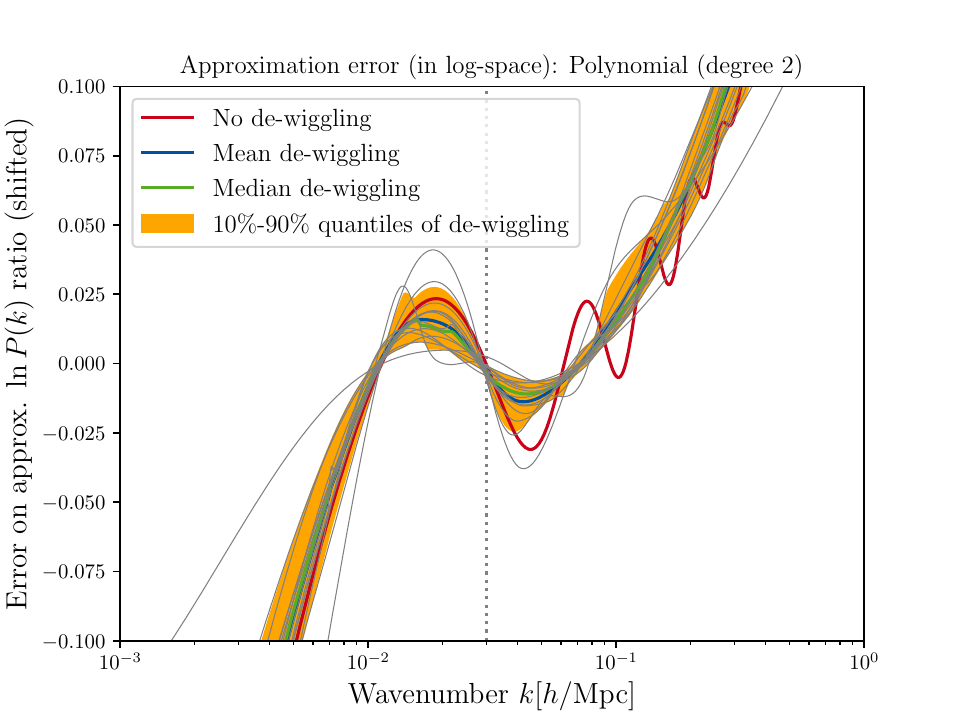} \includegraphics[width=0.49\linewidth]{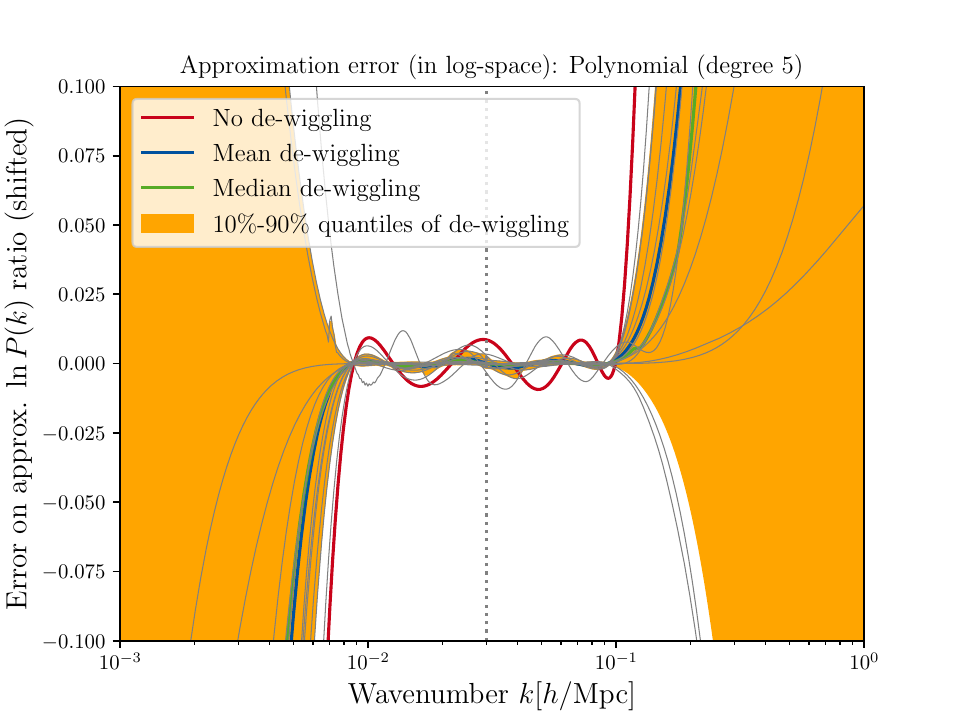}\\
    \includegraphics[width=0.49\linewidth]{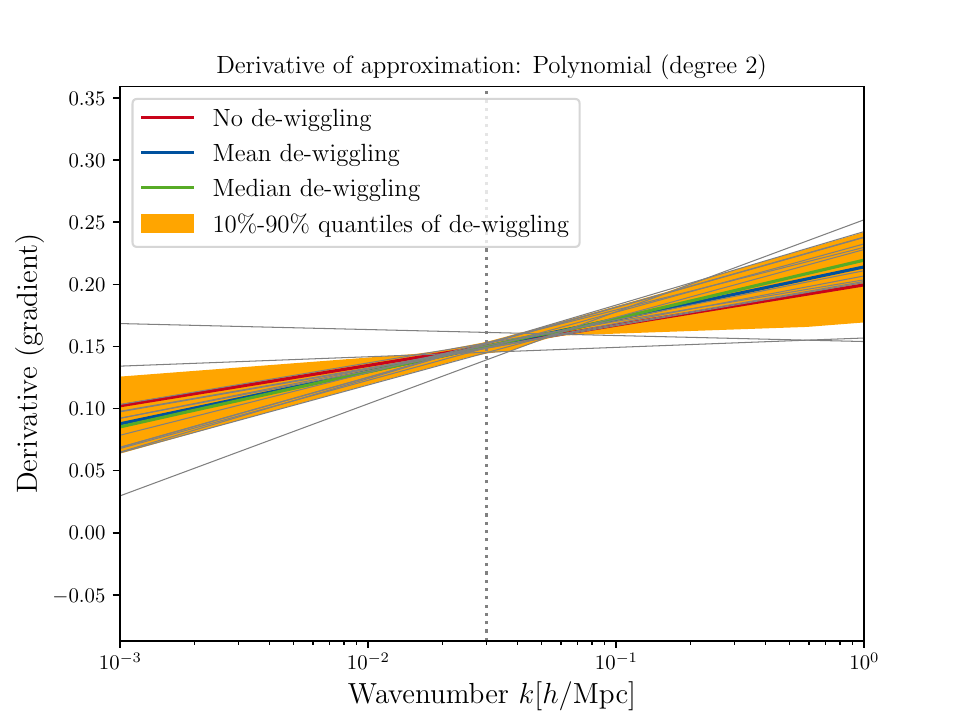} \includegraphics[width=0.49\linewidth]{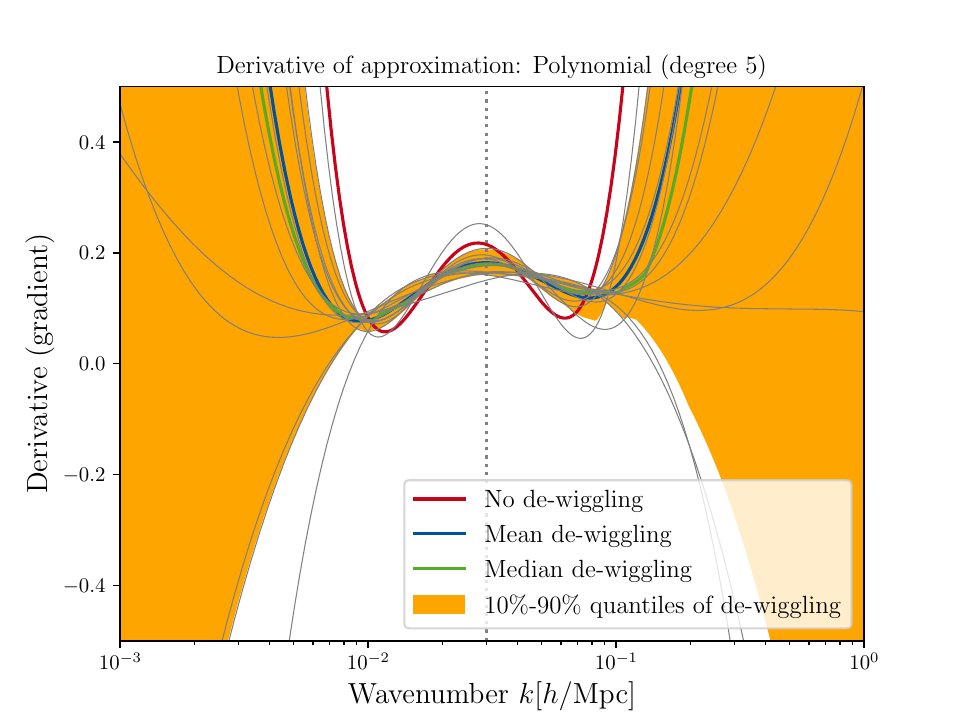}  \\
    \caption{Top: Approximation of the logarithm of the ratio used for \cref{eq:mdef} (approximation in solid, true function in dashed lines). Middle: Relative error of that approximation. Bottom: Derivative of the functional approximation (whose value at the pivot scale $k_p$ is taken as $m$). Left: Polynomial of degree 2. Right: Polynmoial of degree 5.}
    \label{fig:polynomial_derivative}
\end{figure}

\paragraph{Polynomial Derivative}

In this method the ratio $R$ is fitted over a range of scales by a polynomial of degree $d$, and the derivative of this polynomial evaluated at the pivot point $k_p$ is used to determine the slope. We show the results in \cref{fig:polynomial_derivative} for a fit between $k_\mathrm{min}=8 \cdot 10^{-3}$/Mpc and $k_\mathrm{max}=0.1$/Mpc. The results are $m=0.1499 \pm 0.0047$ and $\Delta m = 0.022$ for a polynomial of degree $d=2$, $m = 0.1751 \pm 0.016$ and $\Delta m = 0.07$ for a polynomial of degree $d=3$ (not shown here), and $m=0.1842 \pm 0.025$ and $\Delta m=0.1$ for a polynomial of degree $d=5$. We see that higher order polynomials naturally increase variance, while they also reduce bias (the bias is 3\% for $d=2$, 1\% for $d=3$ (not shown here), and $<1\%$ for $d=5$).

\begin{figure}[h]
    \centering
    
    \includegraphics[width=0.49\linewidth]{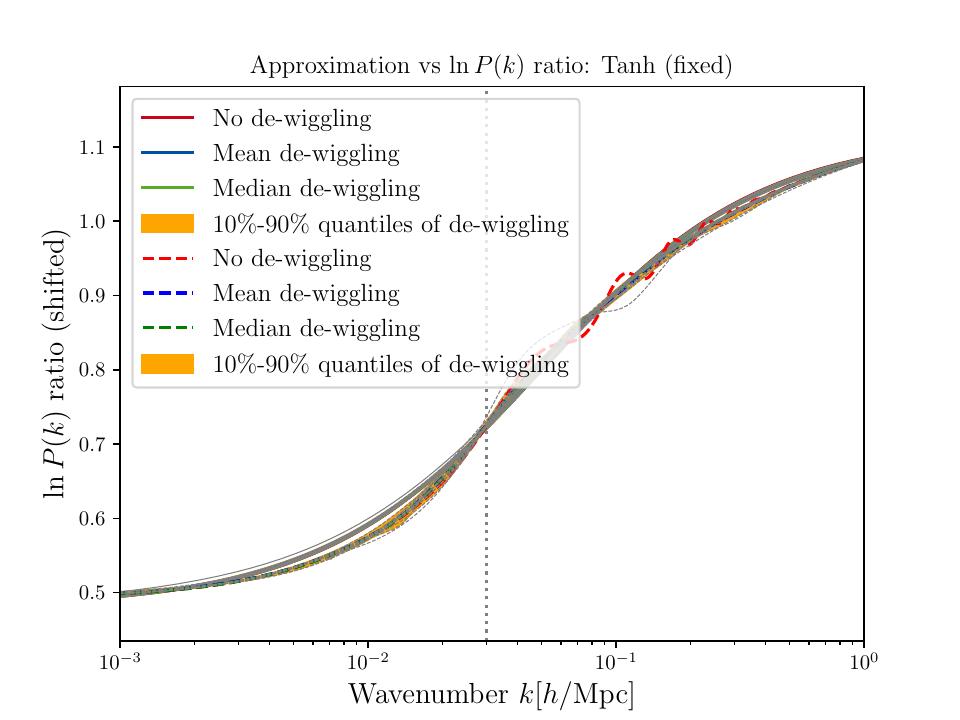} \includegraphics[width=0.49\linewidth]{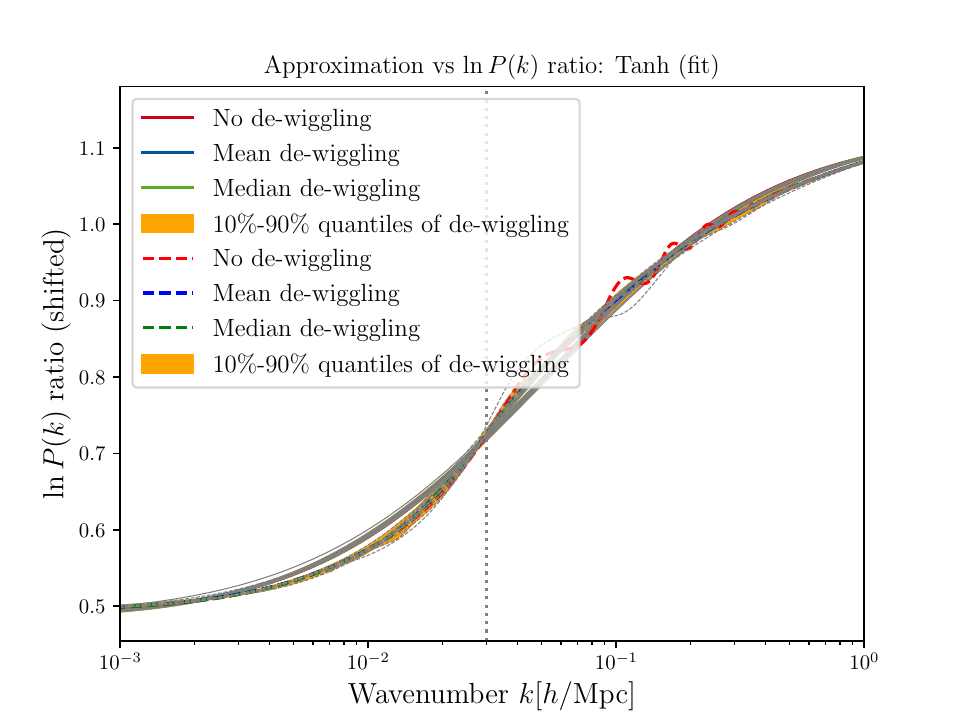}\\
    \includegraphics[width=0.49\linewidth]{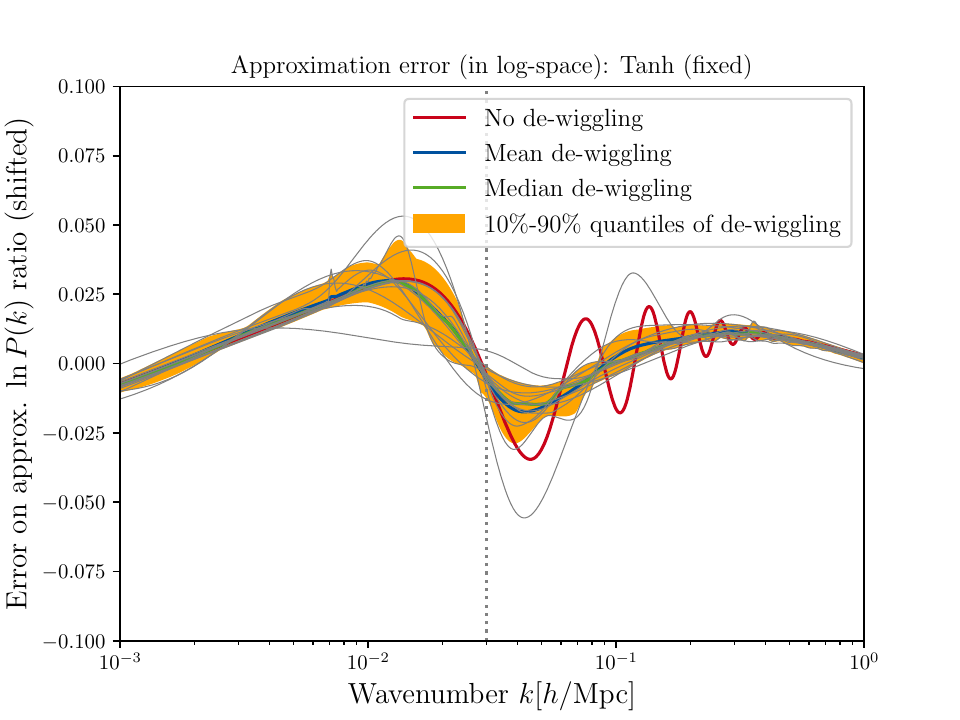} \includegraphics[width=0.49\linewidth]{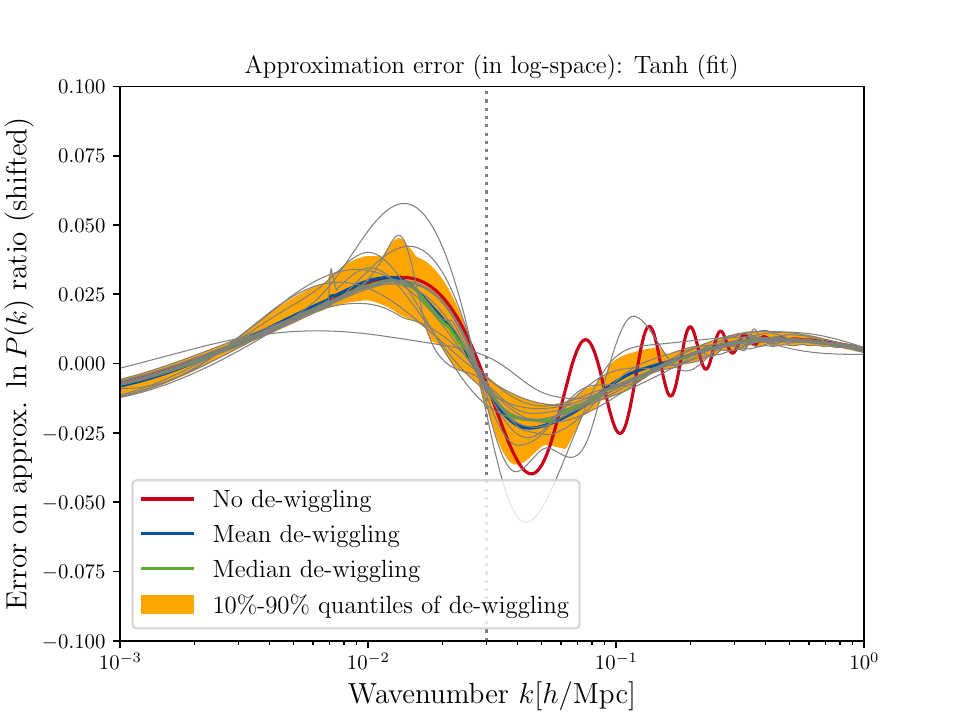}\\
    \includegraphics[width=0.49\linewidth]{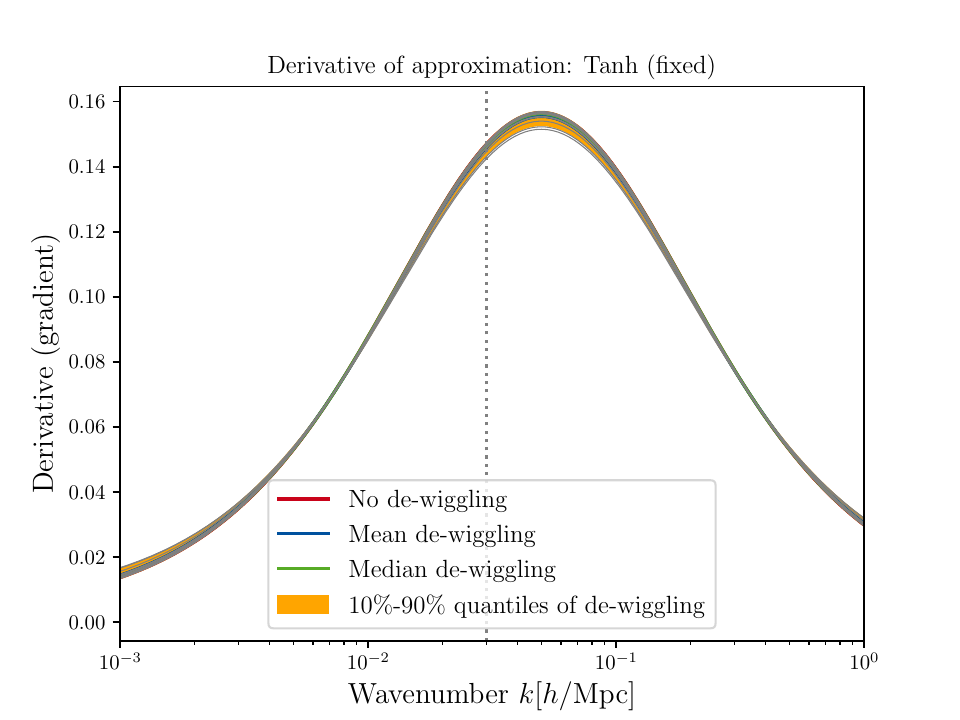} \includegraphics[width=0.49\linewidth]{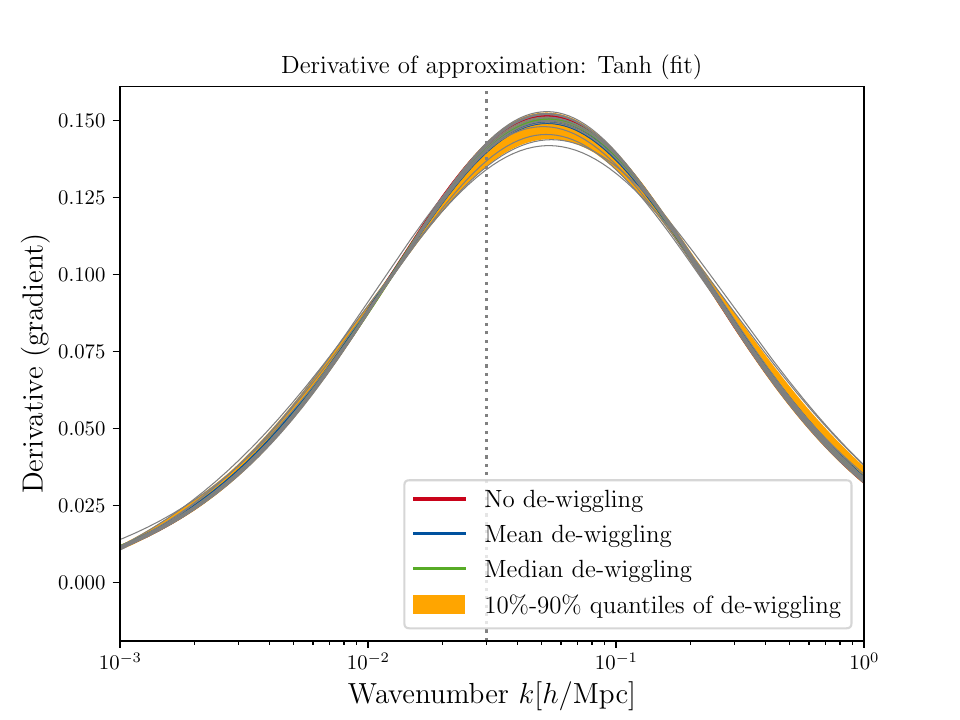} \\
    \caption{Top: Approximation of the logarithm of the ratio used for \cref{eq:mdef} (approximation in solid, true function in dashed lines). Middle: Relative error of that approximation. Bottom: Derivative of the functional approximation (whose value at the pivot scale $k_p$ is taken as $m$). Left: Tanh fitting method (fixed $a$ and $k_t$). Right: Tanh fitting method with all parameters free to vary.}
    \label{fig:tanh_derivative}
\end{figure}

\paragraph{Hyperbolic Tangent}

The shape of the baryonic suppression can be approximated by a hyperbolic tangent curve \cite{brieden2021shapefit}. This is the reason why in the data analysis pipeline a common choice is to rescale the power spectrum template with such a curve. 
This method thus stands out among the others as one particularly close to how the data is analyzed, see \cref{ssec:mtrue} for further discussion on this point. One simply fits the ratio $R$ with a function of the form
\begin{equation}\label{eq:tanh_fit}
    \ln R = p_0+p_1 \cdot (\ln k/k_t)+p_2/a \cdot \tanh[a (\ln k/k_t)]
\end{equation}
Here the $p_i$, $k_t$, and $a$ are the parameters of the family of curves. There are two variations of this method. In the first method, we take $k_t$ and $a$ to be those values advocated for in \cite{brieden2021shapefit}, i.e. $k_t = k_p = 0.03 h/\mathrm{Mpc}$ and $a = 0.6$. In the second method we leave these two also as free parameters of the fit, though importantly we still evaluate the derivative of \cref{eq:mdef} at the same location.

The results for both methods are shown in \cref{fig:tanh_derivative}. Particularly interesting is that in this case even the non-de-wiggled ratios return the same slope as all other methods. This way of computing the derivatives is therefore robust to baryonic oscillations, but it may be prone to bias in the derivative as the bias in $R$ increases steeply away from the pivot point,
see the middle panels of \cref{fig:tanh_derivative}. We find the low variance estimates of $m = 0.1460 \pm 0.0014$ and $\Delta m = 0.0049$ for the fixed parameters case ($a$ and $k_t$), and $m = 0.1402 \pm 0.0026$ and $\Delta m = 0.0087$ when  these are left free.
The approximation has a roughly 5\% maximum bias in either case.

\begin{figure}
    \centering
    \includegraphics[width=0.49\linewidth]{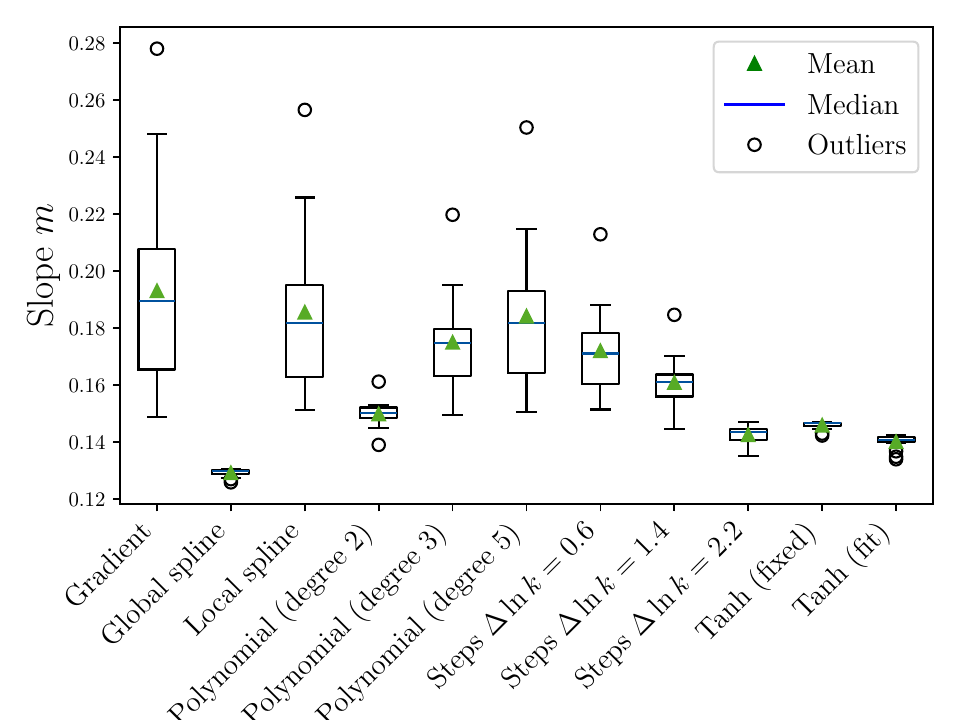}
    \includegraphics[width=0.49\linewidth]{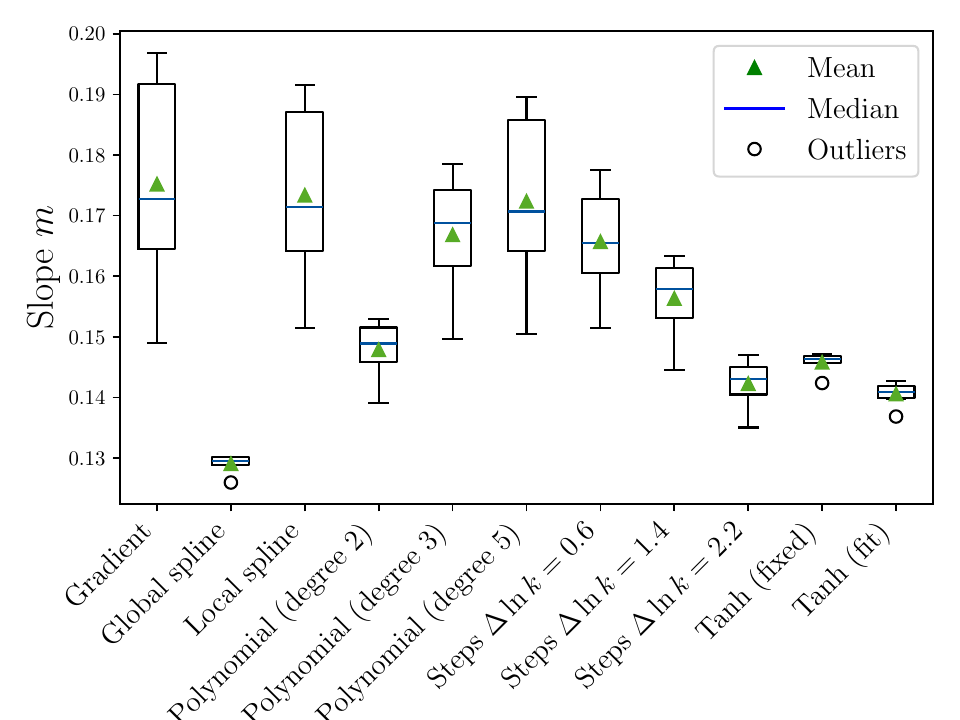}
    \caption{Comparison of the $m$ values found for each different way of computing the derivative listed in \cref{sec:ShapeFit}. Left: using all de-wiggling methods of \cref{tab:dewiggles}. Right: using only the de-wiggling methods of the \enquote*{gold sample} of \cref{tab:dewiggles}. The triangles represent the median, the blue horizontal line the mean, the empty boxes the standard deviation, the error bars the 10-90\% quantiles, and the empty circles outliers beyond that.}
    \label{fig:derivative_comparison}
\end{figure}

\subsubsection{Comparing performances of different derivatives methods}\label{ssec:comparison}

A direct comparison between the different methods for calculating the derivative can be found in \cref{fig:derivative_comparison}. It is notable to see that many of the more unbiased derivative methods (such as e.g., the gradient method, the local spline, the polynomial of degree 5, the steps method with very local support) return values of the slope around $\sim 0.17$ with a large variance, while other
methods (such as e.g., the global spline, the polynomial of degree 2, the steps method with large stepsizes, the tanh methods) return smaller slopes, albeit with less variance. For reference statistical error-bars on $m$ from current surveys are $\sim 0.03-0.05$, i.e., smaller than the observed variance.

It is evident that considering only the gold sample of smoothing methods strongly reduces the spread (notice the reduced $10-90\%$ intervals and the smaller number of outliers). While the gold sample of smoothing methods therefore is more homogeneous in the recovered $m$ values, the remaining scatter is still somewhat large for many of the derivative methods.

\subsection{Matching the way \texorpdfstring{$\boldsymbol{m}$}{m} is extracted from the data}
\label{sec:mfromdata}
The requirement for obtaining unbiased estimates of the underlying cosmological parameters is not identical to the requirement of having  small $m$ variations across different  dewiggling and derivative implementations, as we show below. 
 
As it is customary, algorithms and analysis methods are tested on mock data, where the true underling cosmology is known. In this case it can be checked how biased a particular implementation $I$ is with respect to a certain data analysis pipeline~$p$. The theoretically determined slope $m^\mathrm{theory}$ generally depends on the implementation $I$ and the cosmological model parameters $\theta$ under consideration. On the other hand, the slope extracted from the mock data $m^\mathrm{data}$ depends on the cosmological parameters (and settings) used to generate the mock data $\theta^\mathrm{mock}$ and the slope extraction pipeline~$p$. The requirement of having an un-biased recovery of the cosmological parameters then requires to minimize the recovery bias
\begin{equation}
    B(I |\theta^\mathrm{mock}, p) = |m^\mathrm{theory}(\theta^\mathrm{mock}|I) - m^\mathrm{data}(\theta^\mathrm{mock}|p)|~.
    \label{eq:recoverybias}
\end{equation}

Note that by definition $m^\mathrm{theory}(\theta^\mathrm{fid}|I) = 0$ for all implementations, and usually $m^\mathrm{data}(\theta^\mathrm{fid}|p) = 0$, so to find the best implementation, cosmologies different than the fiducial should be considered. 

The best implementation $I$ will minimize this bias for a range of cosmological parameters $\theta^\mathrm{mock}$.\footnote{ Note that while the performance of any given method $I$ may depend on cosmology, it is customary to assume that this dependence is weak, as long as the cosmologies to be considered are not heavily disfavored by the data.}
In other words, measuring the true underlying value of the parameter $m$ does not matter for cosmological inference: $m^{\rm theory}$ and  $m^{\rm data}$ could both have a large bias, but it is irrelevant   as long as $B$ of \cref{eq:recoverybias} is kept well below the statistical errors.

For example, the data analysis for ShapeFit uses a template that rescales the linear power spectrum via a multiplicative factor (see for example \cite{Maus:2024dzi})
\begin{equation}\label{eq:data_mdef}
    P^\mathrm{adjusted}_\mathrm{template}(k, m, n) = P^\mathrm{fid}_\mathrm{lin}(k) \exp\left\{\frac{m}{a} \tanh\left[a \ln\left(\frac{k}{k_p}\right)\right] + n \ln \left(\frac{k}{k_p}\right)\right\}\,.
\end{equation}
with $a=0.6$ and $k_p=0.03h/\mathrm{Mpc}$. The adjusted template is then rescaled with the usual BAO parameters (e.g., $\alpha_\parallel$ or $\alpha_\perp$) and subsequently compared to the data. Comparing \cref{eq:data_mdef,eq:tanh_fit} it is evident that the method that minimizes the bias of \cref{eq:recoverybias} is most likely\footnote{The template of \cref{eq:data_mdef} is typically transformed to redshift space and subjected to well known effects like redshift space distortions, the Alcock-Paczy\'nski effect, and others. Therefore, while we expect that these additional steps do not change which implementation is most consistent with the data analysis pipeline, we have not conclusively proven this. We leave such a proof using the full data analysis pipeline for future work.} the \textbf{tanh (fixed)} derivative method, which gives closely consistent results regardless of the de-wiggling method.

By using a derivative method that is insensitive to the dewiggling algorithm and that is tuned to the way the data and treated, the current ShapeFit implementation has been found to be unbiased for the cosmologies where it has been tested. 
Different choices that are also consistent with the data analysis pipeline might overall have additional advantages.

\subsection{What is the true value of \texorpdfstring{$\boldsymbol{m}$}{}? Removing ambiguity to  maximize consistency.}\label{ssec:mtrue}

 At first glance, the various ways to de-wiggle the power spectrum from \cref{sec:dewiggling} and to extract the slope $m$ of \cref{sec:ShapeFit} could all appear to be \textit{a priori} equally valid, as they are all just different ways of numerically implementing \cref{eq:mdef}. 
 Yet, because of the BAO peak/power spectrum turnover ambiguity,  for a given cosmological model there  seem to be not one \enquote{true} value of the slope $m$. While unimportant for cosmological inference purposes,
this is unsatisfactory: if we wish to use $m$ like other compressed variables (e.g., $\alpha_\parallel$,$\alpha_\perp$ etc.) it would be very useful to associate a unique value of $m$ to any theory model (given a fiducial cosmology).

Is it possible to find consistency among implementations of dewiggling and derivative? We offer a possible solution next, which involves some post-processing of the ratio $R$ before taking the derivative.
We begin by noting that  since for the cosmologies where it has been tested \textbf{tanh (fixed)} is robust to dewiggling methods and suitable for cosmological inference, we can  take this method as benchmark and adopt as $m^{\rm true}$ the   median of \textbf{tanh (fixed)} method for different de-wiggling algorithms.

\subsubsection{Post-Processing Filters} \label{ssec:postprocessing}

Given that the differences between the smoothing methods of \cref{sec:dewiggling} are typically of the order of 1-5\% and highly oscillatory in nature, one could imagine that a method of further smoothing the de-wiggled power spectrum ratios might aid in getting more consistent slope values. We particularly focus on the windowed average and the Savitzky-Golay filter \cite{esteban2019global} -- The idea is that these filters should typically mostly conserve the overall broadband shape, and simply help reduce the variance between the different de-wiggling methods.

To define the window length of each filter (over which the smoothing is performed) we choose a given physical size $\Delta \ln k$ and convert it into a length in terms of indices as:
\begin{equation}
    N_\mathrm{window} =\left\lfloor N \times \frac{\Delta \ln{k}}{\ln{k_{max}} - \ln{k_{min}}} \right\rfloor \label{eq:window-length}
\end{equation}
where $N$ corresponds to the number of elements in the wavenumber array and $N_\mathrm{window}$ is the number of indices used. 

\begin{figure}
    \centering
    \includegraphics[width=0.5\linewidth]{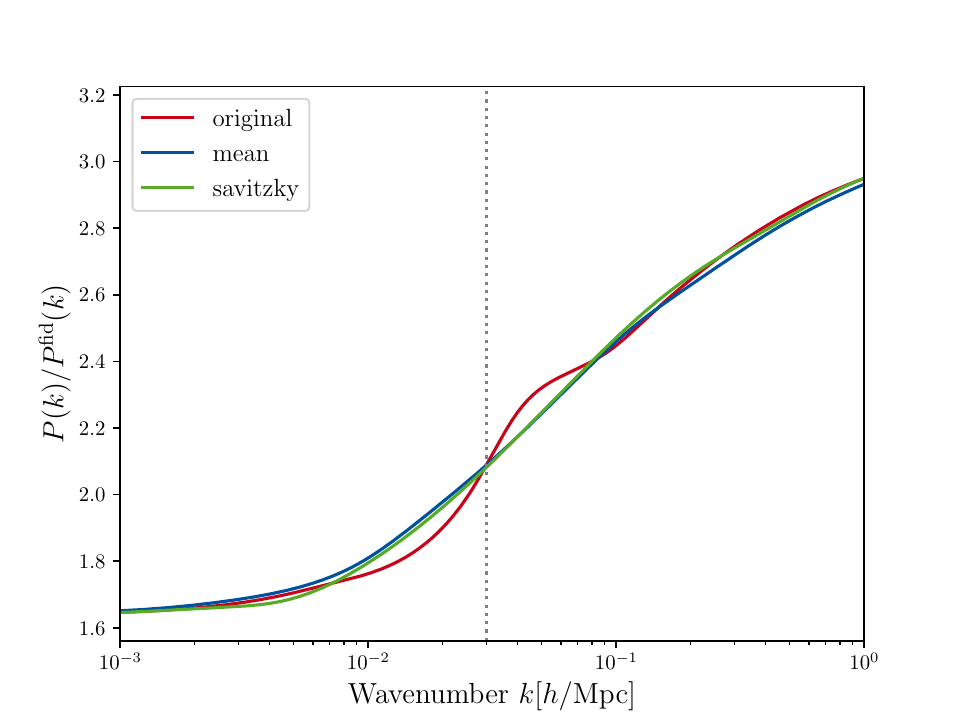}
    \caption{Impact of post-processing smoothing of the ratio $R$ by either a running average (mean) or Savitzky-Golay (savitzky) filtering procedure. We show the example of the univariate spline fit to highlight the impact of the post-processing, which is less visible for other de-wiggling methods.}
    \label{fig:impact_of_smoothing}
\end{figure}

We schematically show the impact of the smoothing method on the ratio between the power spectra in \cref{fig:impact_of_smoothing}. In particular, it is evident that the broadband shape is mostly preserved, while residual oscillations are effectively removed. In what follows, we smooth over 2.5 decades in $\ln k$ for the Savitzky-Golay filter and over 1.3 decades in $\ln k$ for the mean filter. These values are optimized by hand to reduce the impact on the extracted broadband filter while strongly suppressing residual oscillations.

\begin{figure}
    \centering
    \includegraphics[width=1.0\linewidth]{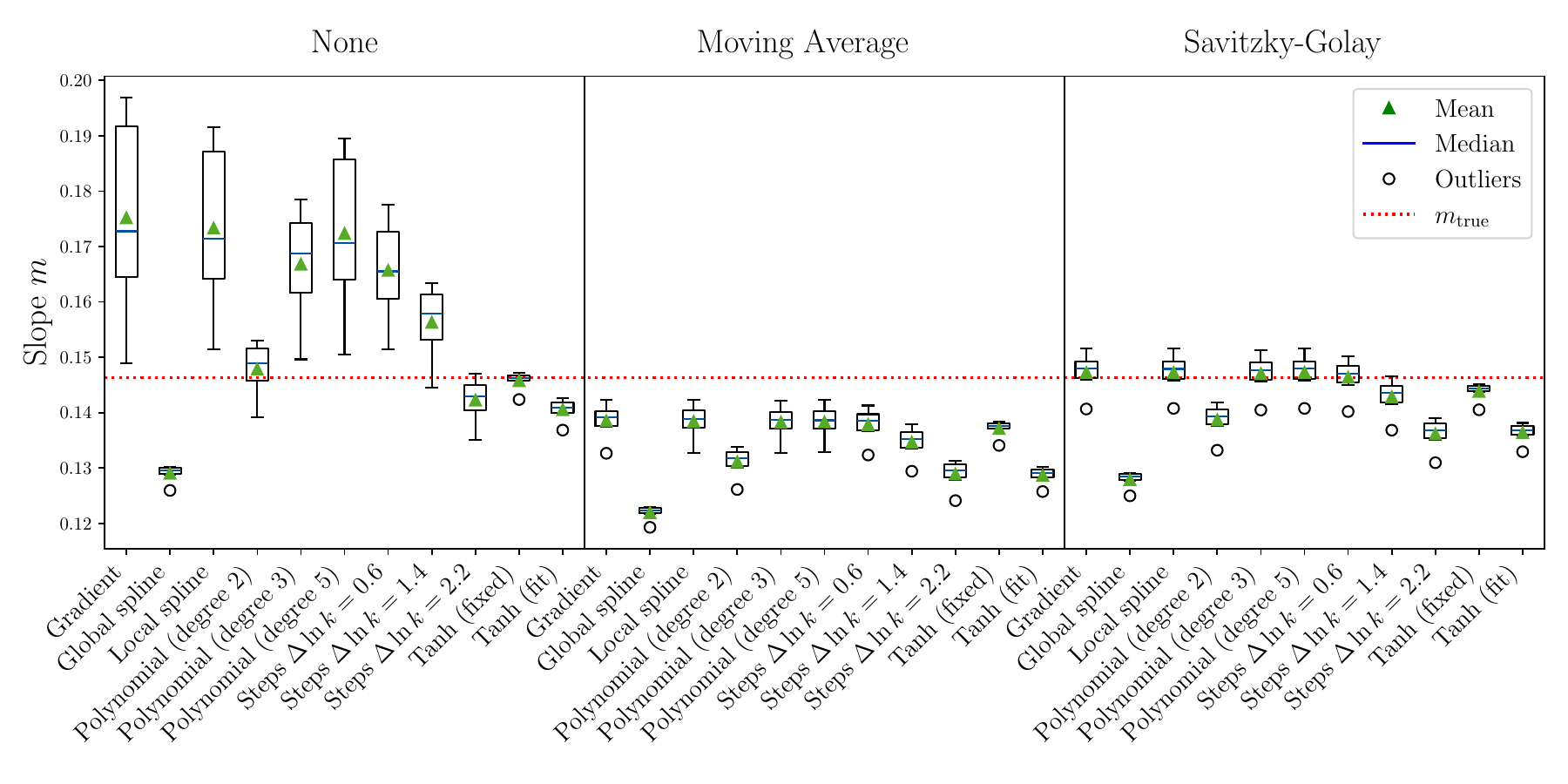}
    \caption{Comparison of the different derivative algorithms for different levels of post-processing. The red line shows the $m_\mathrm{true}$ defined in \cref{ssec:mtrue}. For reference, the statistical error on $m$ from current surveys is $\sigma^\mathrm{stat}_m \sim 0.03-0.05$.}
    \label{fig:comparison_smoothings}
\end{figure}

We compare different post-processing methods in \cref{fig:comparison_smoothings} by their impact on the variance and mean value of the extracted slopes for each different way of computing the derivative. In particular, we observe that the variance of all derivative computations is significantly reduced. There is, however, also a slight impact on the extracted mean value, with the moving average
preferring slightly lower mean values (by $\Delta m\simeq 0.01$) than the Savitzky-Golay filter.

It is evident from \cref{fig:comparison_smoothings} that only the Savitzky-Golay smoothing results in both (1) a broad consistency between most local methods for obtaining a derivative and (2) a mostly un-biased recovery of the slope (for these particular cosmologies, see \cref{sec:results} for different cosmologies).

The resulting systematic error in the context of cosmological inference will be discussed in \cref{sec:results}.

\subsubsection{Robust \texorpdfstring{$\boldsymbol{m^{\rm true}}$}{m\_true} estimate}

\Cref{tab:m_syst} lists the relative $m$ deviations from the  reference \textbf{tanh (fixed)} method. 
The scatter is across  the different dewiggling algorithms.
The derivative/postprocessing combinations  where the  shift is smaller than the scatter are marked in bold.

Not only do we see the broad consistency of various (local) derivative methods with the Savitzky-Golay filter, we also see that they agree nicely on the expected \textit{relative} variance on $m$, which is $\sigma^\mathrm{SG}_{m,\mathrm{syst}} \simeq 0.023 |m|$.
As long as the dewiggled power spectrum is computed with any of the golden sample method and  the ratio $R$ of \cref{eq:mdef} is smoothed using a  Savitzky-Golay filter as described in \cref{ssec:postprocessing}, any local derivative method will yield a robust $m^{\rm true}$ estimate with an associated  scatter (arising  purely from different algorithm choices) of $\sigma^\mathrm{SG}_{m,\mathrm{syst}}\simeq 0.023 |m|$.

\begin{table}[t]
    \centering
    \begin{tabular}{c| c c c}
        Method & $\Delta m/m$ & $\Delta m/m$ & $\Delta m/m$\\
         & (no post-proc.) & (moving average) & (Savitzky-Golay)\\ \hline
        Gradient  & $0.197 \pm 0.118$ & $-0.053 \pm 0.021$ & $\mathbf{0.006 \pm 0.023}$ \\
        Global spline  & $-0.118 \pm 0.010$ & $-0.167 \pm 0.009$ & $-0.126 \pm 0.010$ \\
        Local spline  & $0.185 \pm 0.099$ & $-0.054 \pm 0.021$ & $\mathbf{0.006 \pm 0.023}$ \\
        Polynomial (degree 2) & $\mathbf{0.011 \pm 0.032}$ & $-0.104 \pm 0.017$ & $-0.052 \pm 0.019$ \\
        Polynomial (degree 3) & $0.140 \pm 0.067$ & $-0.055 \pm 0.021$ & $\mathbf{0.005 \pm 0.023}$ \\
        Polynomial (degree 5) & $0.178 \pm 0.096$ & $-0.055 \pm 0.021$ & $\mathbf{0.006 \pm 0.023}$ \\
        Steps $\Delta \ln k = 0.6$ & $0.132 \pm 0.061$ & $-0.058 \pm 0.020$ & $\mathbf{0.000 \pm 0.022}$ \\
        Steps $\Delta \ln k = 1.4$ & $0.068 \pm 0.044$ & $-0.080 \pm 0.019$ & $-0.024 \pm 0.021$ \\
        Steps $\Delta \ln k = 2.2$ & $-0.028 \pm 0.027$ & $-0.119 \pm 0.017$ & $-0.069 \pm 0.018$ \\
        Tanh (fixed)  & $\mathbf{-0.004 \pm 0.011}$ & $-0.063 \pm 0.010$ & $-0.017 \pm 0.011$ \\
        Tanh (fit)  & $-0.039 \pm 0.013$ & $-0.120 \pm 0.010$ & $-0.068 \pm 0.012$ \\
    \end{tabular}
    \caption{Relative $m$  deviation with respect to $m^{\rm true}$, and its scatter across  the different dewiggling algorithms, as a function of  derivative method and post processing procedure.  Recall that $m^{\rm true}$ is taken to be the median of the fixed tanh method (without postprocessing) across the different dewiggling algorithms.
    We mark in bold the  cases where the  shift is smaller than the scatter.}
    
    \label{tab:m_syst}
\end{table}

\begin{table}[t]
    \centering
    \begin{tabular}{c|p{0.7\textwidth}}
        Cosmology & Parameters (others fixed to fiducial) \\ \hline
        Fiducial/Reference & $A_s = 2.1 \cdot 10^{-9}$, $n_s = 0.97$, $\Omega_b h^2 = 0.022$, $\Omega_m = 0.31$, $h = 0.676$ \\
        Showcase cosmology & $A_s = 4 \cdot 10^{-9}$, $\Omega_\mathrm{cdm} h^2 = 0.15$ \\ \hline
        Variations $n_s$ & $n_s = [0.95,1.0]$ \\
        Variations $\Omega_k$ & $\Omega_k = [-0.5 , -0.01, 0.01, 0.5]$ \\
        Variations $A_s$ & $A_s = 2.5\cdot 10^{-9}$\\
        Variations $w_0/w_a$ & $\!\begin{aligned}[t]
            (w_0, w_a) = &[(-1.2,0),(-0.8,0),(-1.2,0.4),(-0.8,0.4),(-1.2,-0.4), \\ &~(-0.8,-0.4),(-0.831,-0.73),(-0.64,-1.3),(-0.727,-1.05)]
        \end{aligned}$
        \\ \hline
        Variations $\Omega_m$ & $(\Omega_\mathrm{cdm} h^2, \Omega_b h^2) = [(0.14, 0.02566), (0.1, 0.01833), (0.1, 0.022), (0.14,0.022)]$\\
        Variations $N_\mathrm{eff}$ & $N_\mathrm{eff} = [1, 2.4, 3.6, 5]$ \\
        Variations $\sum m_\nu$ & $\sum m_\nu = [0.12, 0.4,0.8]\mathrm{eV}$ \\ \hline
        Variations EDE & $\! \begin{aligned}[t] (f_\mathrm{ede},  \Omega_\mathrm{cdm} h^2, \Omega_b h^2, n_s) = &[(0.02, 0.121, 0.0221, 0.975), (0.05, 0.124, 0.0222, 0.98), \\ & ~(0.1, 0.13, 0.0223, 0.99), (0.15, 0.137, 0.0225, 1.0)]\end{aligned}$
    \end{tabular}
    \caption{The different cosmologies under investigation in this paper. The showcase cosmology is used for most plots in \cref{sec:dewiggling,sec:ShapeFit} (unless otherwise specified). 
    The first set of variations is used in \cref{ssec:null_tests}, the second set in \cref{ssec:standard_cosmo}, and the EDE variations in \cref{ssec:ede}.}
    \label{tab:cosmologies}
\end{table}

\section{Systematic error budget on \texorpdfstring{$\boldsymbol{m}$}{m} for cosmological inference from ShapeFit} \label{sec:results}

We have motivated in \cref{ssec:mtrue} a recommendation for using the value of $m$ obtained either by (a) using the \textbf{tanh (fixed)} derivative method (with no post processing) or by (b) using a Savitzky-Golay filter of the ratio in \cref{eq:mdef} and then using any of the local derivative methods. We now assess the performance of these two approaches in the context of cosmological parameters inference where a wide parameter space may be explored, sampling models significantly different from $\Lambda$CDM. We denote the two methods in the following as TANH and SG, respectively.

For the TANH case, we use no post-processing of the ratio in \cref{eq:mdef} and directly apply the fixed tanh derivative method of \cref{ssec:derivatives}. For the SG case, we post-process the ratio in \cref{eq:mdef} with a Savitzky-Golay filter according to \cref{ssec:postprocessing} and apply the steps derivative method of \cref{ssec:derivatives} for $\Delta \ln k = 0.6$. For both cases we report the mean and variance for the smoothing methods in the gold sample and compare it to the systematic uncertainty of \cref{ssec:systematic_budget}. The considered cosmologies are listed in \cref{tab:cosmologies}.

\subsection{Null tests}\label{ssec:null_tests}
\begin{figure}
    \centering
    \includegraphics[width=0.49\linewidth]{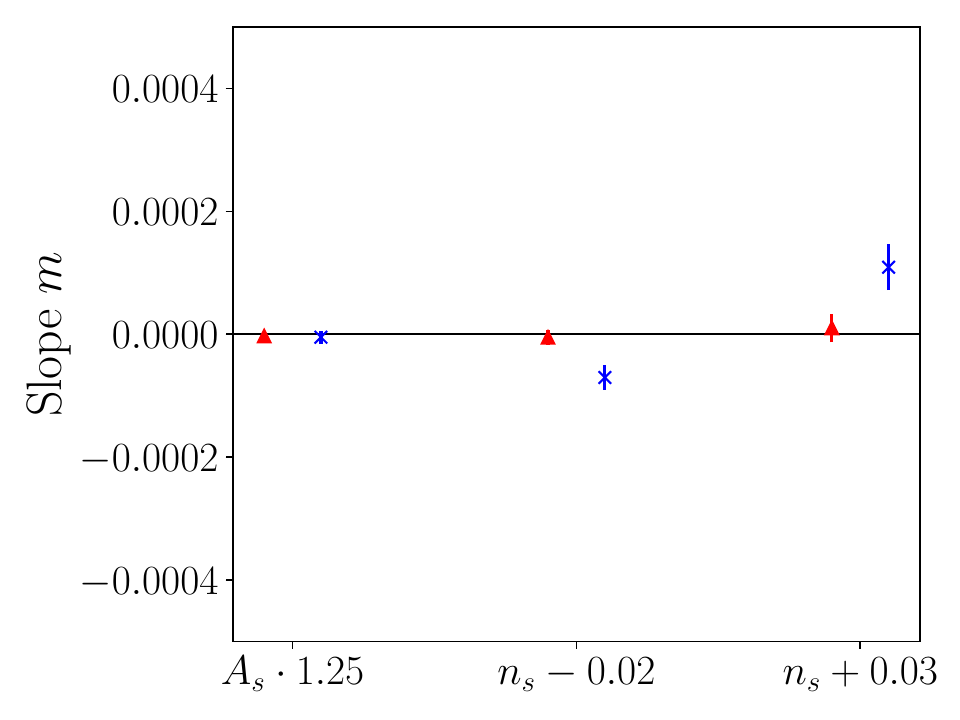}
    \includegraphics[width=0.49\linewidth]{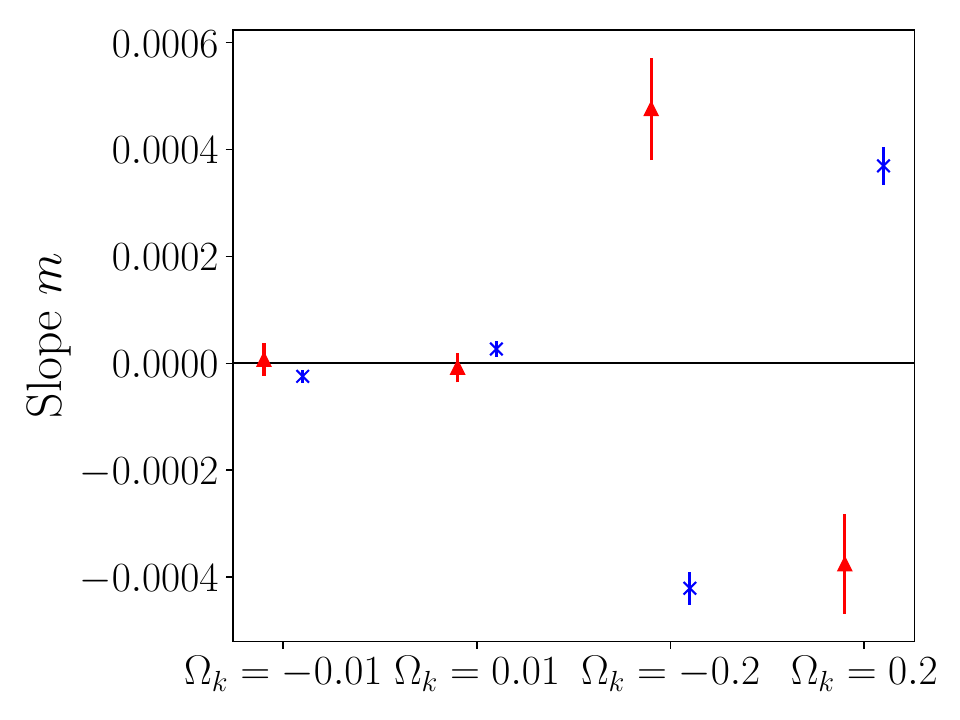}\\
    
    \includegraphics[width=0.7\linewidth]{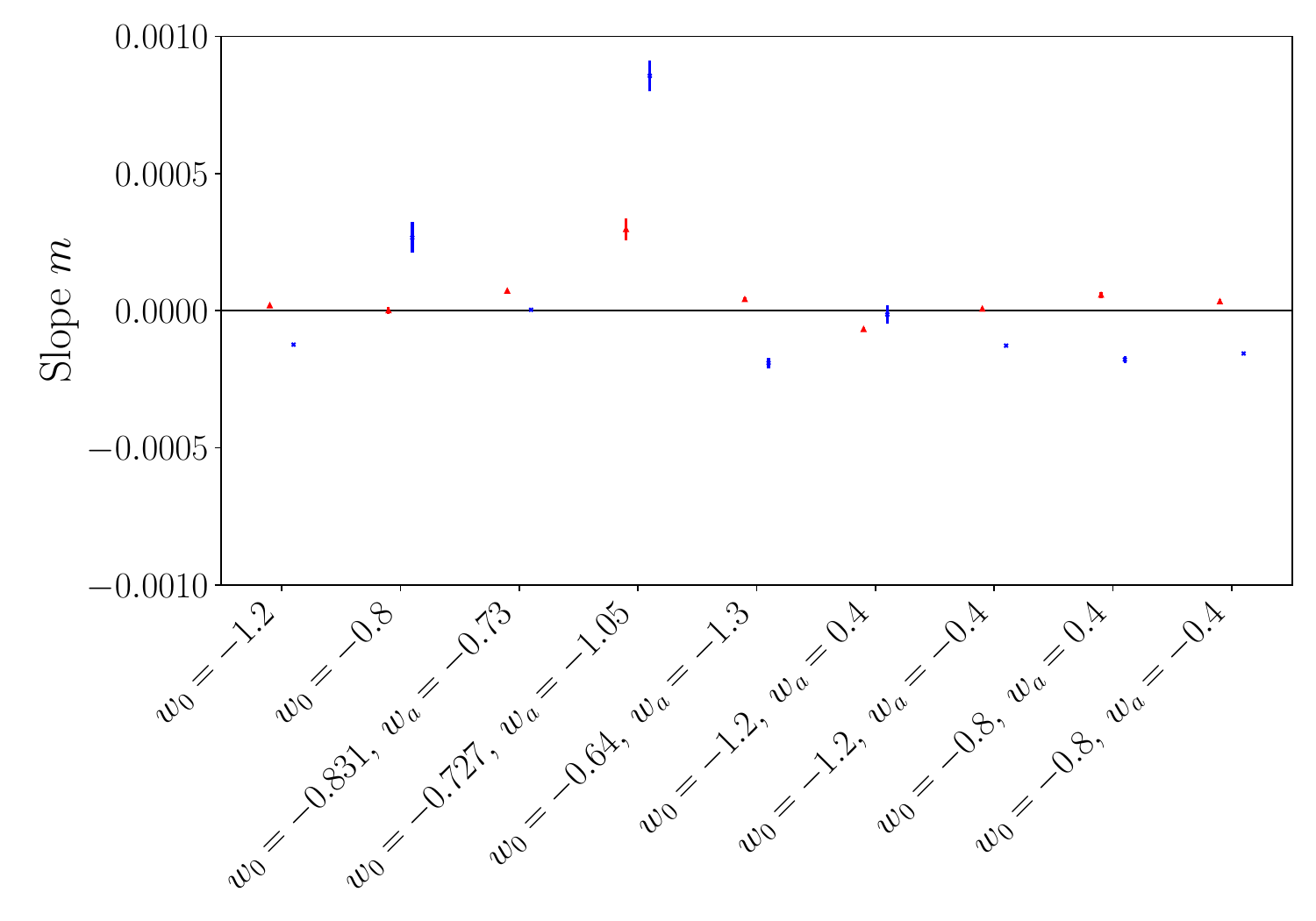}
    \caption{Extracted slopes $m$ for the null tests either with the TANH approach (red) or the SG approach (blue). The marker shows the mean and the line the standard deviation of the different de-wiggling methods. The expected result in each case is a slope $m \approx 0$, since the power spectrum at the pivot point should be unchanged by the respective parameterizations. Top left: Changes in the primordial amplitude and slope, as well as small changes in the curvature. Top right: Large changes in the curvature -- for this case we had to omit the \enquote{EH fit} algorithms as they would need to be adapted to cases of larger curvature (the numerical implementation \emph{currently} assumes $P(k) \propto k^{n_s}$ for $k \ll 0.01/\mathrm{Mpc}$, which is not the case for large curvature). Bottom: Different models of dark energy evolution (excluding also the \enquote{EH fit} algorithm).}
    \label{fig:null_tests}
\end{figure}

We perform a small number of \enquote*{null tests} in order to estimate the residual numerical uncertainty in the limit $m \to 0$. In these tests power spectrum variations are considered --resulting from changes in cosmological parameters-- for which the resulting slope $m$ should be identically zero.
We quantify the residual (small) deviations as a constant term in the systematic error budget, i.e., we parameterize the systematic error budget in $m$ as $\sigma_{\rm m, sys}=\sigma_\mathrm{rel} |m|+\sigma_{m,0}$; while  \cref{ssec:mtrue} found and initial estimate for $\sigma_\mathrm{rel}$ given by $\sigma_\mathrm{rel} \approx \sigma^{SG}_{m,\rm sys}\simeq 0.023$ for a fixed cosmology (see also below for justification that this choice is reasonable for other cosmologies), here we quantify the constant $\sigma_{m,0}$ by examining different cosmologies. The  parameter changes considered here  (in $A_s, n_s$, and background quantities such as curvature and dark energy equation of state parameters) with respect to the fiducial model should leave the shape of the  power spectrum at the pivot point unaltered (hence $m=0$). In \cref{fig:null_tests} we show that the deviations are typically extremely small with $\sigma_{m,0} < 0.001$ (and thus irrelevant for  slopes $|m| \gtrsim 0.005$ due to the 
term $\propto |m|$), but are relevant compared to the intrinsic scatter between different methods. 

The top left panel of \cref{fig:null_tests} shows the considered variations of the primordial power spectrum. For changes in $n_s$ the SG method shows some tiny bias $\sim 3.6\times 10^{-3} \cdot \Delta n_s$\,, which is not of any practical concern given any reasonable $n_s$ variations.
The top right panel of \cref{fig:null_tests} shows results for changes in curvature. For $|\Omega_k| = 0.2$ the power spectrum close to the horizon shows an upturn which is responsible for the large deviations seen. In this case the EH fitting method fails (so it is not included for producing the high $\Omega_k$ points in  this figure).
The bottom panel of \cref{fig:null_tests} shows variations of $w_0$ and $w_a$. Even if the growth of structure is still scale-independent, in these models horizon-scales effects affect the shape of the observed power spectrum at extremely large ($\lesssim 10^{-3}$/Mpc) scales, with minute permille-level leakage into the relevant scales for the most extreme models, see also \cref{app:fig:w0wa} in \cref{app:w0wa}. In this case too the EH fitting method fails and there is an extremely minute bias seen for all derivative methods -- which is in any case only a tiny contribution.
These deviations are smaller than the linear term of the systematic error budget 
as long as $|m| \gtrsim 0.005$.


To summarise, a conservative estimate of the systematic uncertainty at $m=0$ is $\sigma_0=0.001$ which encompassess all the variations found. 
We note that in pure $\Lambda$CDM the differences will typically be even smaller by another order of magnitude. We also note that the EH98 fitting method should not be used in cases of large curvature or non-constant dark energy.

\subsection{Tests on Standard cosmologies}\label{ssec:standard_cosmo}
\begin{figure}
    \centering
    \includegraphics[width=0.49\linewidth]{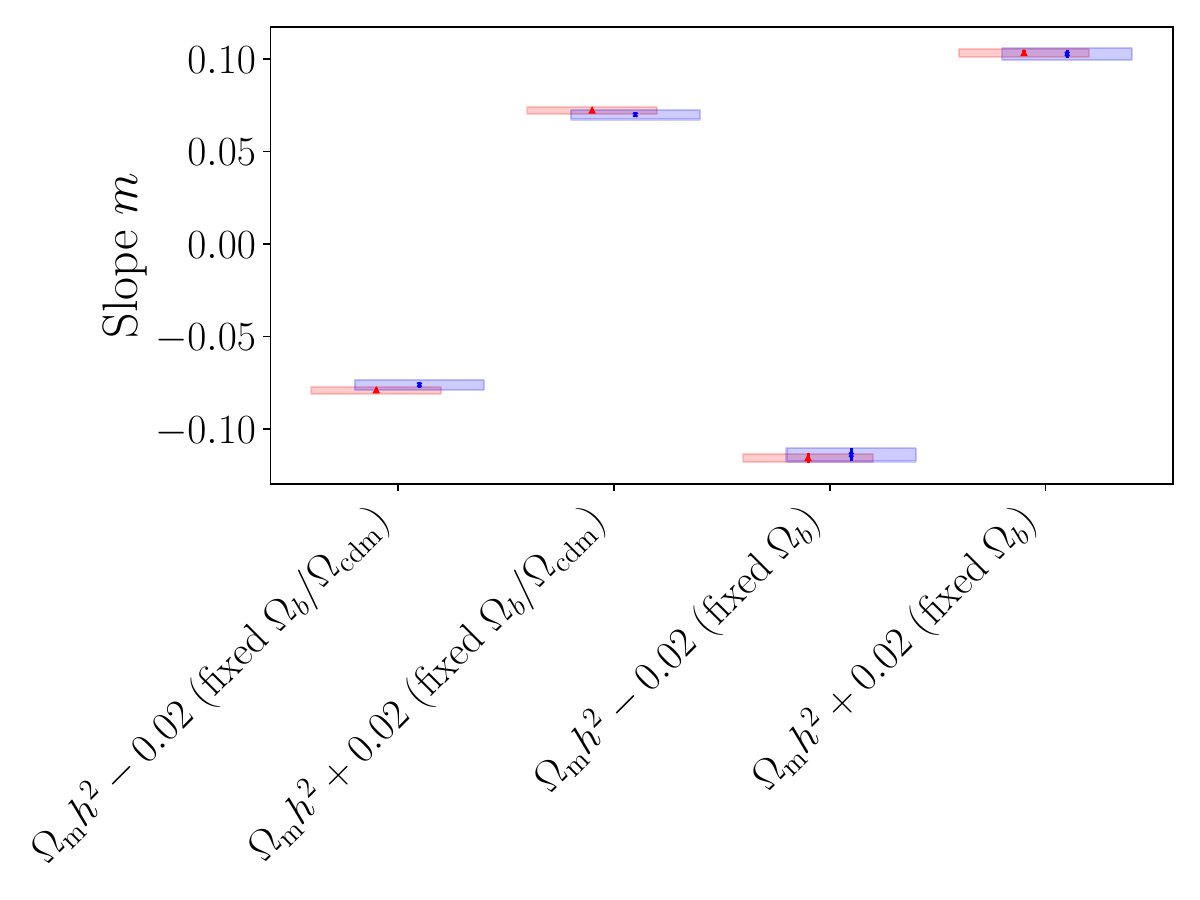}
    \includegraphics[width=0.49\linewidth]{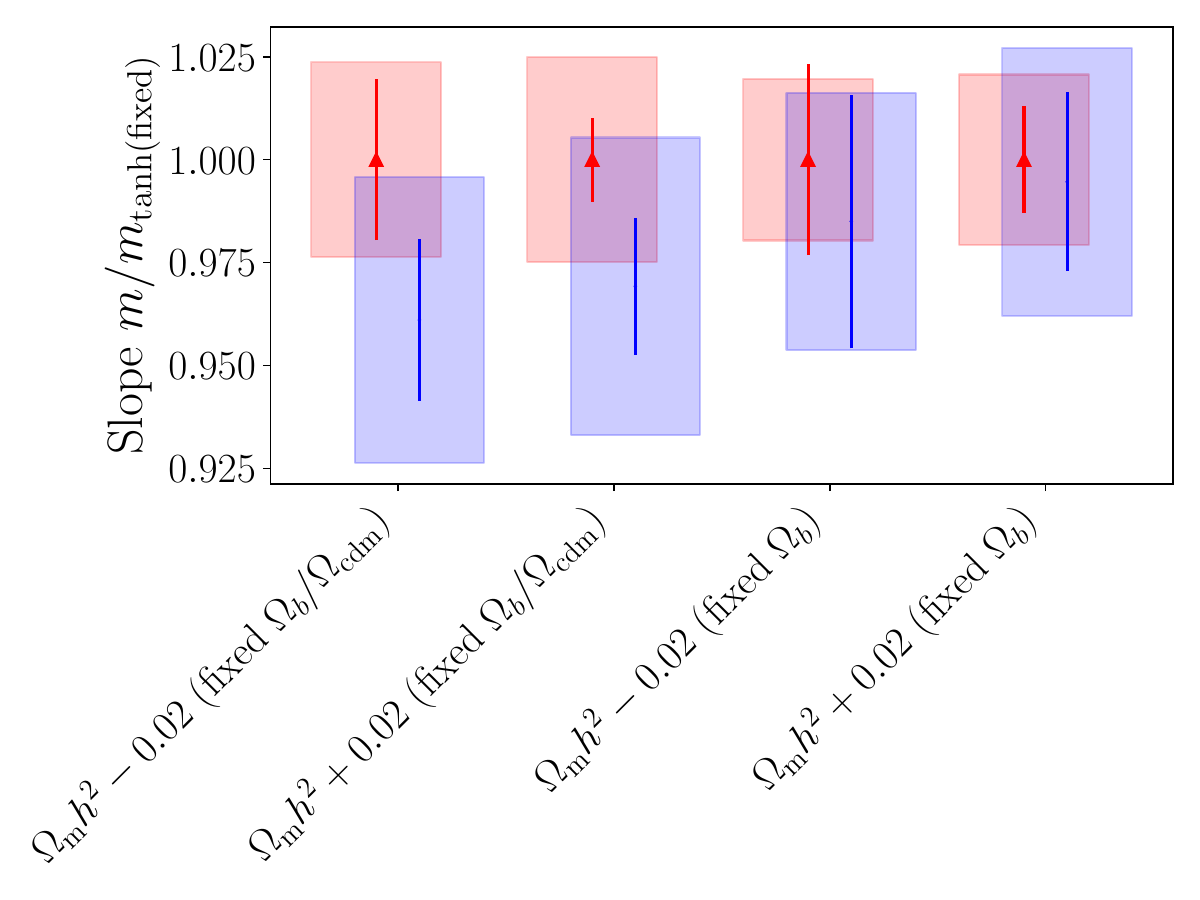}
    \caption{Extracted slopes $m$ for the TANH (red) and SG (blue) approaches for a variety of cosmological scenarios. The marker shows the mean, the line the uncertainty, and the shaded region the assigned systematic uncertainty. We vary $\Omega_m h^2$, leaving $\Omega_b/\Omega_\mathrm{m}$ constant or $\Omega_b$ constant. Left: Absolute. Right: Relative to the TANH method.}
    \label{fig:variations_omegam}
\end{figure}

\begin{figure}
    \centering
    \includegraphics[width=0.49\linewidth]{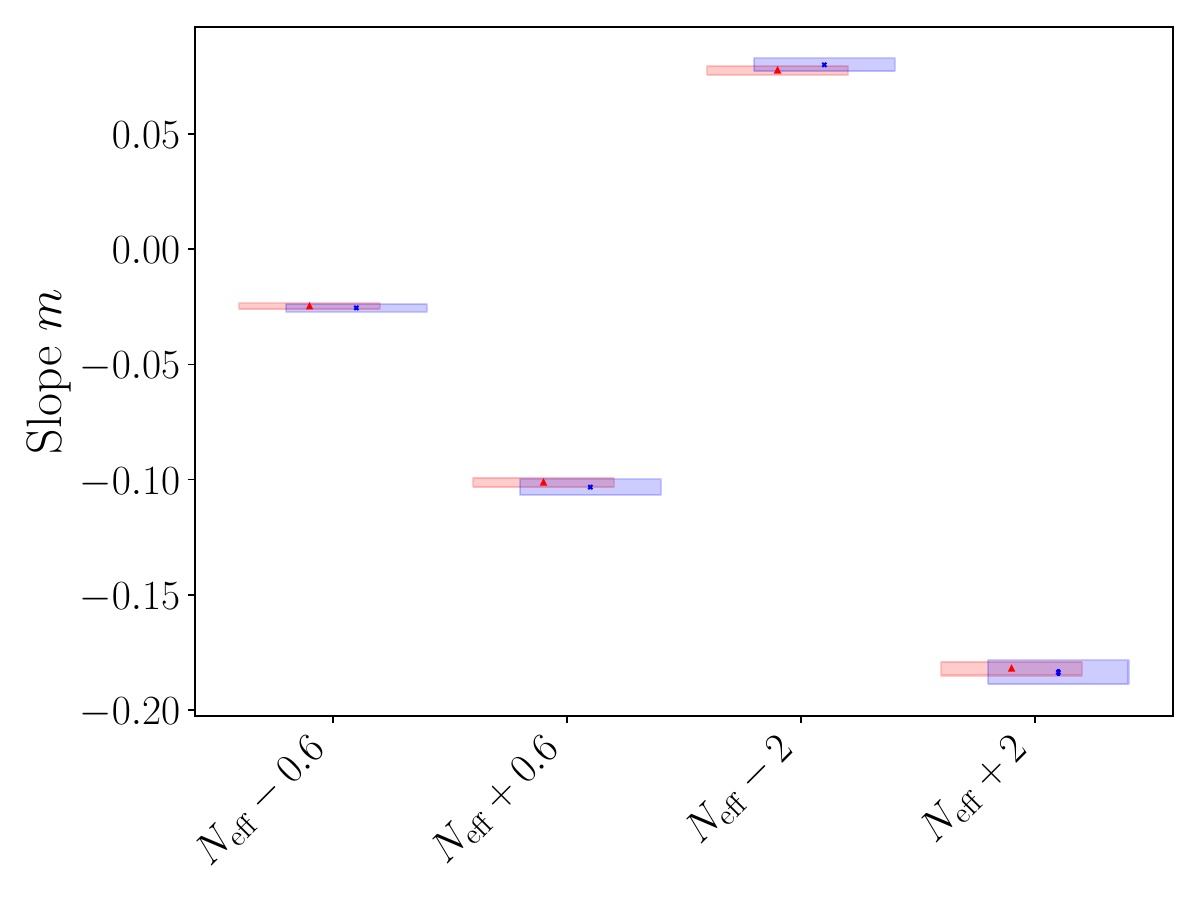}
    \includegraphics[width=0.49\linewidth]{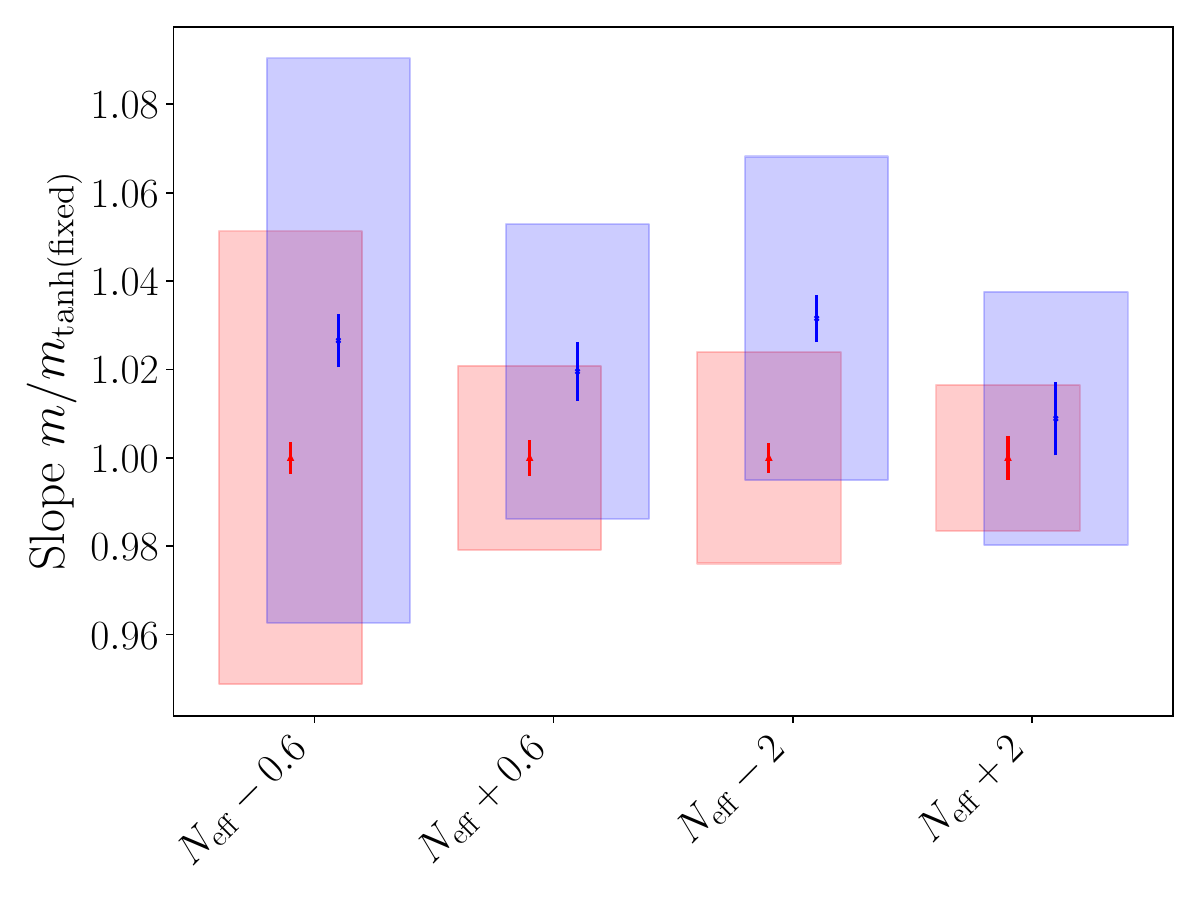}
    \caption{Same as \cref{fig:variations_omegam} but for variations including additional ultra-light relics measured through the effective number of neutrino-like species.}
    \label{fig:variations_neff}
\end{figure}

\begin{figure}
    \centering
    \includegraphics[width=0.49\linewidth]{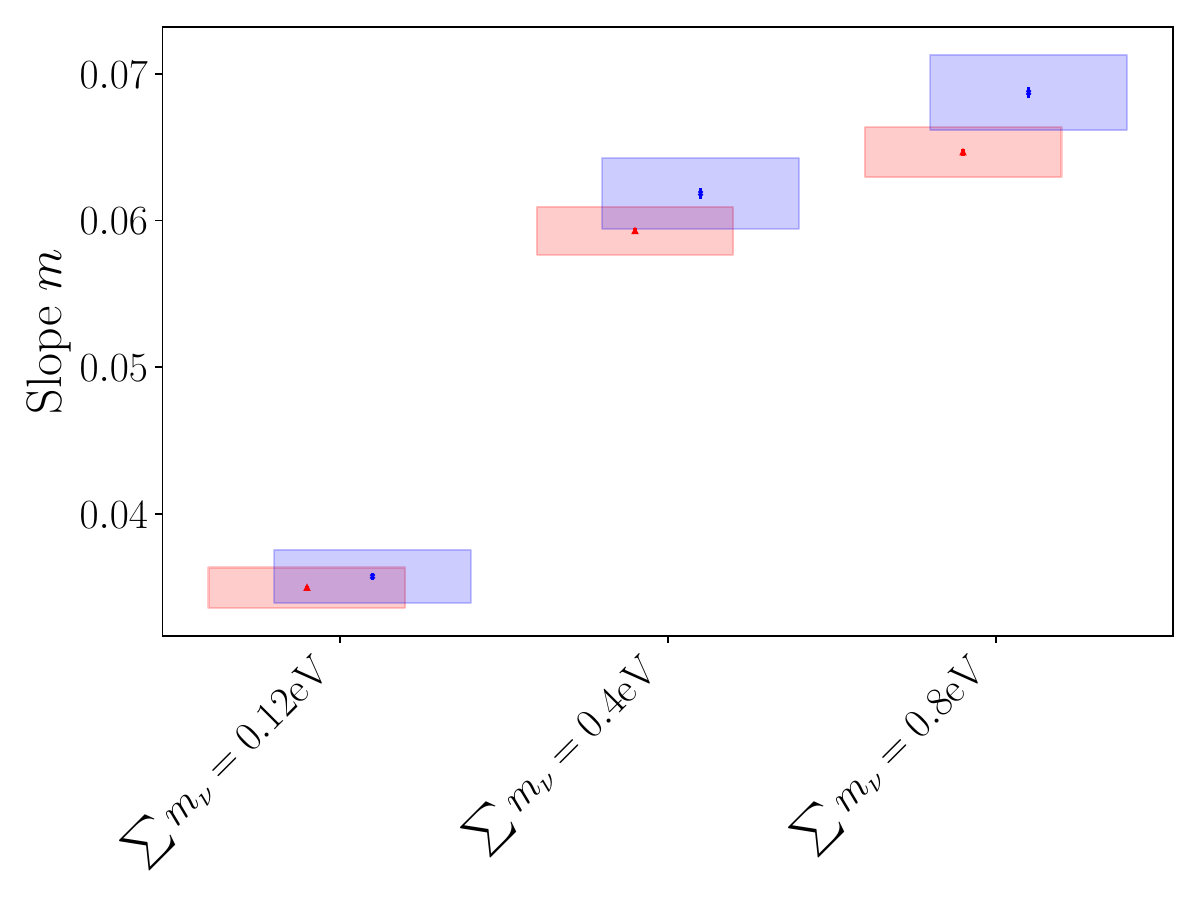}
    \includegraphics[width=0.49\linewidth]{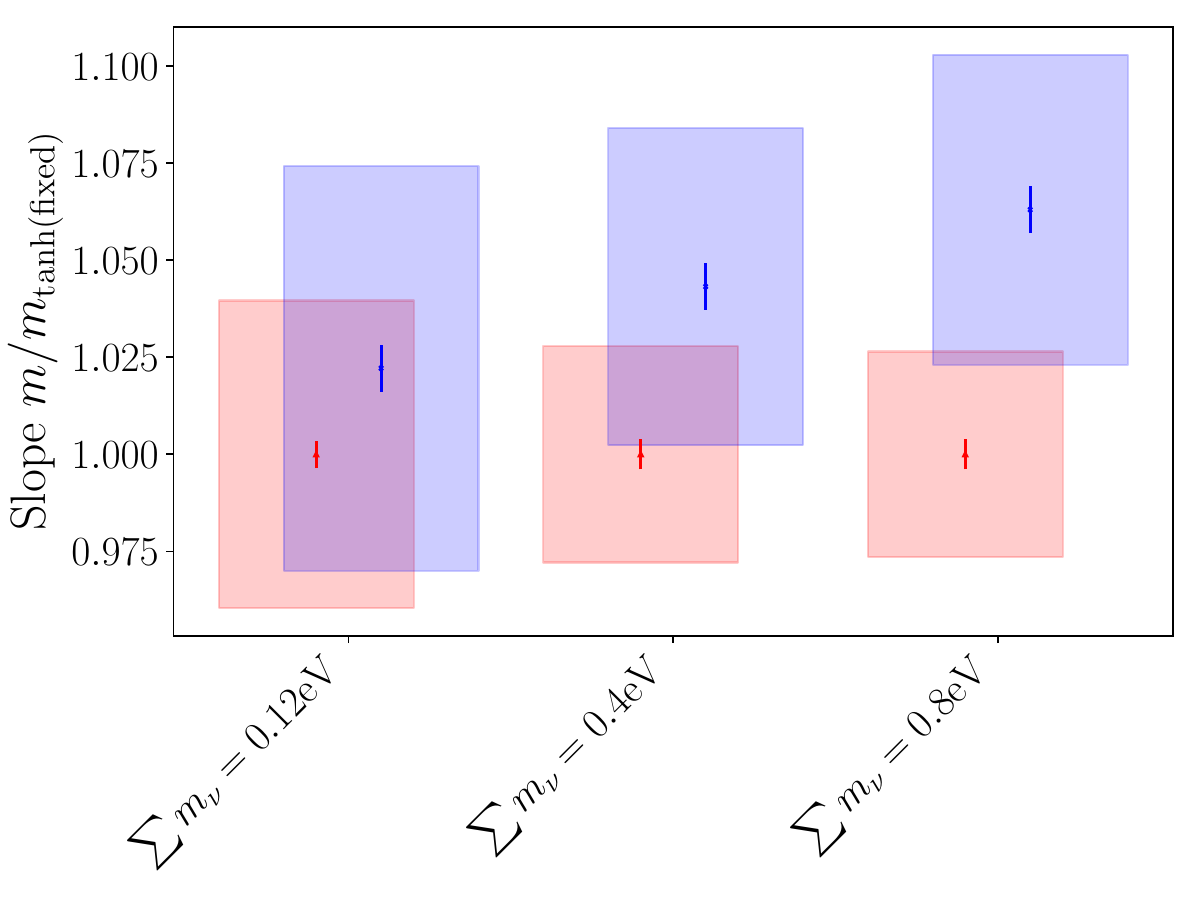}
    \caption{Same as \cref{fig:variations_omegam} but for variations including massive neutrinos with a total neutrino sum mass $\sum m_\nu$.}
    \label{fig:variations_mnu}
\end{figure}

The recovered $m$ values for variations with respect to the fiducial value of $\Omega_m h^2$ are shown in \cref{fig:variations_omegam}, while \cref{fig:variations_neff} corresponds to variations of the effective number of neutrino-like species $N_\mathrm{eff}$ and \cref{fig:variations_mnu} to variations of the total neutrino mass $\sum m_\nu$\,. 
In the figures the error bars shows the variance across the different de-wiggling methods and the shaded region the size of the assigned systematic uncertainty $\sigma_{m, {\rm sys.}}=0.023|m|+0.001$. The EH98 fitting method returns very biased results and therefore is not considered here and will not be considered in comparisons below.

With this caveat, overall, we find a good agreement between the different methods once the systematic uncertainty is taken into account.


\subsection{Early dark energy}\label{ssec:ede}
Early dark energy (EDE) is a hypothetical form of dark energy which is 
relevant in the early universe and diluting away around or shortly after the redshift of matter-radiation equality. EDE contributes to the expansion rate of the universe, affecting the growth of density perturbations and therefore suppressing the growth of structures. A comprehensive discussion on this subject is provided in \cite{poulin2023ups,AxiClass1,AxiClass2}. The EDE model uses an axion-like potential
\begin{equation}
    V(\phi) = m_{\rm ax}^2 f_{\rm ax}^2 [1- \cos(\phi/f_{\rm ax})]^{n_{\rm ax}}
\end{equation}
with parameters $ f_{\rm ax},m_{\rm ax},n_{\rm ax}$, which guide the redshift of the sudden dilution, the overall contribution to the energy density, and the rapidity of the transition, respectively. The parameter $m_{\rm ax}$ is usually replaced with the more phenomenological parameter $f_\mathrm{ede}$, which is defined as the maximum ratio of the energy density of EDE compared to the critical density, see \cite{poulin2023ups} for further discussion.

\begin{figure}
    \centering
    \includegraphics[width=0.65\linewidth]{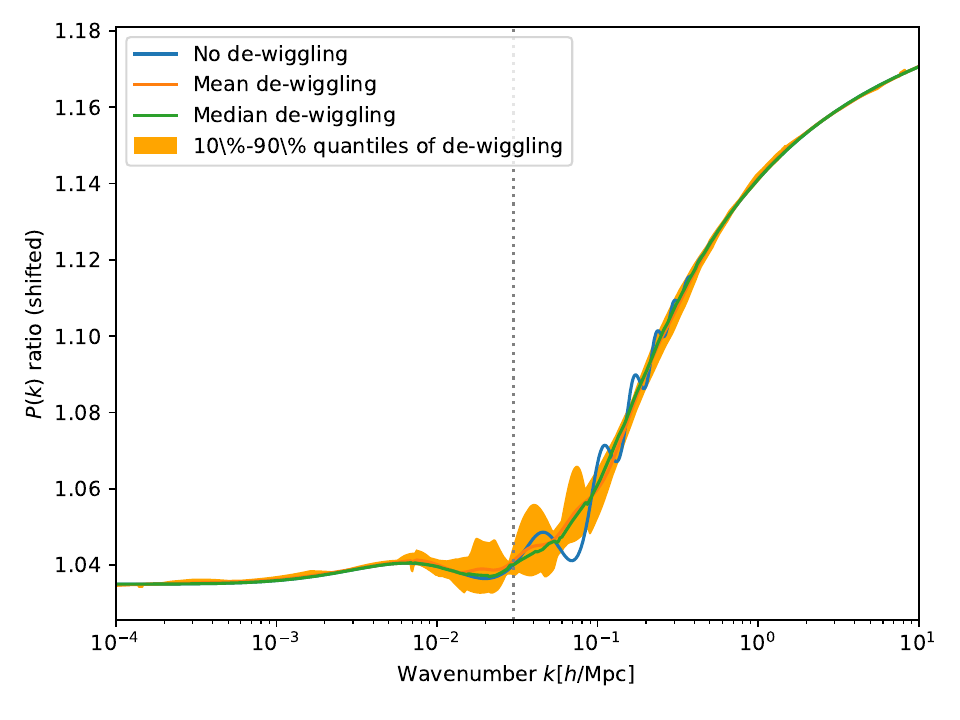}
    \caption{Ratio of the (shifted) power spectra (the $R$ of \cref{eq:mdef}) for an EDE cosmology with $f_\mathrm{ede} = 0.15$.}
    \label{fig:ede_ratio}
\end{figure}

Importantly, the presence of an EDE component causes an additional enhancement in the power spectrum roughly at the same location as the baryon suppression. On large scales, the power spectrum from the turnover to the BAO is fundamentally of a different shape, see also \cref{fig:ede_ratio} and \cite[Fig.~2, Sec. 3]{Klypin:2020tud}.

\begin{figure}
    \centering
    \includegraphics[width=0.49\linewidth]{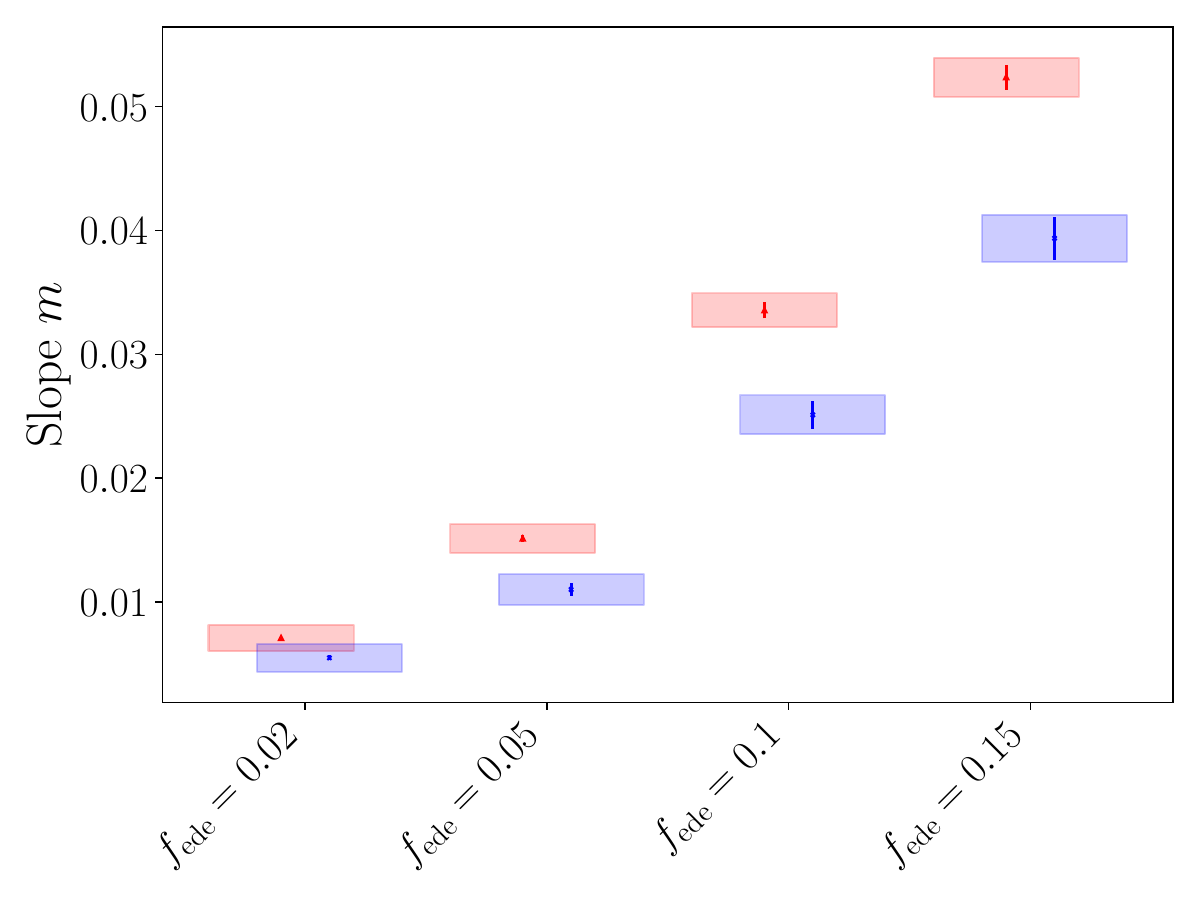}
    \includegraphics[width=0.49\linewidth]{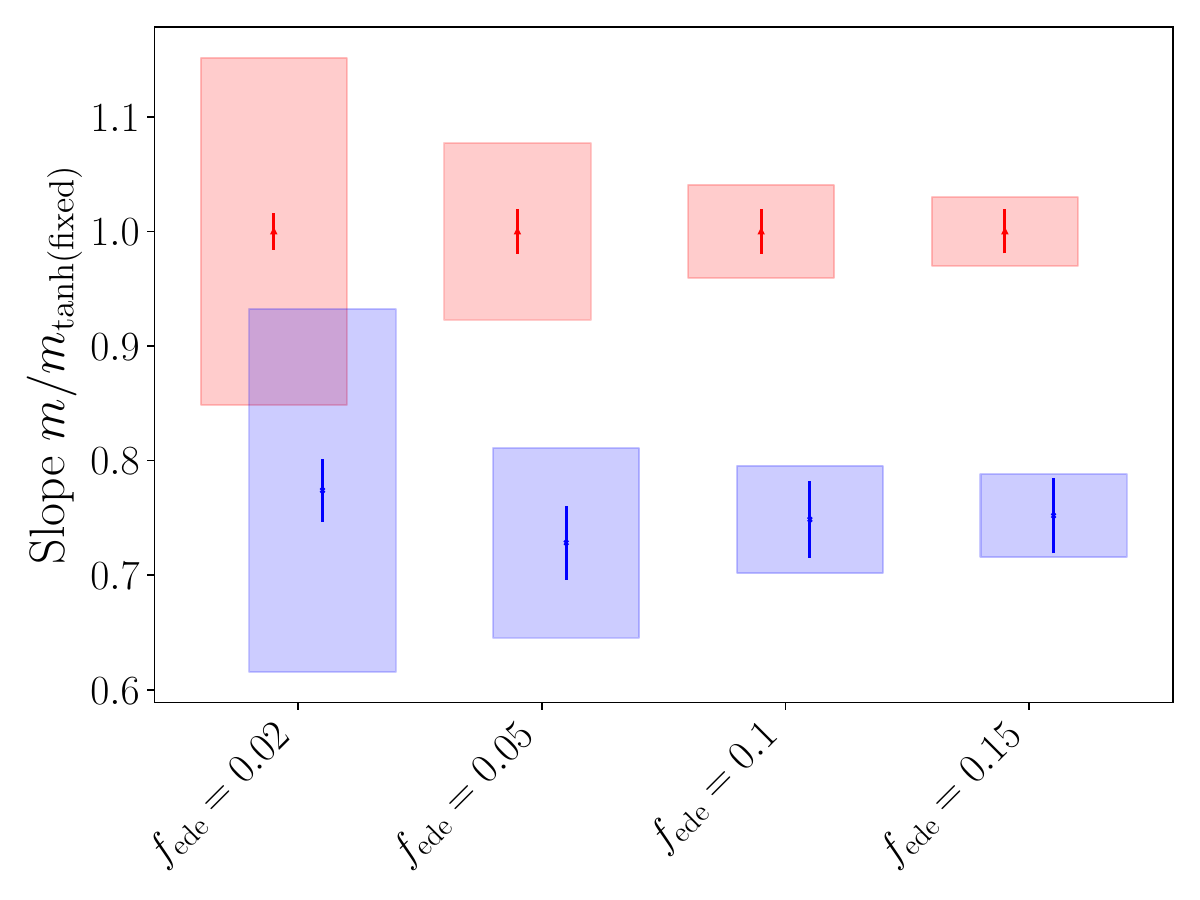}
    \caption{Same as \cref{fig:variations_omegam}, but for EDE with varying $f_\mathrm{ede}$. We also vary $A_s$, $n_s$, $\Omega_m$, $h$, $\Omega_b h^2$ according to the contours, as shown in \cref{tab:cosmologies}. Not changing the other cosmological parameters results in roughly similar (even slightly more discrepant) results.}
    \label{fig:ede_variations}
\end{figure}

\begin{figure}
    \centering
    \includegraphics[width=0.99\linewidth]{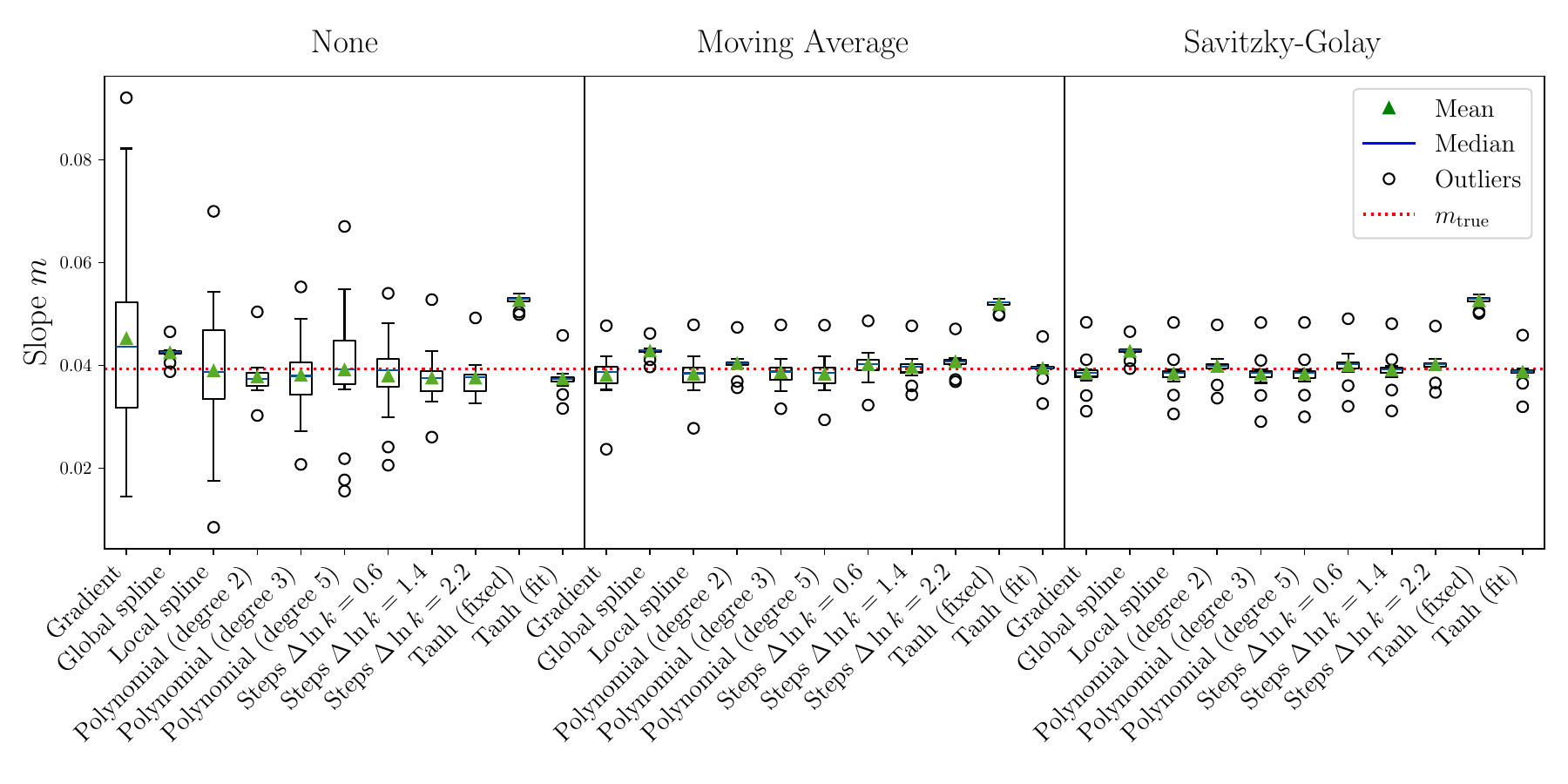}
    \caption{Same as \cref{fig:derivative_comparison} but for a model with $f_\mathrm{ede}=0.15$.}
    \label{fig:slope_comparison_EDE}
\end{figure}

This affects the different methods to measure the derivative $m$ to different degrees.
We show the comparison of the TANH and SG methods for different values of $f_\mathrm{ede}$ in \cref{fig:ede_variations}. The two methods disagree for this case beyond the designated systematic uncertainties due to the shape of the enhancement. From \cref{fig:ede_ratio} it is apparent that this is related to the form of the ratio not being well represented by a hyperbolic tangent function around $k_p = 0.03h/\mathrm{Mpc}$, as the characteristic enhancement/suppression is shifted with respect to the $k \cdot r_d$ expectation in this case (see \cref{eq:mdef}) -- using another shifting parameter other than $r_d$ (which is strongly affected by EDE) might be beneficial here (for example $1/k_\mathrm{eq}$). In \cref{fig:slope_comparison_EDE} we show the result for the SG method as a horizontal dashed red line. As in \cref{fig:derivative_comparison} we show the results for $m$ for different ways to compute the derivative for a case with $f_{\rm ede}=0.15$. If we set aside the \textbf{tanh (fixed)} method, we find  broad agreement between different ways of computing the derivative albeit with more outliers. 
The agreement holds even for 
the \textbf{tanh (fit)} method where the $k_t$ of the method is allowed to be adjusted during the fit (differing from $k_p$). We conclude that the currently used method of obtaining a slope with fixed $k_p$ might not be suitable for early dark energy cases, but adjusting $k_p$ in such cases shows promise.
We leave further exploration of ShapeFit in the EDE model for future work.

\subsection{Systematic error budget on \texorpdfstring{$\boldsymbol{m}$}{m}: a recipe}\label{ssec:systematic_budget}

We propose the following approach to derive $m$ consistently given a theory power spectrum or within a theoretical modeling pipeline:
\begin{samepage}
\begin{enumerate}
    \item Compute the de-wiggled power spectrum with the preferred method 
    (see \cref{sec:dewiggling} for possibilities, we recommend methods in the gold sample)
    \item Compute the ratio required for \cref{eq:mdef}.
    \item Smooth the ratio using a Savitzky-Golay filter as described in \cref{ssec:postprocessing}.
    \item Compute the derivative using any local derivative method (for example using steps with $\Delta \ln k = 0.6$, or the simple gradient).
    \item Associate a systematic uncertainty of $\sigma^{SG}_{m,\mathrm{syst}}$ to the obtained result, quantifying possible  differences in de-wiggling and derivative computation:
    \begin{equation}\label{eq:syst_mdef}
    \sigma^\mathrm{SG}_{m, \mathrm{syst}} = 0.023 |m| + 0.001
\end{equation}
\end{enumerate}
\end{samepage}
It is also possible to use the \textbf{tanh (fixed)} method (see \cref{ssec:derivatives}) as well, adopting a slightly smaller systematic uncertainty of $\sigma^\mathrm{tanh~(fixed)}_{m,\mathrm{syst}} = 0.011 |m| + 0.001$. By definition this smaller systematic uncertainty covers only differences between de-wiggling methods and not different derivative methods. 
Furthermore,
this uncertainty might be under-estimating the systematic biases for cosmologies whose functional shape is not close to a hyperbolic tangent curve, as we discussed in \cref{ssec:ede}.

\begin{figure}
    \centering
    \includegraphics[width=0.5\linewidth]{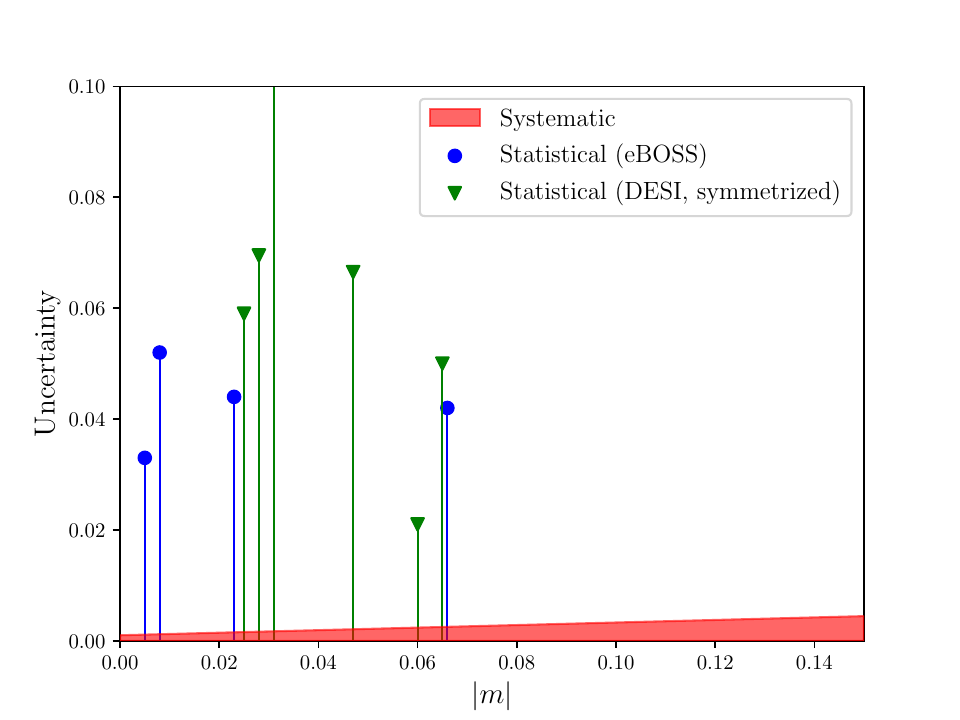}
    \caption{Comparison of statistical and systematic uncertainties for eBOSS (blue, \cite{Brieden:2022lsd}) and DESI (green, \cite{DESI:2024jis}). The systematic uncertainty is that of \cref{eq:syst_mdef}.}
    \label{fig:syst_vs_stat}
\end{figure}
As illustrated in \cref{sec:mfromdata} the systematic error in the value of $m$ of \cref{eq:syst_mdef} may be somewhat conservative for cosmological inference, but this is not a concern for current surveys. 
In fact, note that the systematic error of \cref{eq:syst_mdef}
is about an order of magnitude smaller than the reported statistical uncertainties of $\sigma_{m,stat} \simeq 0.03-0.05$ (see \cite{Brieden:2022lsd,DESI:2024jis}) for $|m| \lesssim 0.1$ as it is currently measured. 
We show a comparison between measured statistical uncertainties and the derived systematic uncertainty in \cref{fig:syst_vs_stat}.

For future applications it is important to keep in mind that in any cosmological data analysis it is common practice to start including systematic contributions to the systematic error budget only if they cross a threshold, usually defined as a fraction $\epsilon$ of the statistical error. For example DESI  takes $\epsilon$ to be 0.25 \citep{DESIsys2024}. 
The (conservative) systematic error on $m$ of \cref{eq:syst_mdef} can then be used to evaluate quickly whether it is one component of the systematic uncertainty to be included in the final error budget (and therefore better quantified specifically for the adopted pipeline) or it is something that can be safely ignored.

\section{Conclusion} \label{sec:conclusion}

Separating the oscillations and the broadband shape of the power spectrum is a very important task both for traditional analysis pipelines (full-modeling) and novel parameter compression schemes (ShapeFit). We have demonstrated that there is a roughly 1-2\% level difference between different proposed methods to de-wiggle the power spectrum, see also \cref{fig:dewiggle_comparison}. Since no method stands out as particularly more well-motivated or well-defined than the others, we argue that this percent-level difference should be seen as an inherent systematic uncertainty to the de-wiggling procedure.

Importantly, the differences between the methods are strongly enhanced when taking the derivative of the power spectra, such as required for the $m$ parameter of ShapeFit (see \cref{eq:mdef}), which is used to compress the broadband information of the power spectrum. Overall, this leads to dramatic differences for the value of the $m$ parameter up to $50\%$. 
These large differences are not a major concern for cosmological inference for current surveys: as long as the theory pipeline is consistent with the data analysis pipeline, there is virtually no bias on the cosmological parameters. However, this result still motivates
ways of taking a derivative that is more robust to the precise  de-wiggling method employed. In this work we have investigated a number of such non-local derivative algorithms and compare the results. We find that there are still large discrepancies between the computed values of~$m$, see \cref{fig:derivative_comparison}.

However, there are two important approaches in which the systematic uncertainty of the computed slopes $m$ is greatly reduced. First, when using a method of computing the derivative that is close to how $m$ would be obtained from the data, the values of $m$ are more consistent, with a relative spread of only around $\sigma_m/m \simeq 1.1\%$. However, it should be noted that in the currently widespread implementation of the method this relies on an approximation of the functional shape of the suppression that might become inaccurate when exploring some cosmologies beyond $\Lambda$CDM that alter early-time physics especially before or at matter-radiation equality, see below. Second, one can post-process the power spectrum ratio of which the derivative is taken by a smoothing procedure, such as for example the Savitzky-Golay filter. In this case we also find a low spread in the values of $\sigma_m/m \simeq 2.3\%$, and also an agreement in the reported mean value with the previous method.

We investigate both approaches in a range of cosmologies and assign a total systematic uncertainty of
\begin{equation}
    \sigma_{m,\mathrm{syst}} = 0.023 |m| + 0.001~,
\end{equation}
with some small constant term derived from a number of null-tests where $m\approx0$ is theoretically expected but not necessarily recovered, see \cref{ssec:null_tests}. We note that the quoted number is the conservative result and a reduced uncertainty of $\sigma_{m,\mathrm{syst}} = 0.011 m + 0.001$ can be obtained with the approach outlined in \cref{ssec:mtrue}. In general the assigned systematic uncertainty $\sigma_{m,\mathrm{syst}}$, is much smaller than current statistical uncertainties, see for example \cref{fig:syst_vs_stat}.

We find that for most simple one- and two-parameter extensions of the $\Lambda$CDM model (including curvature, additional relativistic degrees of freedom, dark energy variations, models with massive neutrinos), the assigned systematic uncertainty very well captures deviations between the two approaches, see \cref{sec:results}. We therefore conclude that present analyses are unaffected. However, there are a number of important caveats:

\begin{itemize}
    \item In cosmologies beyond $\Lambda$CDM for which the baryonic suppression is modified in a non-trivial way, such as for early dark energy cosmologies investigated in \cref{ssec:ede}, further care is needed. We argue that it might be necessary to extract additional information beyond the slope at the pre-defined fixed pivot (such as checking the consistency with a varied pivot analysis). We leave a more detailed investigation into this case for future work.
    \item While the systematic error estimate for the $m$ value may be slightly conservative for cosmological inference, this estimate offers guidance as of when the systematic error in $m$ is negligible and can be ignored, when it is sub dominant so it can simply be propagated in the error budget, or when it can become of concern.
    \item Future analyses with further reductions in the statistical uncertainties might need more careful extraction of $m$ with the objective of  further reducing the systematic error and a) use an un-ambiguous definition of the slope $m$ that can be applied in the same way to the data and to the theoretical modeling of the power spectrum and b) that generalizes well to models with an atypical shape of the baryonic suppression region of the power spectrum.
\end{itemize}

We conclude that for current analyses the differences in the de-wiggling methods are not critical and a sub dominant systematic uncertainty on the slope of the power spectrum $m$ is sufficient for most simple extensions of the $\Lambda$CDM model. Going forward, and depending on the specific application and the model to be constrained, the recipe inevitably will need to be re-fined (both in the extraction from the data and in the theoretical analysis pipeline).

\section*{Acknowledgements}
We thank Hector Gil Marin, Sabino Matarrese, Sergi Novell and Alice Pisani. Funding for this work was partially provided by the Spanish MINECO under project  PID2022-141125NB-I00 MCIN/AEI,
and the ``Center of Excellence Maria de Maeztu 2020-2023'' award to the ICCUB (CEX2019-000918-M funded by 
MCIN/AEI/10.13039/501100011033). This work was supported by the Erasmus+ Programme of the European Union under 2023-1-IT02-KA131-HED-000127536. The content of this publication does not necessarily reflect the official opinion of the European Union. Responsibility for the information and views expressed lies entirely with the authors. Katayoon Ghaemi acknowledges support from the French government under the France 2030 investment plan, as part of the Initiative d’Excellence d’Aix- Marseille Université - A*MIDEX AMX-22-CEI-03.

\bibliographystyle{JHEP}
\bibliography{biblio.bib}

@article{Miller:2001cf,
    author = "Miller, Christopher J. and Nichol, Robert C. and Batuski, David J.",
    title = "{Possible detection of baryonic fluctuations in the large scale structure power spectrum}",
    eprint = "astro-ph/0103018",
    archivePrefix = "arXiv",
    doi = "10.1086/321468",
    journal = "Astrophys. J.",
    volume = "555",
    pages = "68--73",
    year = "2001"
}

@article{DAmico:2019fhj,
    author = "D'Amico, Guido and Gleyzes, J\'er\^ome and Kokron, Nickolas and Markovic, Katarina and Senatore, Leonardo and Zhang, Pierre and Beutler, Florian and Gil-Mar\'\i{}n, H\'ector",
    title = "{The Cosmological Analysis of the SDSS/BOSS data from the Effective Field Theory of Large-Scale Structure}",
    eprint = "1909.05271",
    archivePrefix = "arXiv",
    primaryClass = "astro-ph.CO",
    doi = "10.1088/1475-7516/2020/05/005",
    journal = "JCAP",
    volume = "05",
    pages = "005",
    year = "2020"
}

@article{Troster:2019ean,
    author = {Tr\"oster, Tilman and others},
    title = "{Cosmology from large-scale structure: Constraining $\Lambda$CDM with BOSS}",
    eprint = "1909.11006",
    archivePrefix = "arXiv",
    primaryClass = "astro-ph.CO",
    doi = "10.1051/0004-6361/201936772",
    journal = "Astron. Astrophys.",
    volume = "633",
    pages = "L10",
    year = "2020"
}

@article{bird2012massive,
  title={Massive neutrinos and the non-linear matter power spectrum},
  author={Bird, Simeon and Viel, Matteo and Haehnelt, Martin G},
  journal={Monthly Notices of the Royal Astronomical Society},
  volume={420},
  number={3},
  pages={2551--2561},
  year={2012},
  publisher={Blackwell Publishing Ltd Oxford, UK}
}

@article{Mead:2020vgs,
    author = {Mead, Alexander and Brieden, Samuel and Tr\"oster, Tilman and Heymans, Catherine},
    title = "{hmcode-2020: improved modelling of non-linear cosmological power spectra with baryonic feedback}",
    eprint = "2009.01858",
    archivePrefix = "arXiv",
    primaryClass = "astro-ph.CO",
    doi = "10.1093/mnras/stab082",
    journal = "Mon. Not. Roy. Astron. Soc.",
    volume = "502",
    number = "1",
    pages = "1401--1422",
    year = "2021"
}

@article{Mead:2015yca,
    author = "Mead, Alexander and Peacock, John and Heymans, Catherine and Joudaki, Shahab and Heavens, Alan",
    title = "{An accurate halo model for fitting non-linear cosmological power spectra and baryonic feedback models}",
    eprint = "1505.07833",
    archivePrefix = "arXiv",
    primaryClass = "astro-ph.CO",
    doi = "10.1093/mnras/stv2036",
    journal = "Mon. Not. Roy. Astron. Soc.",
    volume = "454",
    number = "2",
    pages = "1958--1975",
    year = "2015"
}

@article{Smith:2002dz,
    author = "Smith, R. E. and Peacock, J. A. and Jenkins, A. and White, S. D. M. and Frenk, C. S. and Pearce, F. R. and Thomas, P. A. and Efstathiou, G. and Couchmann, H. M. P.",
    collaboration = "VIRGO Consortium",
    title = "{Stable clustering, the halo model and nonlinear cosmological power spectra}",
    eprint = "astro-ph/0207664",
    archivePrefix = "arXiv",
    doi = "10.1046/j.1365-8711.2003.06503.x",
    journal = "Mon. Not. Roy. Astron. Soc.",
    volume = "341",
    pages = "1311",
    year = "2003"
}

@article{Takahashi:2012em,
    author = "Takahashi, Ryuichi and Sato, Masanori and Nishimichi, Takahiro and Taruya, Atsushi and Oguri, Masamune",
    title = "{Revising the Halofit Model for the Nonlinear Matter Power Spectrum}",
    eprint = "1208.2701",
    archivePrefix = "arXiv",
    primaryClass = "astro-ph.CO",
    doi = "10.1088/0004-637X/761/2/152",
    journal = "Astrophys. J.",
    volume = "761",
    pages = "152",
    year = "2012"
}

@article{Klypin:2020tud,
    author = "Klypin, Anatoly and Poulin, Vivian and Prada, Francisco and Primack, Joel and Kamionkowski, Marc and Avila-Reese, Vladimir and Rodriguez-Puebla, Aldo and Behroozi, Peter and Hellinger, Doug and Smith, Tristan L.",
    title = "{Clustering and Halo Abundances in Early Dark Energy Cosmological Models}",
    eprint = "2006.14910",
    archivePrefix = "arXiv",
    primaryClass = "astro-ph.CO",
    doi = "10.1093/mnras/stab769",
    journal = "Mon. Not. Roy. Astron. Soc.",
    volume = "504",
    number = "1",
    pages = "769--781",
    year = "2021"
}

@article{AxiClass1, author = "Smith, Tristan L. and Poulin, Vivian and Amin, Mustafa A.", title = "{Oscillating scalar fields and the Hubble tension: a resolution with novel signatures}", eprint = "1908.06995", archivePrefix = "arXiv", primaryClass = "astro-ph.CO", doi = "10.1103/PhysRevD.101.063523", journal = "Phys. Rev. D", volume = "101", number = "6", pages = "063523", year = "2020" }

@article{AxiClass2, author = "Poulin, Vivian and Smith, Tristan L. and Grin, Daniel and Karwal, Tanvi and Kamionkowski, Marc", title = "{Cosmological implications of ultralight axionlike fields}", eprint = "1806.10608", archivePrefix = "arXiv", primaryClass = "astro-ph.CO", doi = "10.1103/PhysRevD.98.083525", journal = "Phys. Rev. D", volume = "98", number = "8", pages = "083525", year = "2018" }

@article{DESI:2024jis,
    author = "Adame, A. G. and others",
    collaboration = "DESI",
    title = "{DESI 2024 V: Full-Shape Galaxy Clustering from Galaxies and Quasars}",
    eprint = "2411.12021",
    archivePrefix = "arXiv",
    primaryClass = "astro-ph.CO",
    reportNumber = "FERMILAB-PUB-24-0847-PPD",
    month = "11",
    year = "2024"
}

@article{Brieden:2022lsd,
    author = "Brieden, Samuel and Gil-Mar\'\i{}n, H\'ector and Verde, Licia",
    title = "{Model-agnostic interpretation of 10 billion years of cosmic evolution traced by BOSS and eBOSS data}",
    eprint = "2204.11868",
    archivePrefix = "arXiv",
    primaryClass = "astro-ph.CO",
    doi = "10.1088/1475-7516/2022/08/024",
    journal = "JCAP",
    volume = "08",
    number = "08",
    pages = "024",
    year = "2022"
}

@article{Euclid:2023tog,
    author = "Pezzotta, A. and others",
    collaboration = "Euclid",
    title = "{Euclid preparation - XLI. Galaxy power spectrum modelling in real space}",
    eprint = "2312.00679",
    archivePrefix = "arXiv",
    primaryClass = "astro-ph.CO",
    doi = "10.1051/0004-6361/202348939",
    journal = "Astron. Astrophys.",
    volume = "687",
    pages = "A216",
    year = "2024"
}

@article{MoradinezhadDizgah:2020whw,
    author = "Moradinezhad Dizgah, Azadeh and Biagetti, Matteo and Sefusatti, Emiliano and Desjacques, Vincent and Nore{\~n}a, Jorge",
    title = "{Primordial Non-Gaussianity from Biased Tracers: Likelihood Analysis of Real-Space Power Spectrum and Bispectrum}",
    eprint = "2010.14523",
    archivePrefix = "arXiv",
    primaryClass = "astro-ph.CO",
    doi = "10.1088/1475-7516/2021/05/015",
    journal = "JCAP",
    volume = "05",
    pages = "015",
    year = "2021"
}

@article{Philcox:2020xbv,
    author = "Philcox, Oliver H. E. and Sherwin, Blake D. and Farren, Gerrit S. and Baxter, Eric J.",
    title = "{Determining the Hubble Constant without the Sound Horizon: Measurements from Galaxy Surveys}",
    eprint = "2008.08084",
    archivePrefix = "arXiv",
    primaryClass = "astro-ph.CO",
    doi = "10.1103/PhysRevD.103.023538",
    journal = "Phys. Rev. D",
    volume = "103",
    number = "2",
    pages = "023538",
    year = "2021"
}

@article{Jiang:2025ylr,
    author = "Jiang, Jun-Qian and Piao, Yun-Song",
    title = "{Can the sound horizon-free measurement of $H_0$ constrain early new physics?}",
    eprint = "2501.16883",
    archivePrefix = "arXiv",
    primaryClass = "astro-ph.CO",
    month = "1",
    year = "2025"
}

@article{SECADA1999278,
title = {Numerical evaluation of the Hankel transform},
journal = {Computer Physics Communications},
volume = {116},
number = {2},
pages = {278-294},
year = {1999},
issn = {0010-4655},
doi = {https://doi.org/10.1016/S0010-4655(98)00108-8},
url = {https://www.sciencedirect.com/science/article/pii/S0010465598001088},
author = {José D. Secada}
}

@article{Bahr-Kalus:2023ebd,
    author = "Bahr-Kalus, Benedict and Parkinson, David and Mueller, Eva-Maria",
    title = "{Measurement of the matter-radiation equality scale using the extended baryon oscillation spectroscopic survey quasar sample}",
    eprint = "2302.07484",
    archivePrefix = "arXiv",
    primaryClass = "astro-ph.CO",
    doi = "10.1093/mnras/stad1867",
    journal = "Mon. Not. Roy. Astron. Soc.",
    volume = "524",
    number = "2",
    pages = "2463--2476",
    year = "2023",
    note = "[Erratum: Mon.Not.Roy.Astron.Soc. 526, 3248--3249 (2023)]"
}

@Inbook{Ivanov:2022mrd,
author="Ivanov, Mikhail M.",
editor="Bambi, Cosimo
and Modesto, Leonardo
and Shapiro, Ilya",
title="Effective Field Theory for Large-Scale Structure",
bookTitle="Handbook of Quantum Gravity",
year="2023",
publisher="Springer Nature Singapore",
address="Singapore",
pages="1--48",
isbn="978-981-19-3079-9",
doi="10.1007/978-981-19-3079-9_5-1",
url="https://doi.org/10.1007/978-981-19-3079-9_5-1"
}

@ARTICLE{class,
       author = {{Blas}, Diego and {Lesgourgues}, Julien and {Tram}, Thomas},
        title = "{The Cosmic Linear Anisotropy Solving System (CLASS). Part II: Approximation schemes}",
      journal = {\jcap},
         year = 2011,
        month = jul,
       volume = {2011},
       number = {7},
          eid = {034},
        pages = {034},
          doi = {10.1088/1475-7516/2011/07/034},
archivePrefix = {arXiv},
       eprint = {1104.2933},
 primaryClass = {astro-ph.CO},
       adsurl = {https://ui.adsabs.harvard.edu/abs/2011JCAP...07..034B}
}

@article{Sherwin:2012nh,
    author = "Sherwin, Blake D. and Zaldarriaga, Matias",
    title = "{The Shift of the Baryon Acoustic Oscillation Scale: A Simple Physical Picture}",
    eprint = "1202.3998",
    archivePrefix = "arXiv",
    primaryClass = "astro-ph.CO",
    doi = "10.1103/PhysRevD.85.103523",
    journal = "Phys. Rev. D",
    volume = "85",
    pages = "103523",
    year = "2012"
}

@article{Sanchez:2008iw,
    author = "Sanchez, Ariel G. and Baugh, Carlton M. and Angulo, Raul",
    title = "{What is the best way to measure baryonic acoustic oscillations?}",
    eprint = "0804.0233",
    archivePrefix = "arXiv",
    primaryClass = "astro-ph",
    doi = "10.1111/j.1365-2966.2008.13769.x",
    journal = "Mon. Not. Roy. Astron. Soc.",
    volume = "390",
    pages = "1470--1490",
    year = "2008"
}

@article{Padmanabhan:2009yr,
    author = "Padmanabhan, Nikhil and White, Martin",
    title = "{Calibrating the Baryon Oscillation Ruler for Matter and Halos}",
    eprint = "0906.1198",
    archivePrefix = "arXiv",
    primaryClass = "astro-ph.CO",
    doi = "10.1103/PhysRevD.80.063508",
    journal = "Phys. Rev. D",
    volume = "80",
    pages = "063508",
    year = "2009"
}

@article{Crocce:2007dt,
    author = "Crocce, Martin and Scoccimarro, Roman",
    title = "{Nonlinear Evolution of Baryon Acoustic Oscillations}",
    eprint = "0704.2783",
    archivePrefix = "arXiv",
    primaryClass = "astro-ph",
    doi = "10.1103/PhysRevD.77.023533",
    journal = "Phys. Rev. D",
    volume = "77",
    pages = "023533",
    year = "2008"
}

@article{Prada:2014bra,
    author = {Prada, Francisco and Sc\'occola, Claudia G. and Chuang, Chia-Hsun and Yepes, Gustavo and Klypin, Anatoly A. and Kitaura, Francisco-Shu and Gottl\"ober, Stefan and Zhao, Cheng},
    title = "{Hunting down systematics in baryon acoustic oscillations after cosmic high noon}",
    eprint = "1410.4684",
    archivePrefix = "arXiv",
    primaryClass = "astro-ph.CO",
    doi = "10.1093/mnras/stw312",
    journal = "Mon. Not. Roy. Astron. Soc.",
    volume = "458",
    number = "1",
    pages = "613--623",
    year = "2016"
}

@article{Porto:2016pyg,
    author = "Porto, Rafael A.",
    title = "{The effective field theorist\textquoteright{}s approach to gravitational dynamics}",
    eprint = "1601.04914",
    archivePrefix = "arXiv",
    primaryClass = "hep-th",
    doi = "10.1016/j.physrep.2016.04.003",
    journal = "Phys. Rept.",
    volume = "633",
    pages = "1--104",
    year = "2016"
}

@article{Ivanov:2019hqk,
    author = "Ivanov, Mikhail M. and Simonovi\'c, Marko and Zaldarriaga, Matias",
    title = "{Cosmological Parameters and Neutrino Masses from the Final Planck and Full-Shape BOSS Data}",
    eprint = "1912.08208",
    archivePrefix = "arXiv",
    primaryClass = "astro-ph.CO",
    reportNumber = "INR-TH-2019-023, CERN-TH-2019-217",
    doi = "10.1103/PhysRevD.101.083504",
    journal = "Phys. Rev. D",
    volume = "101",
    number = "8",
    pages = "083504",
    year = "2020"
}

@article{Simon:2022csv,
    author = "Simon, Th\'eo and Zhang, Pierre and Poulin, Vivian",
    title = "{Cosmological inference from the EFTofLSS: the eBOSS QSO full-shape analysis}",
    eprint = "2210.14931",
    archivePrefix = "arXiv",
    primaryClass = "astro-ph.CO",
    doi = "10.1088/1475-7516/2023/07/041",
    journal = "JCAP",
    volume = "07",
    pages = "041",
    year = "2023"
}

@article{DESI:2024hhd,
    author = "Adame, A. G. and others",
    collaboration = "DESI",
    title = "{DESI 2024 VII: Cosmological Constraints from the Full-Shape Modeling of Clustering Measurements}",
    eprint = "2411.12022",
    archivePrefix = "arXiv",
    primaryClass = "astro-ph.CO",
    reportNumber = "FERMILAB-PUB-24-0854-PPD",
    month = "11",
    year = "2024"
}

@article{DESI:2024mwx,
    author = "Adame, A. G. and others",
    collaboration = "DESI",
    title = "{DESI 2024 VI: Cosmological Constraints from the Measurements of Baryon Acoustic Oscillations}",
    eprint = "2404.03002",
    archivePrefix = "arXiv",
    primaryClass = "astro-ph.CO",
    reportNumber = "FERMILAB-PUB-24-0154-PPD",
    month = "4",
    year = "2024"
}

@article{eBOSS:2020yzd,
    author = "Alam, Shadab and others",
    collaboration = "eBOSS",
    title = "{Completed SDSS-IV extended Baryon Oscillation Spectroscopic Survey: Cosmological implications from two decades of spectroscopic surveys at the Apache Point Observatory}",
    eprint = "2007.08991",
    archivePrefix = "arXiv",
    primaryClass = "astro-ph.CO",
    doi = "10.1103/PhysRevD.103.083533",
    journal = "Phys. Rev. D",
    volume = "103",
    number = "8",
    pages = "083533",
    year = "2021"
}

@article{2dFGRS:2005yhx,
    author = "Cole, Shaun and others",
    collaboration = "2dFGRS",
    title = "{The 2dF Galaxy Redshift Survey: Power-spectrum analysis of the final dataset and cosmological implications}",
    eprint = "astro-ph/0501174",
    archivePrefix = "arXiv",
    doi = "10.1111/j.1365-2966.2005.09318.x",
    journal = "Mon. Not. Roy. Astron. Soc.",
    volume = "362",
    pages = "505--534",
    year = "2005"
}

@article{SDSS:2005xqv,
    author = "Eisenstein, Daniel J. and others",
    collaboration = "SDSS",
    title = "{Detection of the Baryon Acoustic Peak in the Large-Scale Correlation Function of SDSS Luminous Red Galaxies}",
    eprint = "astro-ph/0501171",
    archivePrefix = "arXiv",
    reportNumber = "FERMILAB-PUB-05-057-A-CD",
    doi = "10.1086/466512",
    journal = "Astrophys. J.",
    volume = "633",
    pages = "560--574",
    year = "2005"
}

@article{schoneberg2022bao+,
  title={BAO+ BBN revisited—growing the Hubble tension with a 0.7 km/s/Mpc constraint},
  author={Sch{\"o}neberg, Nils and Verde, Licia and Gil-Mar{\'\i}n, H{\'e}ctor and Brieden, Samuel},
  journal={Journal of Cosmology and Astroparticle Physics},
  volume={2022},
  number={11},
  pages={039},
  year={2022},
  publisher={IOP Publishing}
}

@article{brieden2021shapefit,
  title={ShapeFit: extracting the power spectrum shape information in galaxy surveys beyond BAO and RSD},
  author={Brieden, Samuel and Gil-Mar{\'\i}n, H{\'e}ctor and Verde, Licia},
  journal={Journal of Cosmology and Astroparticle Physics},
  volume={2021},
  number={12},
  pages={054},
  year={2021},
  publisher={IOP Publishing}
}

@article{brieden2021model,
  title={Model-independent versus model-dependent interpretation of the SDSS-III BOSS power spectrum: Bridging the divide},
  author={Brieden, Samuel and Gil-Mar{\'\i}n, H{\'e}ctor and Verde, Licia},
  journal={Physical Review D},
  volume={104},
  number={12},
  pages={L121301},
  year={2021},
  publisher={APS}
}

@article{brieden2022model,
  title={Model-agnostic interpretation of 10 billion years of cosmic evolution traced by BOSS and eBOSS data},
  author={Brieden, Samuel and Gil-Mar{\'\i}n, H{\'e}ctor and Verde, Licia},
  journal={Journal of Cosmology and Astroparticle Physics},
  volume={2022},
  number={08},
  pages={024},
  year={2022},
  publisher={IOP Publishing}
}

@article{audren2013conservative,
  title={Conservative constraints on early cosmology with MONTE PYTHON},
  author={Audren, Benjamin and Lesgourgues, Julien and Benabed, Karim and Prunet, Simon},
  journal={Journal of Cosmology and Astroparticle Physics},
  volume={2013},
  number={02},
  pages={001},
  year={2013},
  publisher={IOP Publishing}
}

@article{brinckmann2019montepython,
  title={MontePython 3: boosted MCMC sampler and other features},
  author={Brinckmann, Thejs and Lesgourgues, Julien},
  journal={Physics of the Dark Universe},
  volume={24},
  pages={100260},
  year={2019},
  publisher={Elsevier}
}

@article{vlah2016perturbation,
  title={Perturbation theory, effective field theory, and oscillations in the power spectrum},
  author={Vlah, Zvonimir and Seljak, Uro{\v{s}} and Chu, Man Yat and Feng, Yu},
  journal={Journal of Cosmology and Astroparticle Physics},
  volume={2016},
  number={03},
  pages={057},
  year={2016},
  publisher={IOP Publishing}
}

@article{hinton2016extraction,
  title={Extraction of Cosmological Information from WiggleZ},
  author={Hinton, Samuel},
  journal={arXiv preprint arXiv:1604.01830},
  year={2016}
}

@article{blas2016time,
  title={Time-sliced perturbation theory II: baryon acoustic oscillations and infrared resummation},
  author={Blas, Diego and Garny, Mathias and Ivanov, Mikhail M and Sibiryakov, Sergey},
  journal={Journal of Cosmology and Astroparticle Physics},
  volume={2016},
  number={07},
  pages={028},
  year={2016},
  publisher={IOP Publishing}
}

@article{chudaykin2020nonlinear,
  title={Nonlinear perturbation theory extension of the Boltzmann code CLASS},
  author={Chudaykin, Anton and Ivanov, Mikhail M and Philcox, Oliver HE and Simonovi{\'c}, Marko},
  journal={Physical Review D},
  volume={102},
  number={6},
  pages={063533},
  year={2020},
  publisher={APS}
}

@article{esteban2019global,
  title={Global analysis of three-flavour neutrino oscillations: synergies and tensions in the determination of $\theta$23, $\delta$CP, and the mass ordering},
  author={Esteban, Ivan and Gonz{\'a}lez-Garc{\'\i}a, Maria Concepti{\'o}n and Hernandez-Cabezudo, Alvaro and Maltoni, Michele and Schwetz, Thomas},
  journal={Journal of High Energy Physics},
  volume={2019},
  number={1},
  pages={1--35},
  year={2019},
  publisher={Springer}
}

@ARTICLE{Hamann2010,
       author = {{Hamann}, Jan and {Hannestad}, Steen and {Lesgourgues}, Julien and {Rampf}, Cornelius and {Wong}, Yvonne Y.~Y.},
        title = "{Cosmological parameters from large scale structure - geometric versus shape information}",
      journal = {\jcap},
         year = 2010,
        month = jul,
       volume = {2010},
       number = {7},
          eid = {022},
        pages = {022},
          doi = {10.1088/1475-7516/2010/07/022},
archivePrefix = {arXiv},
       eprint = {1003.3999},
 primaryClass = {astro-ph.CO},
       adsurl = {https://ui.adsabs.harvard.edu/abs/2010JCAP...07..022H}
}

@phdthesis{Wallisch:2018rzj,
    author = "Wallisch, Benjamin",
    title = "{Cosmological Probes of Light Relics}",
    eprint = "1810.02800",
    archivePrefix = "arXiv",
    primaryClass = "astro-ph.CO",
    doi = "10.17863/CAM.30368",
    school = "Cambridge U.",
    year = "2018"
}

@article{Eisenstein:2006nj,
    author = "Eisenstein, Daniel J. and Seo, Hee-jong and White, Martin J.",
    title = "{On the Robustness of the Acoustic Scale in the Low-Redshift Clustering of Matter}",
    eprint = "astro-ph/0604361",
    archivePrefix = "arXiv",
    doi = "10.1086/518755",
    journal = "Astrophys. J.",
    volume = "664",
    pages = "660--674",
    year = "2007"
}

@article{Seo:2005ys,
    author = "Seo, Hee-Jong and Eisenstein, Daniel J.",
    title = "{Baryonic acoustic oscillations in simulated galaxy redshift surveys}",
    eprint = "astro-ph/0507338",
    archivePrefix = "arXiv",
    doi = "10.1086/491599",
    journal = "Astrophys. J.",
    volume = "633",
    pages = "575--588",
    year = "2005"
}

@book{dodelson2020modern,
  title={Modern cosmology},
  author={Dodelson, Scott and Schmidt, Fabian},
  year={2020},
  publisher={Academic press}
}

@article{Lai:2024bpl,
    author = "Lai, Y. and others",
    title = "{A comparison between Shapefit compression and Full-Modelling method with PyBird for DESI 2024 and beyond}",
    eprint = "2404.07283",
    archivePrefix = "arXiv",
    primaryClass = "astro-ph.CO",
    month = "4",
    year = "2024"
}

@article{Maus:2024dzi,
    author = "Maus, M. and others",
    title = "{An analysis of parameter compression and full-modeling techniques with Velocileptors for DESI 2024 and beyond}",
    eprint = "2404.07312",
    archivePrefix = "arXiv",
    primaryClass = "astro-ph.CO",
    month = "4",
    year = "2024"
}

@article{BOSS:2013sqq,
    author = "Kirkby, David and others",
    collaboration = "BOSS",
    title = "{Fitting Methods for Baryon Acoustic Oscillations in the Lyman-{\textbackslash{}alpha} Forest Fluctuations in BOSS Data Release 9}",
    eprint = "1301.3456",
    archivePrefix = "arXiv",
    primaryClass = "astro-ph.CO",
    doi = "10.1088/1475-7516/2013/03/024",
    journal = "JCAP",
    volume = "03",
    pages = "024",
    year = "2013"
}

@ARTICLE{2010MNRAS.404...60R,
       author = {{Reid}, Beth A. and {Percival}, Will J. and {Eisenstein}, Daniel J. and {Verde}, Licia and {Spergel}, David N. and {Skibba}, Ramin A. and {Bahcall}, Neta A. and {Budavari}, Tamas and {Frieman}, Joshua A. and {Fukugita}, Masataka and {Gott}, J. Richard and {Gunn}, James E. and {Ivezi{\'c}}, {\v{Z}}eljko and {Knapp}, Gillian R. and {Kron}, Richard G. and {Lupton}, Robert H. and {McKay}, Timothy A. and {Meiksin}, Avery and {Nichol}, Robert C. and {Pope}, Adrian C. and {Schlegel}, David J. and {Schneider}, Donald P. and {Stoughton}, Chris and {Strauss}, Michael A. and {Szalay}, Alexander S. and {Tegmark}, Max and {Vogeley}, Michael S. and {Weinberg}, David H. and {York}, Donald G. and {Zehavi}, Idit},
        title = "{Cosmological constraints from the clustering of the Sloan Digital Sky Survey DR7 luminous red galaxies}",
      journal = {\mnras},
         year = 2010,
        month = may,
       volume = {404},
       number = {1},
        pages = {60-85},
          doi = {10.1111/j.1365-2966.2010.16276.x},
archivePrefix = {arXiv},
       eprint = {0907.1659},
 primaryClass = {astro-ph.CO},
       adsurl = {https://ui.adsabs.harvard.edu/abs/2010MNRAS.404...60R}
}

@article{Epperson01041987,
    author = {James F. Epperson},
    title = {On the Runge Example},
    journal = {The American Mathematical Monthly},
    volume = {94},
    number = {4},
    pages = {329--341},
    year = {1987},
    publisher = {Taylor \& Francis},
    doi = {10.1080/00029890.1987.12000642},
    URL = {https://doi.org/10.1080/00029890.1987.12000642}
}

@article{Eisenstein:1997jh,
    author = "Eisenstein, Daniel J. and Hu, Wayne",
    title = "{Power spectra for cold dark matter and its variants}",
    eprint = "astro-ph/9710252",
    archivePrefix = "arXiv",
    reportNumber = "IASSNS-AST-97-61",
    doi = "10.1086/306640",
    journal = "Astrophys. J.",
    volume = "511",
    pages = "5",
    year = "1997"
}

@article{eisenstein1998baryonic,
  title={Baryonic features in the matter transfer function},
  author={Eisenstein, Daniel J and Hu, Wayne},
  journal={The Astrophysical Journal},
  volume={496},
  number={2},
  pages={605},
  year={1998},
  publisher={IOP Publishing}
}

@article{poulin2023ups,
  title={The Ups and Downs of Early Dark Energy solutions to the Hubble tension: a review of models, hints and constraints circa 2023},
  author={Poulin, Vivian and Smith, Tristan L and Karwal, Tanvi},
  journal={Physics of the Dark Universe},
  pages={101348},
  year={2023},
  publisher={Elsevier}
}

@ARTICLE{PTchallenge,
       author = {{Brieden}, Samuel and {Gil-Mar{\'\i}n}, H{\'e}ctor and {Verde}, Licia},
        title = "{PT challenge: validation of ShapeFit on large-volume, high-resolution mocks}",
      journal = {\jcap},
         year = 2022,
        month = jun,
       volume = {2022},
       number = {6},
          eid = {005},
        pages = {005},
          doi = {10.1088/1475-7516/2022/06/005},
archivePrefix = {arXiv},
       eprint = {2201.08400},
 primaryClass = {astro-ph.CO},
       adsurl = {https://ui.adsabs.harvard.edu/abs/2022JCAP...06..005B}
}

@ARTICLE{DESIsys2024,
       author = {{DESI Collaboration} and {Adame}, A.~G. and {Aguilar}, J. and {Ahlen}, S. and {Alam}, S. and {Alexander}, D.~M. and {Alvarez}, M. and {Alves}, O. and {Anand}, A. and {Andrade}, U. and et al.},
        title = "{DESI 2024 V: Full-Shape Galaxy Clustering from Galaxies and Quasars}",
      journal = {arXiv e-prints},
         year = 2024,
        month = nov,
          eid = {arXiv:2411.12021},
        pages = {arXiv:2411.12021},
          doi = {10.48550/arXiv.2411.12021},
archivePrefix = {arXiv},
       eprint = {2411.12021},
 primaryClass = {astro-ph.CO},
       adsurl = {https://ui.adsabs.harvard.edu/abs/2024arXiv241112021D}
}
\appendix
\section{Smoothing and dewiggling methods}\label{app:smoothing}
In this appendix we present in detail the smoothing and dewiggling methods used in the main text of the paper. 
 \subsection{Numerical smoothing}\label{ssec:smoothing}

Given that the oscillations extend equally above and below the broadband, one can expect a simple numerical smoothing method to smooth out these oscillations. The expectation is that, averaging the power spectrum over a large enough distance, the oscillations are averaged out, while the broadband remains at least approximately intact.

\begin{figure}[h]
    \centering
    \includegraphics[width=0.5\linewidth]{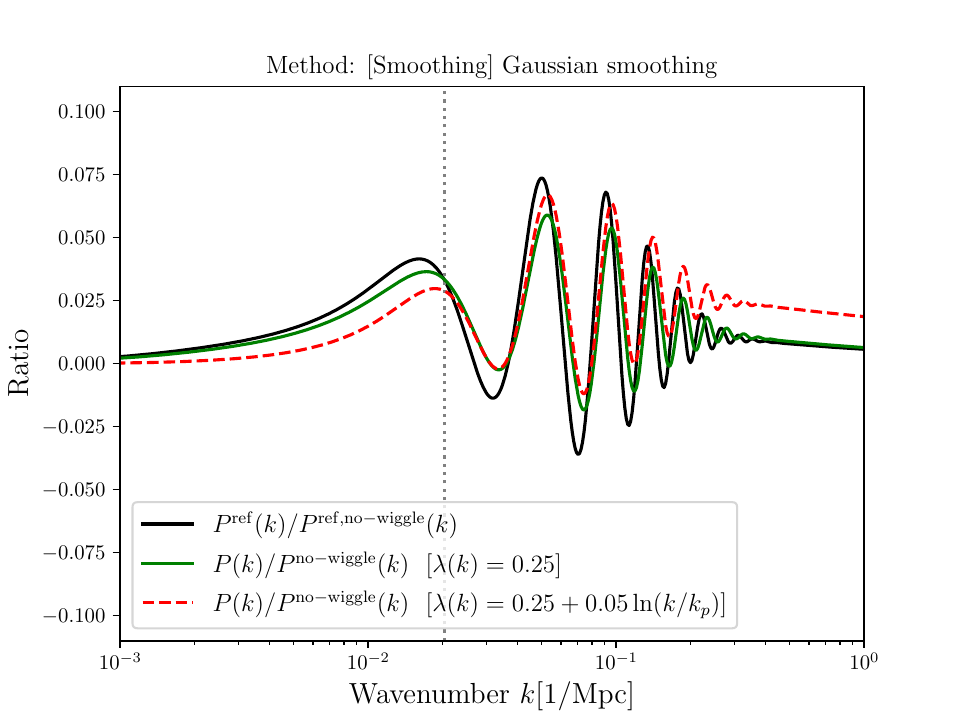}
    \caption{The ratio of the power spectra divided by the dewiggled power spectra for the simple Gaussian smoothing method applied to the fiducial/reference (black) and showcase (green) cosmologies.
    The dashed grey line represents the pivot wavenumber $k_p$ at which the ShapeFit slope (see \cref{sec:ShapeFit}) is  evaluated (translated from $h$/Mpc to 1/Mpc).}
    \label{fig:wiggle_simple_gauss}
\end{figure}
The \textbf{Simple Gaussian} method, 
proposed in \cite{vlah2016perturbation}, involves a simple convolution of the power spectrum with a Gaussian function of the type
\begin{equation}
    f(k, k_0) = \frac{1}{\sqrt{
    2\pi \lambda^2}} \exp\left(-\frac{(\ln k-\ln k_0)^2}{2\lambda^2}\right)~,
\end{equation}
yielding a broadband de-wiggled power spectrum $P^\mathrm{no-wiggle}(k_0) = \int f(k,k_0) P(k) d\ln k$ where $\lambda$ is the width of the Gaussian.  Here for illustrative purposes we set $\lambda=0.25/\mathrm{Mpc}$ and show the resulting wiggle/no-wiggle split in \cref{fig:wiggle_simple_gauss}. Since the averaging length is somewhat small, we do not see a large impact on the broadband shape, while the oscillations are mostly removed which is the goal here. We also show a case of scale dependence with $\lambda \to \lambda(k)$ as advertised in Ref.~\cite{vlah2016perturbation} in \cref{fig:wiggle_simple_gauss} as well, noting that this results in further suppression of the first peak and/or a larger offset at large $k$ due to stronger smoothing there.

The main disadvantage of this method (and similar ones) is that small wiggles can remain in the power spectrum even after the averaging (see e.g, discussion in \cite{vlah2016perturbation}); if the smoothing scale is increased too much, the broadband begins to be strongly affected.

\subsection{Fitting smooth functions} \label{ssec:fitting}
The idea of this approach is to fit sufficiently smooth functions to the overall power spectrum. Such functions are optimally able to fit the overall broadband shape but lack enough freedom to also follow the oscillations. If the oscillations cannot be fit, a maximum likelihood estimation (like the least-square difference algorithm) will typically balance the amplitude of the oscillations below and above the fit, which is a good approximation of the true broadband. The 
big issue that these methods face is that typically the broadband spectrum is difficult to model with simple elementary functions at sufficient accuracy, but just going to higher-order expansions risks starting to fit the oscillations as well. We discuss how the specific algorithms avoid this problem for each algorithm below.

\begin{figure}[h]
    \centering
    \includegraphics[width=0.5\linewidth]{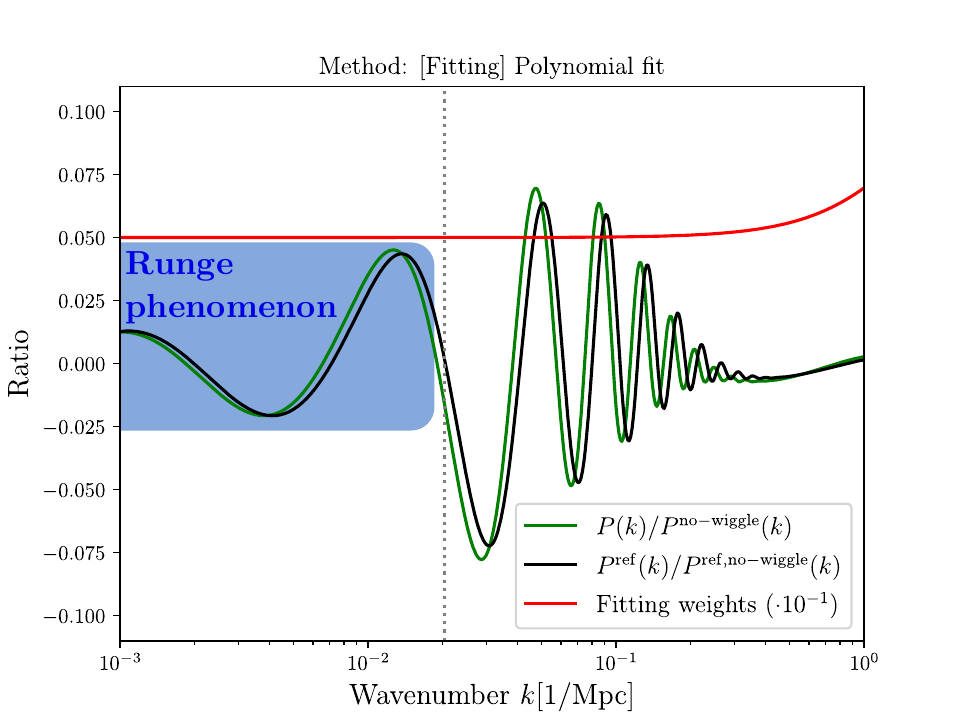}
    \caption{Same as \cref{fig:wiggle_simple_gauss} but for the polynomial fitting method. The red line represents the weights used for the fit according to \cref{eq:poly_weights}. The blue region highlights the scales in which Runge's phenomenon becomes relevant.}
    \label{fig:polynomial_fit}
\end{figure}

\paragraph{Polynomial fit}
This method has been initially proposed in \cite[app.~A]{hinton2016extraction} for the WiggleZ survey. The final selected method involved fitting a polynomial of degree $n$ to the power spectrum. Of course a high-order polynomial will naturally also fit the oscillations. To counteract this issue, the authors down-weighted the wavenumbers according to the formula
\begin{equation} \label{eq:poly_weights}
    w = 1 - \alpha \exp[-k^2/(2\sigma^2)]~,
\end{equation}
which for a fine-tuned choice of 
$n$, the Gaussian width $\sigma$, and the weighting amplitude $\alpha$, down-weights the region containing the BAO oscillations (and the larger scales) compared to the overall broadband fit. We choose the same parameters as \cite[app.~A]{hinton2016extraction}, namely $n=13$, $\sigma=1/\mathrm{Mpc}$, $\alpha=0.5$ and fit the polynomial in log-log space. We have checked that up-weighting smaller wavenumbers in \cref{eq:poly_weights} below the BAO oscillations does not yield any significant improvement.

The resulting wiggle component of the reference and showcase power spectra (as per  \cref{tab:cosmologies})  are shown in \cref{fig:polynomial_fit}. 
The polynomial representation introduces unphysical oscillations 
even where no BAO are present due to the well known Runge-phenomenon \cite{Epperson01041987}: polynomial approximations often do not converge towards the true function but are subject to a type of \enquote*{ringing} around the true function. The corresponding wavenumbers for which this is relevant are highlighted in blue in \cref{fig:polynomial_fit}.
Increasing the polynomial order to reduce these ringing artifacts, however, also allows the polynomial to track the physical BAO oscillations.
Therefore we do not consider this method as a good de-wiggling method (see also \cref{tab:dewiggles}).

\begin{figure}[h]
    \centering
    \includegraphics[width=0.49\linewidth]{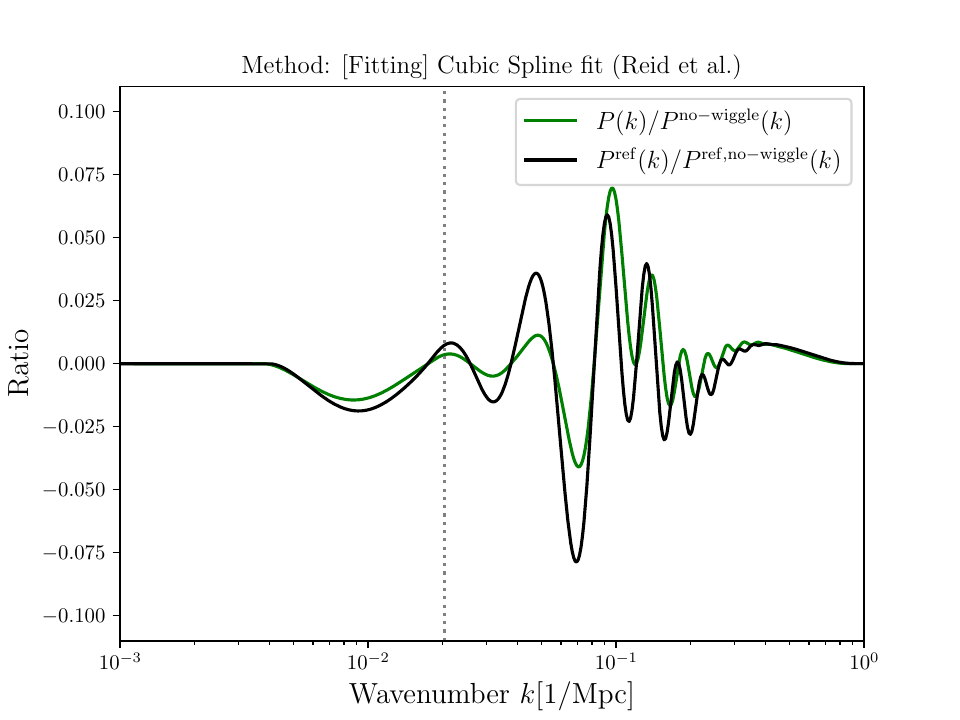}    \includegraphics[width=0.49\linewidth]{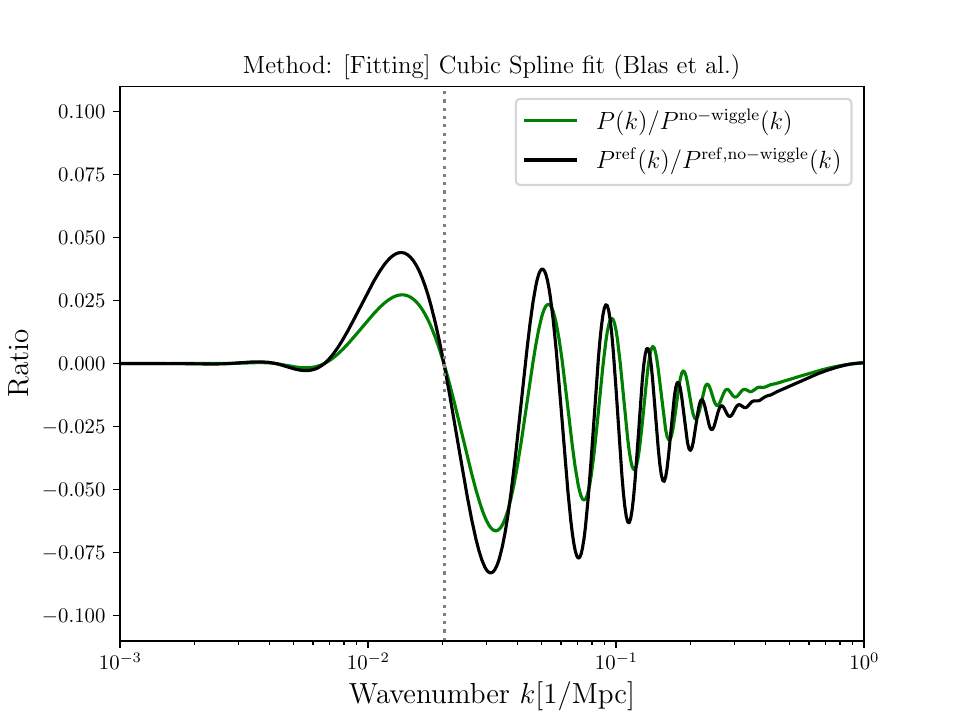}
    \caption{Same as \cref{fig:wiggle_simple_gauss}, but for the cubic fitting methods.
    Left: The method adopted in \cite{2010MNRAS.404...60R}, with the modification of also including additional points left and right of the interval [$10^{-3}$/Mpc, $1$/Mpc]. Right: The method adopted in \cite[app.~B]{blas2016time}.}    \label{fig:cubic_fit}
\end{figure}

\paragraph{Cubic Spline fit}
The authors of \cite[app.~B]{blas2016time} propose a simple approach of fitting a cubic spline function (see \cref{app:ssec:cubic}) to the power spectrum, choosing only a small number of points ($\sim 20$) to the right and left of BAO scales, as well as an additional pivot point at $k=0.03 h/\mathrm{Mpc}$. 
Using cubic interpolation between the pivot point and the left/right sides beyond the BAO ensures that the wiggles cannot be traced. Defining exactly where the BAO start and end is crucial for this approach and the choice depends on cosmology.
In our case, we select for the fiducial model the region [$10^{-2}$/Mpc, 0.45/Mpc] and rescale it by $r_d/r_d^\mathrm{fid}$ for other  models.

Such a cubic spline approach is very similar to the one originally adopted in \cite{2010MNRAS.404...60R}, using different node points. We also implement the method of \cite{2010MNRAS.404...60R} for reference (which uses a selection of 8 node points that lie mostly inside the BAO region). Using just these 8 points does not result in a good overall fit. Instead we also include 20 points in the region outside the BAO interval [$10^{-3}$/Mpc, $1$/Mpc] in the fit in order to force the de-wiggled power spectrum to coincide with the original power spectrum there.

As evident in \cref{fig:cubic_fit}, neither method is entirely satisfactory. With our implementation of the algorithm of \cite{2010MNRAS.404...60R} (left panel), the fit is forced to have the linear and de-wiggled power spectra coincide at all 8 nodes within the BAO region, resulting in a highly distorted fit there (and small shallow residual wiggles in the no-wiggle power spectrum) -- the amplitude of the wiggle/no-wiggle ratio does not follow the expected shape of the BAO.
It is likely that the choice of nodes in \cite{2010MNRAS.404...60R} was optimized for a particular cosmology and would have to be changed when considering any other cosmology.

On the other hand, the approach of \cite[app.~B]{blas2016time} using only a single pivot point in the BAO region works significantly better. Here too, however, the result is highly dependent on the interval chosen to contain the BAO wiggles, as well as the number of points outside the BAO region. This is also true in particular for the subtraction of the first peak and therefore the slope $m$ according to \cref{eq:mdef}. For example, doubling the number of points from 20 to 40 outside the BAO region gives slope differences $|\Delta m|$ up to 0.02 and moving the left edge
of the BAO region from $10^{-2}$/Mpc to $5 \cdot 10^{-2}$/Mpc gives $|\Delta m|$ up to 0.05 (see also \cref{ssec:derivatives}).

\begin{figure}[h]
    \centering
    \includegraphics[width=0.49\linewidth]{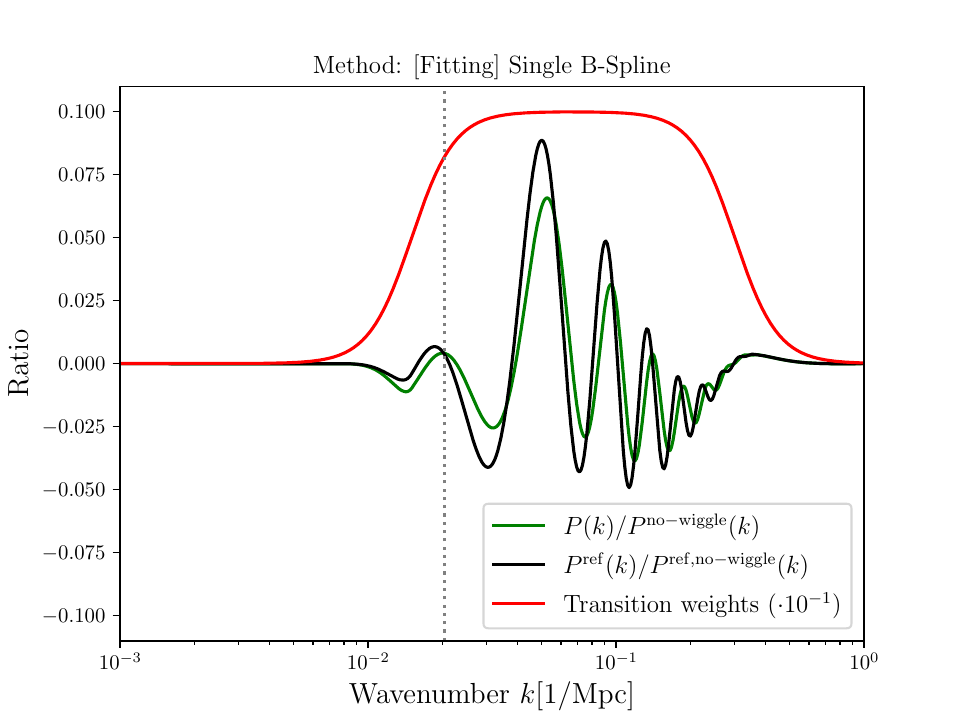}
    \includegraphics[width=0.49\linewidth]{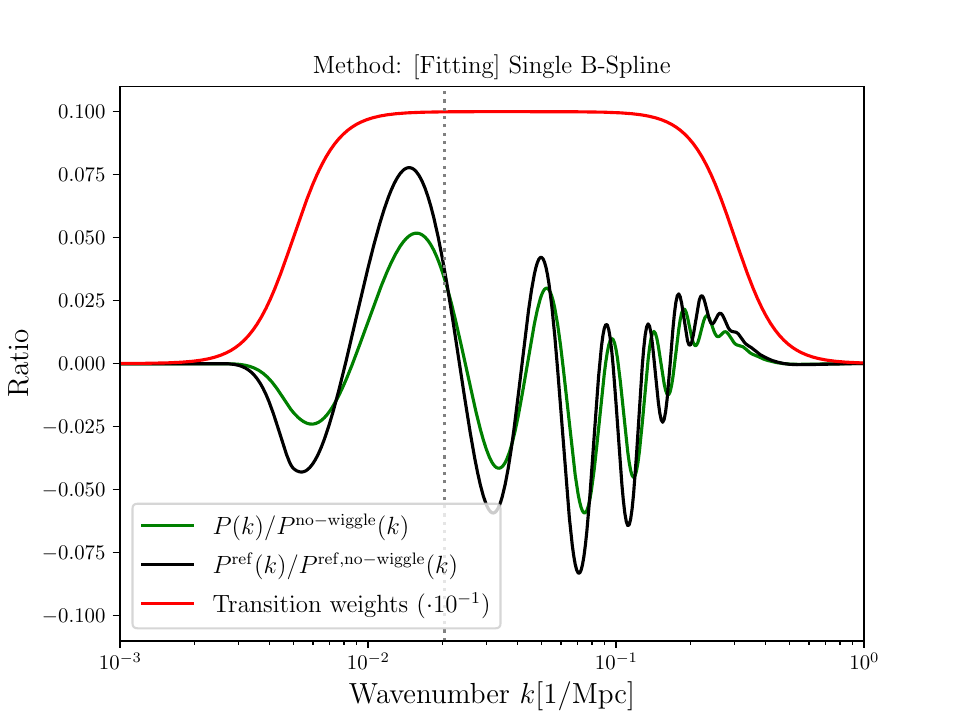}\\
    \includegraphics[width=0.49\linewidth]{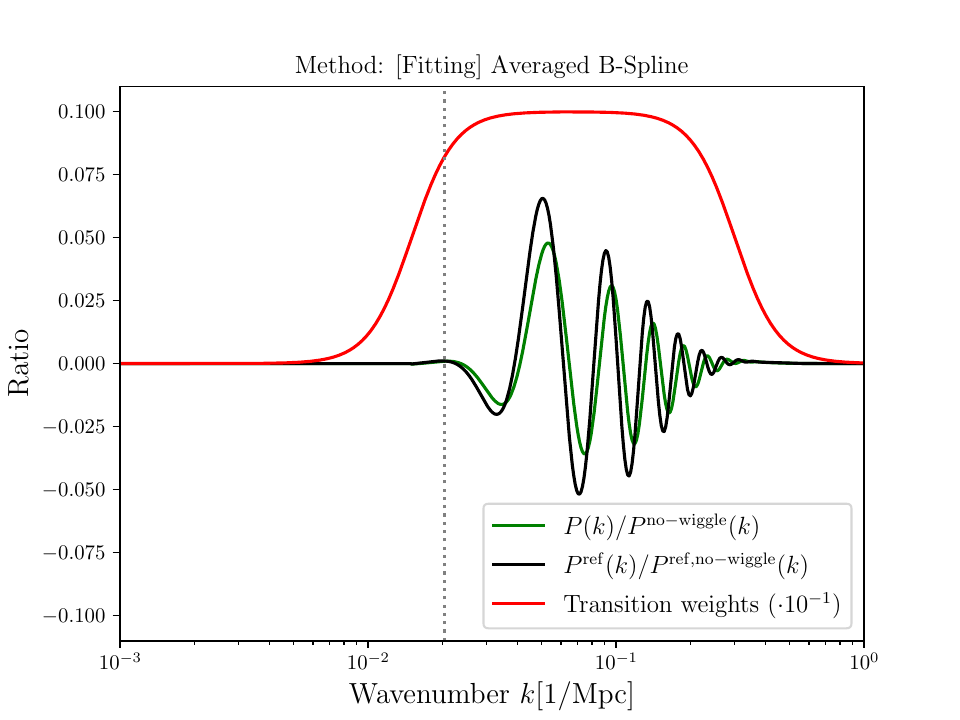}
    \includegraphics[width=0.49\linewidth]{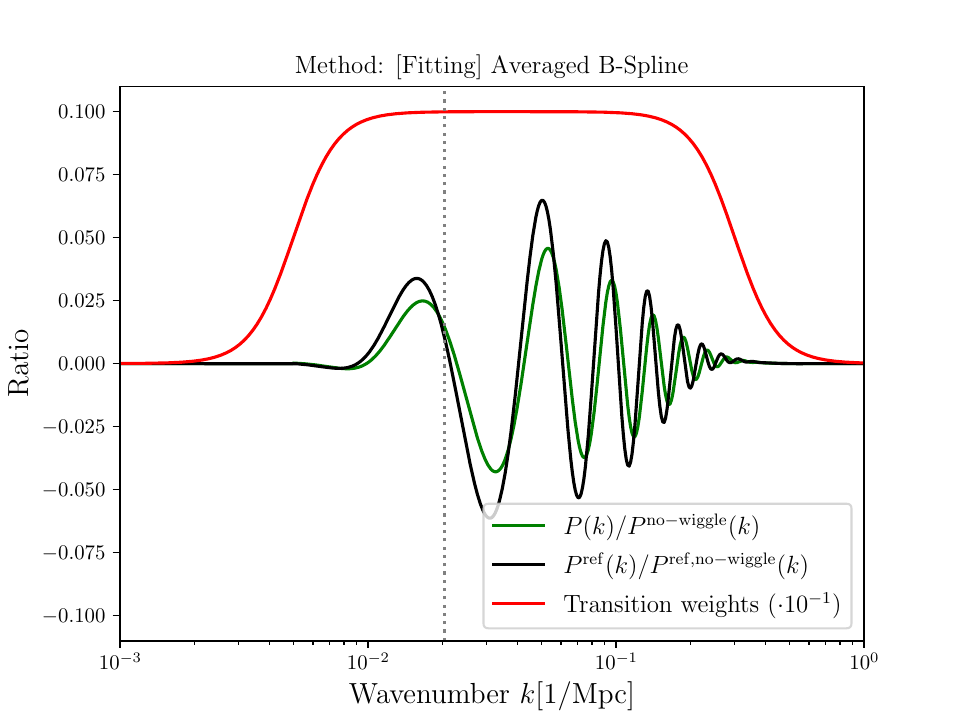}
    \caption{Same as \cref{fig:wiggle_simple_gauss}, but for spline fitting methods. The red lines denote the weights: the fitting region corresponds to the flat portion of the red line. Top: The method for a single spline. Bottom: The method for the average of multiple splines. Left: A narrow replacement window (which would impact the slope at $k_p$). Right: A broad replacement window (which would not impact the slope at $k_p$).
    }
    \label{fig:bspline_fit}
\end{figure}

\paragraph{B-Spline fit}
In \cite[app.~A]{vlah2016perturbation} the authors provide several different de-wiggling methods, one of which is based on a fit with B-splines (generalizing the cubic spline to higher polynomial degrees $d$ and a different number of knots $n_k$\,, see \cref{app:ssec:univariate_spline} for more details).
We show the case of $d=2, n_k=8$ in the top panels of \cref{fig:bspline_fit}.

The authors of \cite[app.~A]{vlah2016perturbation} combine the spline approximations from multiple combinations of $d,n_k$, imposing additional conditions in the weighed sum. Here for simplicity we use equal weights. The bottom panels of  \cref{fig:bspline_fit} correspond to the case when we include $d=\{2,3,4,5\}$ with $n_k=\{2d+4,2d+6\}$ and simply average them with equal weights.

For best results, we find that the B-spline fit must be restricted to a fitting region (which includes the BAO) between some $k_\mathrm{start}$ and $k_\mathrm{end}$\,.
Since the B-spline is not guaranteed to be continuous with the original power spectrum at the edges of the fitting interval, the method of \cref{app:replacement} must be applied to ensure continuity. 
In particular the fitting region, $[k_\mathrm{start}$, $k_\mathrm{end}]$, is padded with a transition region defined by $k_\mathrm{min}<k_\mathrm{start}$, and $k_\mathrm{max}>k_\mathrm{end}$ where a \enquote{smooth replacement} according to \cref{app:replacement} is performed. In the bottom panels of \cref{fig:bspline_fit} we choose $\ln k_\mathrm{min} = \ln k_\mathrm{start} - 3 \Delta$ where $\Delta = 0.4$ is the \enquote{replacement width} of \cref{app:replacement} and $\ln k_\mathrm{max} = \ln k_\mathrm{end} + 2 \Delta$.

\Cref{fig:bspline_fit} compares wider and narrower B-spline fitting regions. For a single spline the method reacts quite strongly to wider fitting regions
(smaller $k_\mathrm{start}$):  
the fit begins to diverge from the broadband for the wider fitting region top right panel) -- the oscillations are not symmetric around $0$ in the ratio. In contrast, the average of multiple splines remains mostly stable even with a much wider fitting window (bottom right panel). For further comparisons, we choose a combination of the more robust average of multiple splines together with a wider fitting region (for which the wiggle/de-wiggle ratio at the pivot wavenumber of \cref{sec:ShapeFit} is not impacted by the replacement method of \cref{app:replacement}).

\begin{figure}[h]
    \centering
    \includegraphics[width=0.49\linewidth]{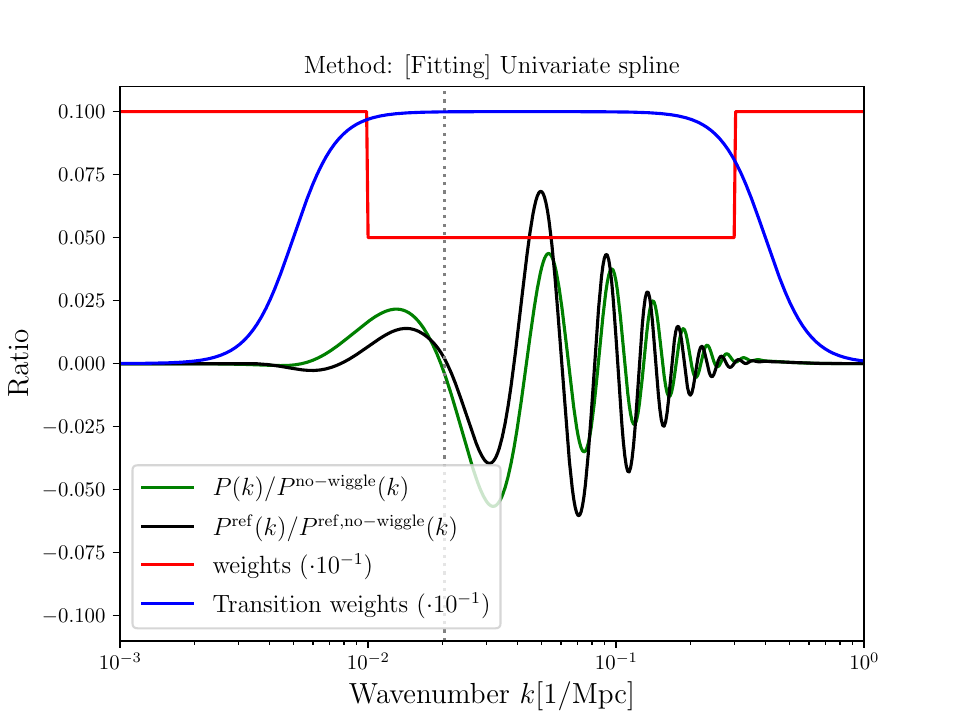}
    \includegraphics[width=0.49\linewidth]{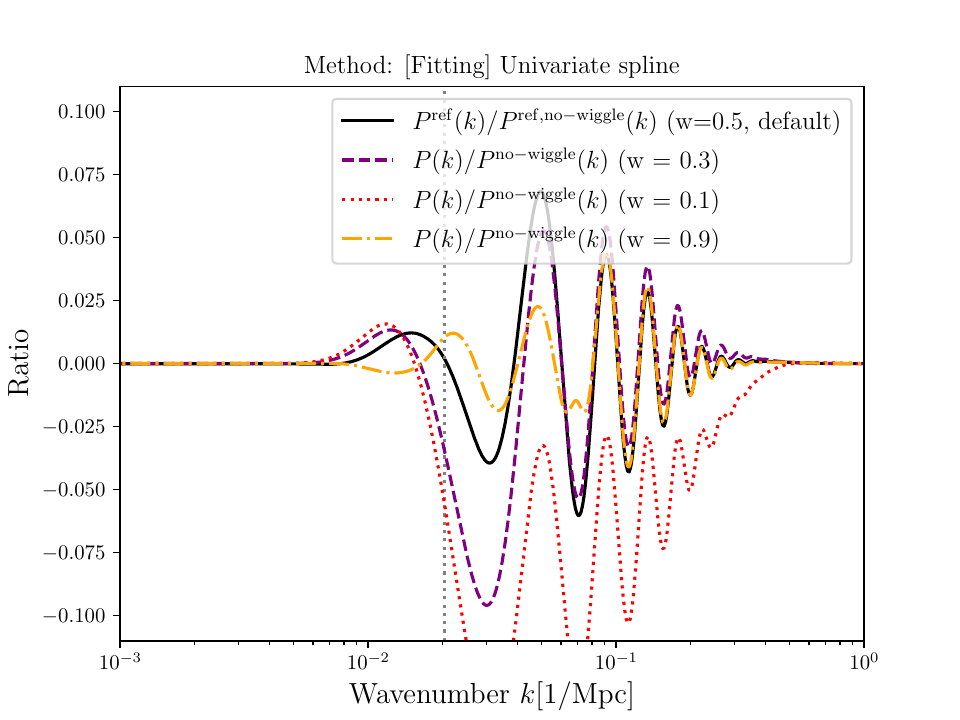}
    \caption{Same as \cref{fig:wiggle_simple_gauss}, but for a univariate spline fitting method. Left:  The wiggle/no-wiggle ratio for the univariate spline fitting method applied to a given model and fiducial power spectrum. In red we show the used weights, and in blue the weights used for the replacement scheme of \cref{app:replacement}. Right: Variations on the fiducial cosmology for different weight suppressions $w$ ranging from $w=0.1$ to $w=0.9$.}
    \label{fig:univariate_fit}
\end{figure}
\paragraph{Univariate spline fit}
This idea is simply based on fitting the overall power spectrum with a univariate spline, the concept of which is explained in detail in \cref{app:ssec:univariate_spline}. For this case we weigh the different parts of the power spectrum according to weights that are unity outside the BAO range, and suppressed by a factor $w=0.5$ inside the BAO range\footnote{We take the range here to lie between $k_\mathrm{start} = 0.01/\mathrm{Mpc}$ and $k_\mathrm{end} = 0.3/\mathrm{Mpc}$ for the fiducial cosmology, rescaled by $r_d/r_d^\mathrm{fid}$ for other cosmologies.} and use a smoothing strength $\mathfrak{s}=0.01$. Finally, we only employ the de-wiggling algorithm in the aforementioned range, using the technique of \cref{app:replacement} with $\Delta=0.4$ to ensure a smooth transition.

In \cref{fig:univariate_fit} (left panel) we show the resulting wiggle/no-wiggle decomposition for the power spectrum. There is not a strong dependence on the precise covered wavenumber range. However, there is a strong dependence on the suppression weight $w$, as we show in the right panel of \cref{fig:univariate_fit}.

\begin{figure}[h]
    \centering
    \includegraphics[width=0.49\linewidth]{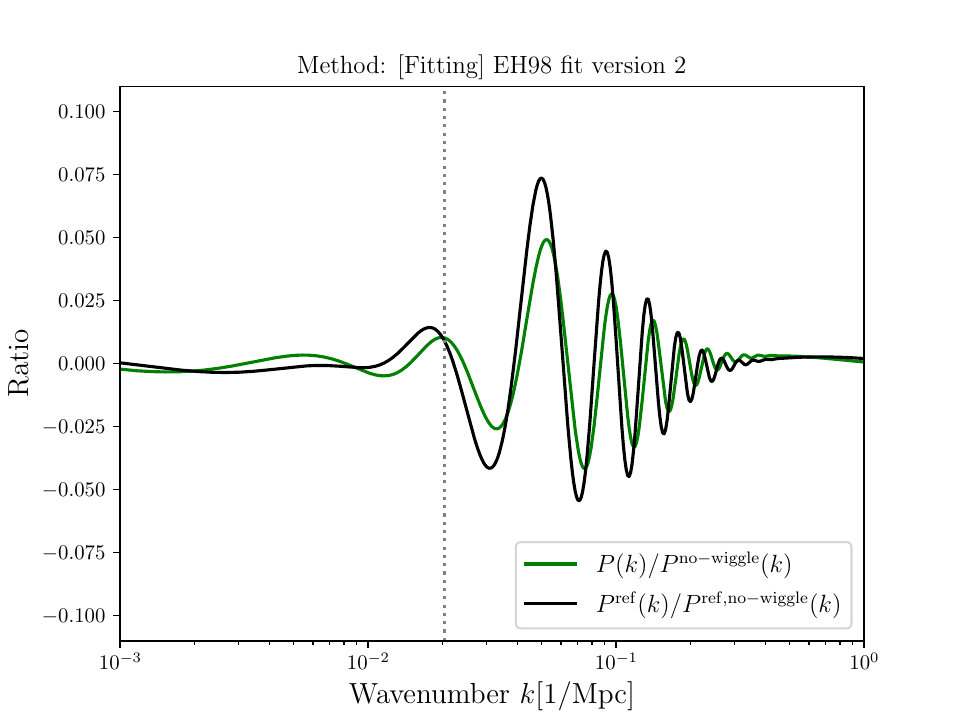}
    \caption{Same as \cref{fig:wiggle_simple_gauss}, but for the EH fitting method.}
    \label{fig:eh_fit}
\end{figure}

\paragraph{EH fit}
The authors of \cite[app.~B]{blas2016time} also propose an approach based on the EH98 formula as a fitting function, by generalizing its parameter dependencies, as well as including an additional correction. 
By slightly re-formulating their formula to be closer to the original proposal of \cite{eisenstein1998baryonic,Eisenstein:1997jh},\footnote{In particular, we have re-parameterized  $c_1 = \sqrt{d_1} d_2$, $c_2 = d_3/\sqrt{d_1}$, and $c_3 = (1-\sqrt{d_1})/\sqrt{d_1}$, and $c_{\geq 4} = c_{\geq 4}$ (where the $c_i$ are the coefficients in \cite[app.~B]{blas2016time}). Given that the mapping between the $\{c_i\}$ and $\{d_i\}$ is one-to-one, we don't observe any differences between the formulations.}
our fitting formula reads
\begin{align}
    P(k) &\approx A k^{n} T(k)^2 (1+\Delta(k)) \\
    T(k) & = L(k) \cdot [L(k)+C(k) \kappa^2]^{-1}\label{eq:EH_fit_tk}\\
    L(k) & = \ln(e+\sqrt{d_1} d_2 \kappa)\label{eq:EH_fit_lk}\\
    C(k) & = 14.4 + 325 / (1+60.5 \kappa^{1.08})\\
    \kappa &= d_3 k \cdot \left(\sqrt{d_1}+\frac{(1-\sqrt{d_1})}{1+(d_4 k)^4}\right)^{-1}\label{eq:EH_fit_kappa}\\
    \Delta(k) & = d_5 \tanh\left(1+\log(d_6 \kappa)/d_7\right)
\end{align}
with the parameters $A$, $n$, and $d_i$ being fitted (in log-log space) to the power spectrum.

The resulting wiggle/no-wiggle decomposition is shown in \cref{fig:eh_fit}, and the wiggles seem to be well characterized. There are some very small residual oscillations for $k < 0.01/\mathrm{Mpc}$, but this is largely irrelevant for most applications.

\subsection{Inflections}\label{ssec:inflections}

The algorithms in this category are based on fitting a smooth function through the inflection points of the oscillations to determine the broadband behavior. The idea is that in the absence of the broadband component the inflection points coincide with the zero-points of the oscillations. Although conceptually simple, the actual implementation must carefully avoid a number of common pitfalls.
In particular, the inflection points of the power spectrum are typically biased with respect to the true zeros of the BAO due to the broadband slope.\footnote{To see this very quickly, consider a function $f(x)$ to consist of an oscillation plus some broadband $f(x) = \sin(x) + b(x)$. If the broadband is un-curved $b(x) = a x + b$, then the inflection points $x_n = n \pi$ (determined by $f''(x_n)=0$) coincide with the zeros of the oscillations. Therefore, the function evaluated at the inflection points follows the broadband $f(x_n) = b(x_n)$. If one now has a curved broadband instead, $b(x) = \gamma x^2$, then the inflection points are $x_{2n} = 2n\pi + \arcsin{2 \gamma}$ and $x_{2n+1} = 2 n \pi + [\pi-\arcsin{2 \gamma}]$ and therefore do not coincide with the zeros of the oscillations (which are still $n \pi$, of course). Therefore, the function evaluated at the inflection points does not follow the broadband, for example $f(x_{2n}) = b(x_{2n}) + \sin(\arcsin(2\gamma)+2n\pi) = b(x_{2n})+2\gamma \neq b(x_{2n})$.} This creates a circular problem: to accurately determine the zero-points of the oscillations (to remove the wiggles), the de-wiggled power spectrum already needs to be known. The circularity problem is typically solved 
by first adopting  approximations of the broadband, enabling a nearly un-biased determination of the zeros of the oscillations, which in turn allows for the determination of the true broadband.

In addition one must ensure that the numerically determined inflections are robust with respect to the finite wavenumber sampling of the provided power spectrum.

One of the first implementations\footnote{Due to Mario Ballardini} (which we dub \enquote{Cubic Inflections}) is implemented by default in the \texttt{MontePython} code \cite{audren2013conservative, brinckmann2019montepython}, while a second version (which we dub \enquote{EH inflections})\footnote{Implemented by Samuel Brieden for ShapeFit\cite{brieden2021shapefit}.} is one of the original options in ShapeFit (see \cite{brieden2021shapefit}). 

Both algorithms use estimates of the broadband to get approximate zero-points of the oscillations by dividing of the true function by the approximation (assuming the oscillations are multiplicative). The \enquote{Cubic Inflections} algorithm uses a cubic spline approximation of the overall power spectrum as an approximate broadband, while the \enquote{EH inflections} algorithm uses the EH98 formula.
The \enquote{EH inflections} algorithm also uses a simpler helper algorithm, dubbed here \enquote{Gradient inflections}\footnote{A more recent option in ShapeFit also provided by Samuel Brieden. }.
In all cases, the range containing the BAO features, $[k_{\rm start},k_{\rm end}]$, is an input. A suitable range containing the BAO is estimated by multiplying a fiducial range appropriately with the sound horizon scale.

We present and compare these algorithms below.

\begin{figure}[h]
    \centering
    \includegraphics[width=0.49\linewidth]{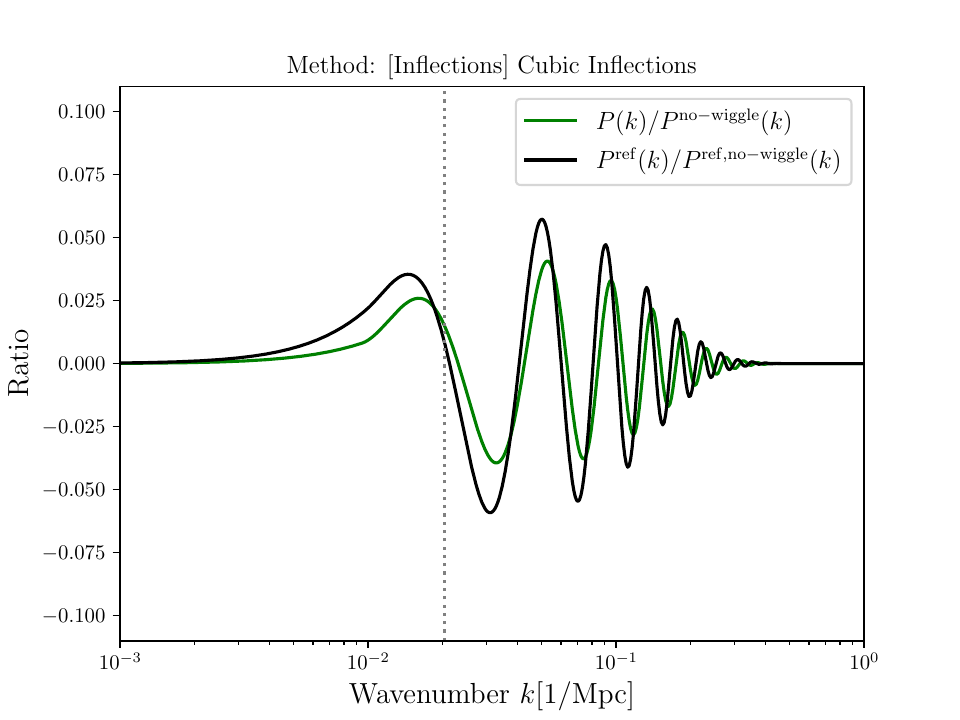}
    \includegraphics[width=0.49\linewidth]{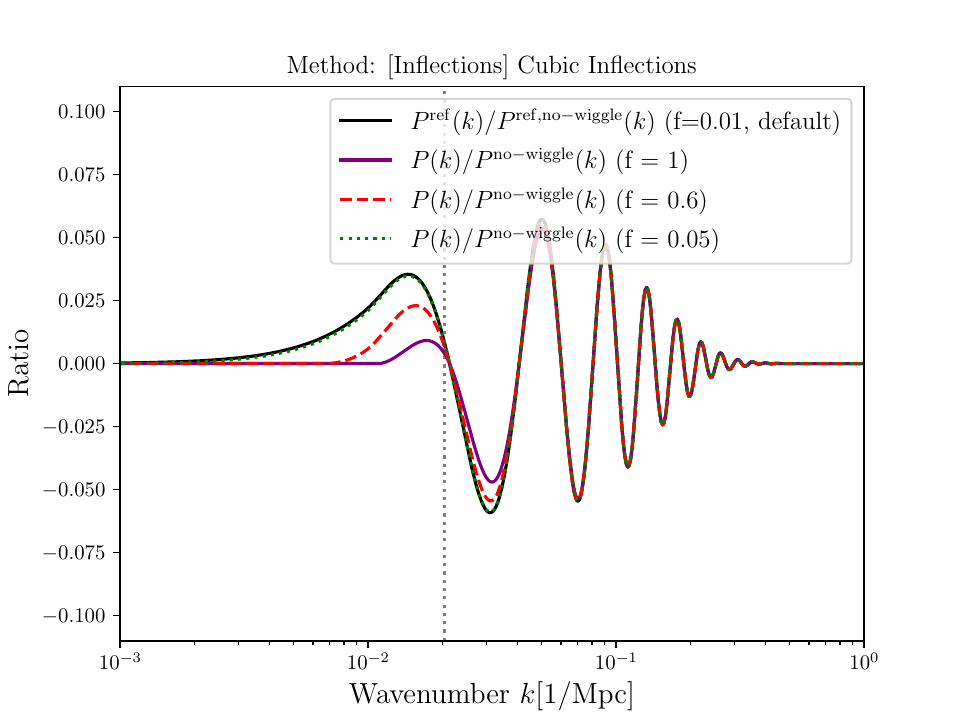}
    \caption{Same as \cref{fig:wiggle_simple_gauss}, but for the Cubic Inflections method. Left: The wiggle/no-wiggle ratio for the cubic inflections method applied to a given model and fiducial power spectrum. Right: Variations with different starting wavenumber wavenumbers (original wavenumber multiplied by $f\leq1$).}
    \label{fig:cubic_inflections}
\end{figure}

\paragraph{Cubic Inflections}

This algorithm uses the following steps:
\begin{enumerate}
    \item A cubic spline is used to fit the power spectrum outside the range containing the BAO features, essentially providing a smooth cubic approximation in the BAO region and a close-to-exact approximation outside of it.
    \item The ratio of the power spectrum and the smooth cubic approximation is used to define approximately what should constitute as the wiggles.
    \item The second derivative of the wiggles is then interpolated using a cubic univariate spline (see \cref{app:ssec:univariate_spline}) of smoothing strength $\mathfrak{s}=1$. This essentially fits the second derivatives with a function that ensures a certain degree of smoothness as in principle the second derivative of a Cubic spline function can be quite discontinuous between the different knots (see \cref{app:ssec:cubic}).
    \item The zeroes of the smoothed second derivative of the wiggles are calculated. These are the inflection points of the oscillations. Note that any zeros outside the BAO region are removed.
    \item A spline is fitted through the inflection points as well as the wave numbers from regions outside the BAO range, giving a smooth function (which captures the deviation of the true broadband from the cubic approximation).
    \item The cubic approximation is then multiplied by this smooth function to recover the true broadband.
\end{enumerate}

We show the results of this algorithm in \cref{fig:cubic_inflections}. As implemented here, we do not use the default settings of  \texttt{MontePython} code \cite{audren2013conservative, brinckmann2019montepython} whereby the beginning of the BAO region $k_{\rm start} = 0.028/\mathrm{Mpc}$ coincides with the turnover of the power spectrum. This would force the wiggle-to-nowiggle ratio to be zero at this wavenumber. Here we allow for additional freedom to $k_{\rm start}$ through a multiplicative rescaling factor $f<1$. Additionally, we scale the starting/ending wavenumbers of the BAO region proportionally to the sound horizon of the given cosmology.

This choice is motivated by the fact that the peak of the power spectrum coincides with the first BAO wiggle (see also \cref{app:illdefined}), so allowing the first peak to also be removed can be a reasonable choice. We show in \cref{fig:cubic_inflections} (right panel) that indeed such an approach converges for $f \to 0.01$, which we therefore take as our fiducial value.

\begin{figure}[h]
    \centering
    \includegraphics[width=0.49\linewidth]{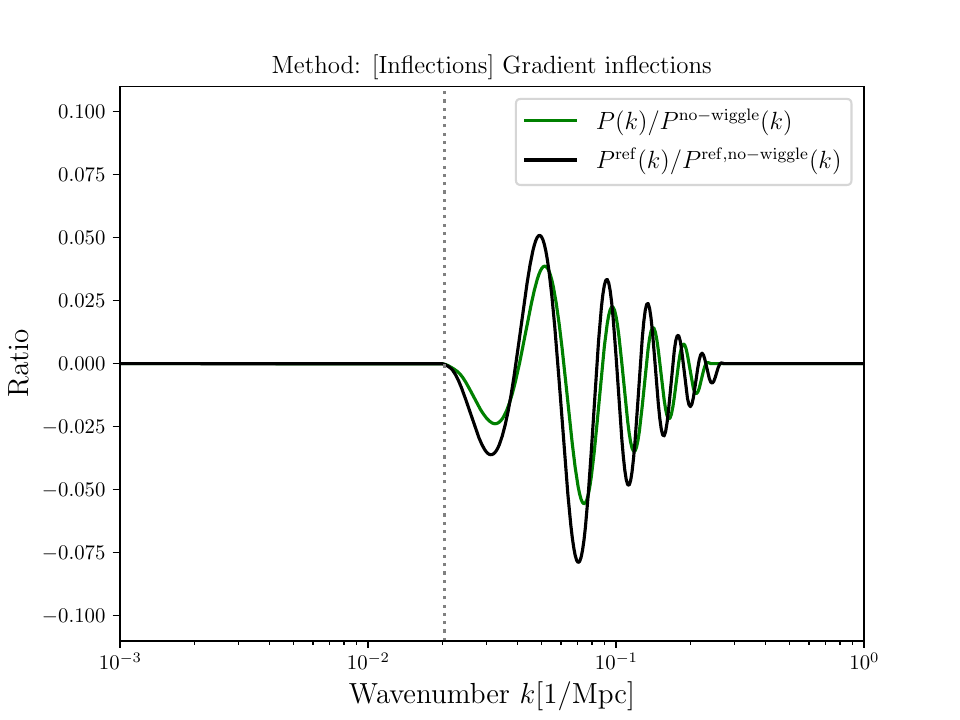} \\  \includegraphics[width=0.49\linewidth]{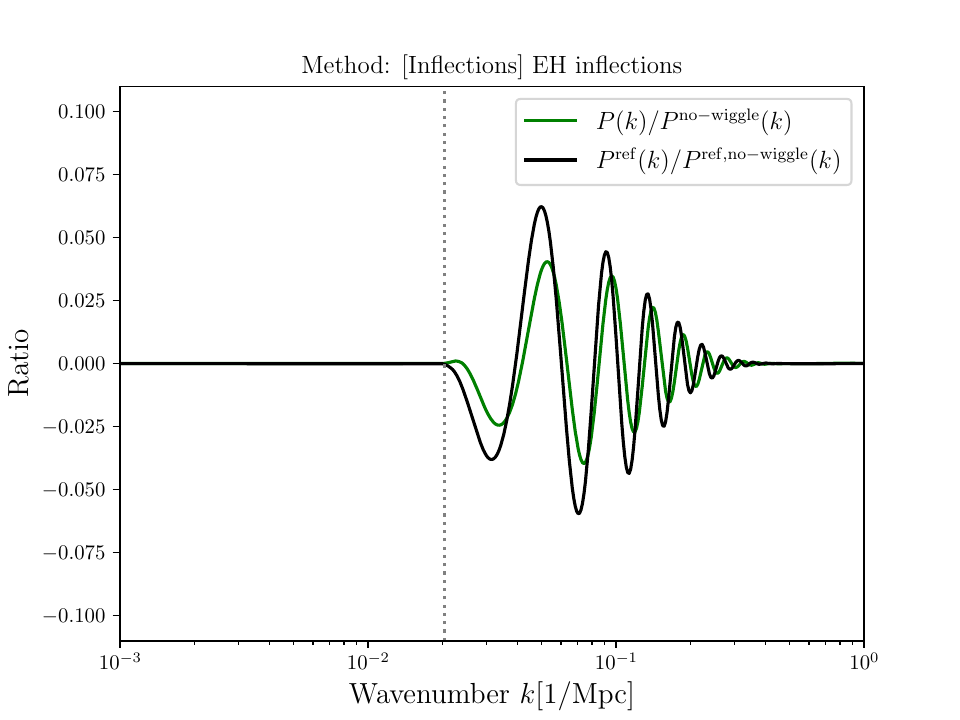}    \includegraphics[width=0.49\linewidth]{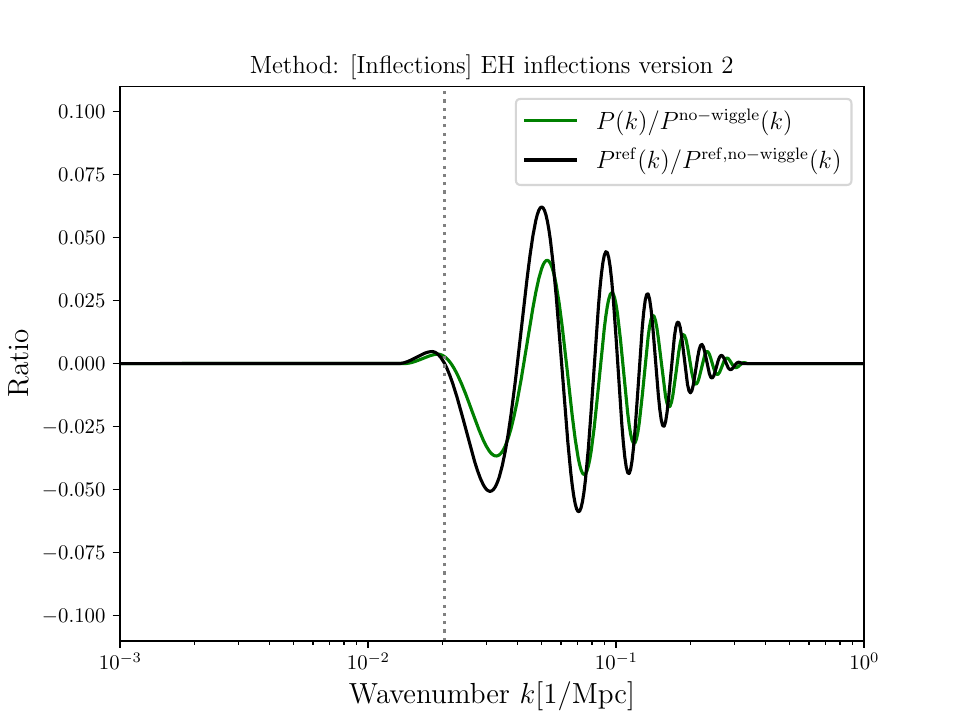}
    \caption{Same as \cref{fig:wiggle_simple_gauss}, but for the different inflections methods. Top: Just the 'gradient inflections' algorithm. Bottom left: The first version of the 'EH inflections' algorithm. Bottom right: The second version of the 'EH inflections' algorithm.}
    \label{fig:EH_inflections}
\end{figure}

\paragraph{Gradient Inflections}

This  is a helper algorithm  which is used for the \enquote{EH inflections} methods discussed below.
Given a properly normalized wiggly spectrum, instead of determining the zeros of the second derivatives (as above), this method attempts to find the peaks of the gradient. The steps are as follows:

\begin{enumerate}
    \item The derivative of the power spectrum is computed with a naive forward difference method.
    \item The maxima and minima of this gradient are computed (and only those in the predetermined BAO range are kept), effectively giving the (possibly biased) zeroes of the wiggles.
    \item The parts before and after the BAO region as well as the intermediate power spectrum at the location of the zeros of the wiggles are interpolated separately for the maxima and the minima of the gradient, and the two functions are averaged.
\end{enumerate}

There are two different implemented possibilities for removing the broadband from the power spectrum together with the \enquote{Gradient Inflections} method to remove the remaining wiggles. We discuss these two methods below.

\paragraph{EH Inflections}

For this method the \enquote{Gradient Inflections} algorithm is applied to the ratio of the power spectrum and the EH98 formula. The idea is that the rescaled EH98 formula might not precisely reproduce the power spectrum (or its amplitude), but it reproduces the broadband slope closely enough to give an almost unbiased estimate of the inflection points.

\begin{enumerate}
    \item First, for a chosen fiducial cosmology the power spectrum, $P^{\rm fid}(k)$, is divided by the one given by the EH98 formula EH98${}^{\rm fid}$, and the \enquote{Gradient Inflections} method is applied to this ratio in order to give a very rough approximate fiducial broadband power spectrum. See \cref{eq:nowiggle_v1a} where $G[x]$ is the \enquote{Gradient Inflections} method.
    Hence the ratio $P^\mathrm{no-wiggle,fid}(k)/ \mathrm{EH98}^{\rm fid}(k)$ defines a EH98-to-broadband correction factor for the fiducial cosmology.
    \item For evaluations for any other cosmology, the power spectrum is divided both by the EH98 model and by the fiducial EH98-to-broadband correction factor. Therefore, this ratio will contain the wiggles over an approximately flat baseline (the flatter the closer to the fiducial cosmology). Applying the \enquote{Gradient Inflections} method to this ratio then removes these wiggles with quite high accuracy. The final result is multiplied back by the fiducial no-wiggle power spectrum to restore the correct amplitude, see \cref{eq:nowiggle_v1b}.
\end{enumerate}%
\begin{subequations}
    \begin{align}
        P^\mathrm{no-wiggle,fid}(k) = G\left[\frac{P^\mathrm{fid}(k)}{\mathrm{EH98}^\mathrm{fid}(k)}\right] \cdot \mathrm{EH98}^\mathrm{fid}(k)  \label{eq:nowiggle_v1a}\\
        P^\mathrm{no-wiggle}(k) \approx P^\mathrm{no-wiggle,fid}(k) \cdot G\left[\frac{P(k)/\mathrm{EH98}(k)}{P^\mathrm{no-wiggle,fid}(k)/\mathrm{EH98}^\mathrm{fid}(k)}\right] \label{eq:nowiggle_v1b}
    \end{align}
\end{subequations}

\paragraph{EH Inflections version 2}

The second version of the same algorithm differs only by a few small improvements. 
\Cref{eq:nowiggle_v1a} remains the same, see \cref{eq:nowiggle_v2a}. However, 
compared to \cref{eq:nowiggle_v1b}, in \cref{eq:nowiggle_v2b} the wavenumbers are multiplied by $s = r^\mathrm{fid}_d/r_d$ to shift the power spectrum and align the BAO wiggles with those of the fiducial power spectrum. Thus the ratio $P(k/s)/P^{\rm fid}(k)$ is already mostly smooth. Normalizing this by the ratio of the EH98 transfer functions $\mathrm{EH98}(k)/\mathrm{EH98}^{\rm fid}$ yields a function optimally close to unity with only small residual differences arising from a) imperfect compensation of the terms and b) the general broadband shape deviation from the EH98 formula. The \enquote{Gradient Inflections} algorithm is then applied to this quantity. The overall correct amplitude and shape are then recovered by multiplying with the corresponding terms as shown in \cref{eq:nowiggle_v2b}.

\begin{subequations}
    \begin{align}
        P^\mathrm{no-wiggle,fid}(k) &= G\left[\frac{P^\mathrm{fid}(k)}{\mathrm{EH98}^\mathrm{fid}(k)}\right] \cdot \mathrm{EH98}^\mathrm{fid}(k) \label{eq:nowiggle_v2a}\\
        P^\mathrm{no-wiggle}(k) &\approx \left\{G\left[\frac{P(k /s)/\mathrm{EH98}(k)}{P^\mathrm{fid}(k)/\mathrm{EH98}^\mathrm{fid}(k)}\right] \cdot \frac{\mathrm{EH98}(k)}{\mathrm{EH98}^\mathrm{fid}(k)} \right\}\bigg|_{k=k s} \cdot P^\mathrm{no-wiggle,fid}(k s) \label{eq:nowiggle_v2b}
    \end{align}
\end{subequations}

The results for each of the algorithms are displayed in \cref{fig:EH_inflections}. Interestingly, it is obvious that the BAO oscillation coinciding with the first peak is not being removed (the ratio is close to zero around and before $k_p$). This means that the broadband shape is possibly only recovered after the first peak.

\FloatBarrier
\subsection{Correlation function peak removal}\label{ssec:peakremoval}

The final type of wiggle-removal technique discussed here uses the fact that in the correlation function the BAO peak is well localized and therefore typically easier to remove.  The transformation between power spectrum and correlation function can be generally written as \cite{dodelson2020modern}
\begin{equation}\label{eq:pk_to_xi}
   \xi(r) = \int d\ln k \frac{\sin(kr)}{kr} \left(\frac{P(k) k^3}{ 2\pi^2 }\right)
   \end{equation}
with the inverse transform
\begin{equation}
\label{eq:xi_to_pk}
     P(k) =  \int d\ln r \frac{\sin(kr)}{kr} \left(4\pi \xi(r) r^3\right)
\end{equation}
This type of transformation can either be interpreted as a sine transform or as a Hankel transform~\cite{SECADA1999278}, since the Bessel function of the first kind has $J_{1/2}(x) = \sin(x) \cdot \sqrt{2/(\pi x)}$.

The challenge for this method is to achieve sufficient accuracy in the transformations \cref{eq:pk_to_xi,eq:xi_to_pk} without sacrificing speed which is achieved by utilizing implementations based on the fast Fourier transform.

\begin{figure}[h]
    \centering
    \includegraphics[width=0.49\linewidth]{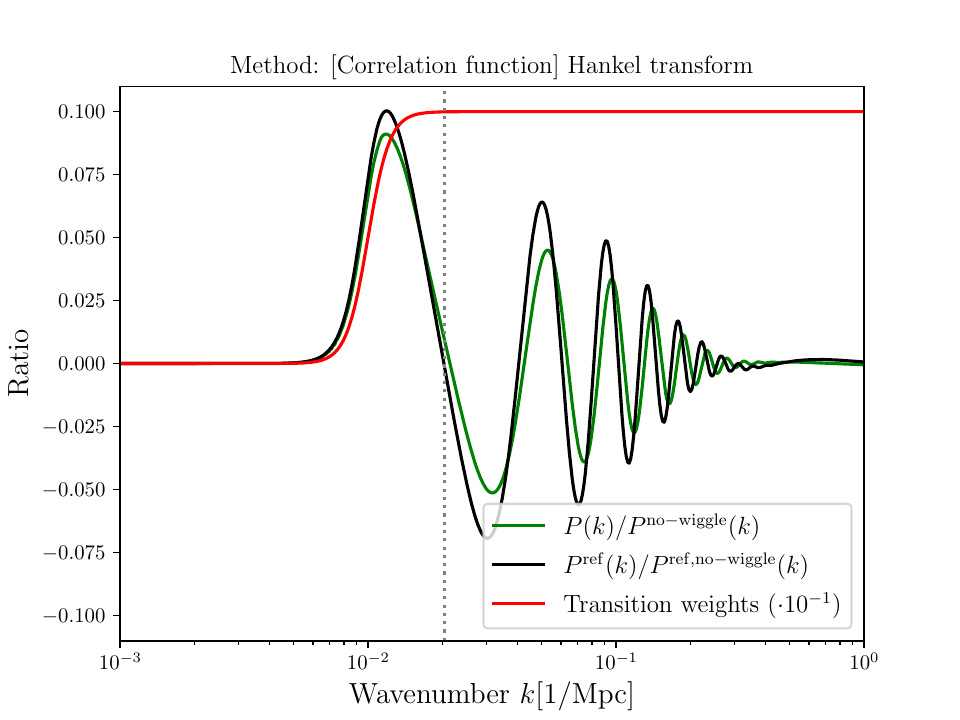} 
    \caption{Same as \cref{fig:wiggle_simple_gauss}, but for the Hankel transform method.}
    \label{fig:hankel_transform}
\end{figure}
\paragraph{Hankel transform}
In \cite[Sec.~2.2.1]{BOSS:2013sqq} a wiggle-removal algorithm was presented for the BOSS survey. We implement it here by logarithmically sampling $k$, computing $k^{3/2}P(k)$, and obtaining the corresponding correlation function via the fast Hankel transform.
We then fit a linear combination of $r^k$ with $k \in \{-3,-2,-1,0,1\}$ to the correlation function between pre-defined bounds outside of the peak range, such as $50-86 \mathrm{Mpc}$ and $150-190 \mathrm{Mpc}$. The resulting interpolated correlation function is then  transformed back into a power spectrum using \cref{eq:xi_to_pk}. Naturally the range of the scales that lie outside the peak range need to be adjusted to the cosmology 
(this could be done automatically, although for our tests a manual scaling by a factor of $1/s = r_d/r_d^\mathrm{fid}$ works well enough). To avoid high-frequency oscillations induced by the discrete sampling in the transformation, we smooth the final result with a univariate spline (see \cref{app:ssec:univariate_spline}) with $\mathfrak{s}=10^{-2}$. Finally, we replace the original power spectrum according to \cref{app:replacement}, taking $k_\mathrm{max} \to \infty$, $k_\mathrm{min}=10^{-2}/\mathrm{Mpc}$ and using a shorter width of $\Delta=0.2$.

The problem for this method is that while the higher peaks are correctly captured, the oscillations at and below the first peak diverge (at least in the present numerical implementation) -- and we note that this algorithm was not necessarily designed to capture the $P(k)$ at smaller wavenumbers than the first peak, see also the top left panel of \cite[Fig.~2]{BOSS:2013sqq}.

\begin{figure}[h]
    \centering
    \includegraphics[width=0.49\linewidth]{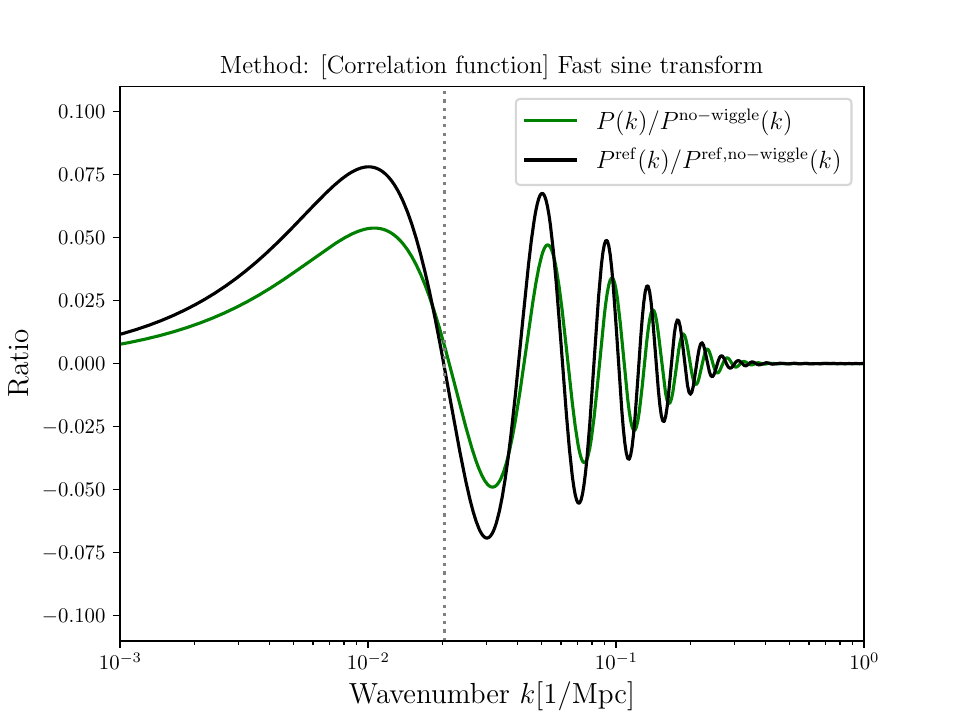} 
    \caption{Same as \cref{fig:wiggle_simple_gauss}, but for the Fast sine transform method.}
    \label{fig:fast_sine_transform}
\end{figure}
\paragraph{Fast sine transform}
This algorithm is one of the most frequently used in cosmology. It has been proposed already in \cite{Hamann2010} and continued to be developed in the subsequent years, for example in \cite[App.~D]{Wallisch:2018rzj} or recently in \cite[Sec.~4.2]{chudaykin2020nonlinear} (this is the version we use), and represents the state-of-the-art, as it is used for example in \cite{Maus:2024dzi,Lai:2024bpl} for the DESI analysis. Ref. \cite{Lai:2024bpl} also includes a comparison to a subset of the other algorithms presented in this work (the \textbf{EH Inflections} and \textbf{Polynomial fit} algorithms). 

Following \cite[Sec.~4.2]{chudaykin2020nonlinear} we sample $\ln(kP(k))$ logarithmically between $k_\mathrm{min}=7\cdot 10^{-5}/\mathrm{Mpc}$ and $k_\mathrm{max}=7/\mathrm{Mpc}$ with $2^{16}$ points. Then a fast discrete sine transform is used, of which the even and odd parts are fit separately with linear combinations of $r^k$ with $k \in \{1,2,3,4,5,6\}$ on  two ranges of scales that exclude the peak: $50-120\mathrm{Mpc}$ and $240-500\mathrm{Mpc}$ in this case.\footnote{This is different to the Hankel transform above due to different tilt and different transformation. However, we also perform the appropriate scaling with $1/s = r_d/r_d^\mathrm{fid}$ in this case.} We additionally tilt the even and odd transforms by $r^{1.3}$ and $r^{1.5}$, respectively, during the fitting of the linear combination of powerlaws, to up-weigh important features. Finally, the resulting polynomial fits -- now with the peak removed -- are transformed back (after removing the $r^{1.3}$ and $r^{1.5}$ tilt) and can be directly used as the power spectrum $P(k)$.

We show the resulting wiggle/no-wiggle ratio in \cref{fig:fast_sine_transform}. It is evident that a large peak is present at wavenumbers around $10^{-2}$/Mpc and smaller. Such a peak can be identified with the plateau of the BAO towards $k \to 0$. Whether that plateau should be identified as a \enquote*{peak} or broadband is a matter of definition, as we demonstrate in \cref{app:illdefined} using a toy example. 

\section{Smooth replacement}\label{app:replacement}
Given that not all algorithms find a power spectrum in the entire range of wavenumbers, sometimes the de-wiggled power spectrum is substituted by the original power spectrum outside of a pre-defined range. It is simply assumed that the original power spectrum outside of that range will show sufficiently small wiggles that these can be effectively ignored.

If one would just replace the power spectrum inside the given range, one would naturally find discontinuities at the edge of the replacement from slight mismatches in the power spectrum amplitude between the true wiggly power spectrum and the de-wiggled power spectrum. In order to minimize such discontinuities, a smooth replacement is performed instead, where the output smoothly interpolates between the true and the de-wiggled power spectrum.

The formula used for such cases is simply
\begin{equation}
    \ln P(k) = \mathfrak{t}(k) \cdot \ln P^\mathrm{original}(k) + [1-\mathfrak{t}(k)] \cdot \ln P^\mathrm{de-wiggled}(k)
\end{equation}

In order to ensure a smooth transition, one simply has to choose a smooth transition function $\mathfrak{t}(k)$. We work in logarithmic space since in this space the power spectrum differences do typically not vary over orders of magnitude. A given smooth transition as a function of $\ln k$ can be written as 
\begin{equation}
    t_\pm(k | k_0,\Delta) = \frac{1}{2} \pm \frac{1}{2} \tanh([\ln k - \ln k_0]/\Delta)~,
\end{equation}
where $\ln k_0$ and $\Delta$ are adjustable parameters. The former is the center point of the transition and the latter is the width of the transition. This transition $t_\pm(k)$ would smoothly interpolate from one power spectrum to the other (favoring either the original or the de-wiggled power spectrum at low $k$ depending on the sign). However, we want to use the original power spectrum in two regimes: below and above the interval where the de-wiggling algorithm works reliably. In this case we simply use the combination 
\begin{equation}
    \mathfrak{t}(k) = t_{-}(k| k_\mathrm{min}, \Delta) + t_{+}(k | k_\mathrm{max}, \Delta)
\end{equation}
with $k_\mathrm{min}$, $k_\mathrm{max}$, and $\Delta$ being chosen for each usage case to optimize the consistency between the methods. We call the parameter $\Delta$ the \enquote{replacement width} for convenience.

\section{Plateau Peak/Wiggle decomposition}\label{app:illdefined}
Roughly speaking, the BAO are expected to follow a form like $\sin(k r_d)/(k r_d)$ \cite{Eisenstein:1997jh} which asymptotes towards unity at $k \to 0$. Whether this first plateau should be considered as oscillation or part of the broadband then makes a crucial difference in whether such a peak around $k \simeq \pi/(2 r_d) \simeq 0.01/\mathrm{Mpc}$ is apparent in the wiggle to no-wiggle ratio, and more indirectly how large the peak at  $k \simeq 3 \pi/(2 r_d) \simeq 0.03/\mathrm{Mpc}$ should be. Both of these effects also have a crucial impact on the slope at $k_p \simeq \pi /r_d \simeq 0.03 h/\mathrm{Mpc} \simeq 0.02/\mathrm{Mpc}$. Overall, the oscillation-broadband decomposition is not well defined in this range between $k \in [\pi/(2 r_d) , 3\pi/(2 r_d)]$, therefore owing to the large differences between the different methods in this range, see also \cref{fig:dewiggle_comparison}.

\begin{figure}
    \centering
    \includegraphics[width=0.49\linewidth]{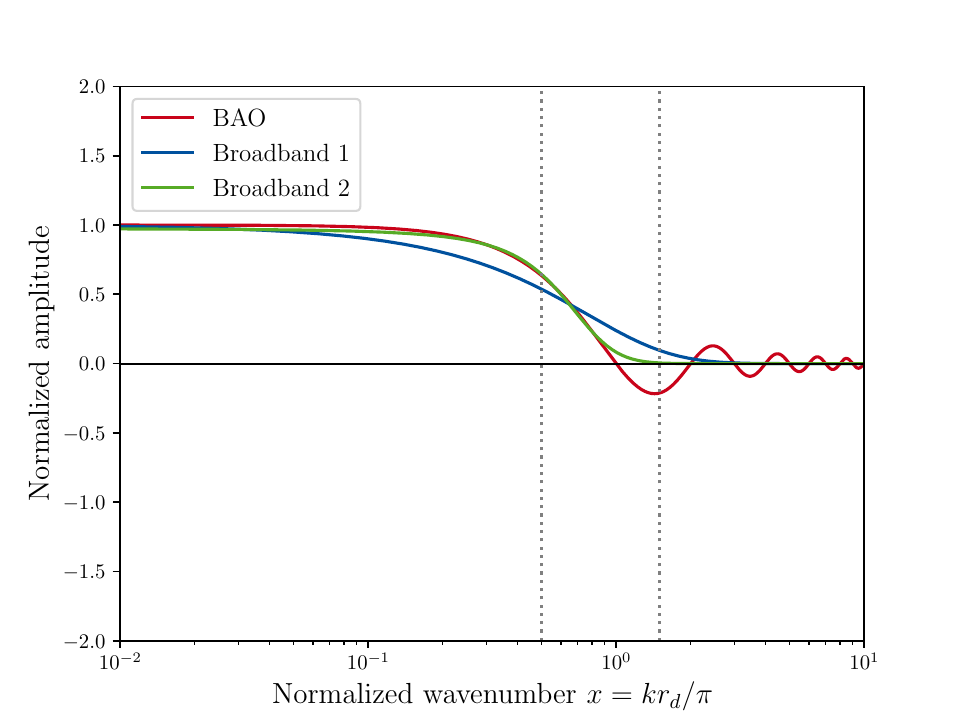}
    \includegraphics[width=0.49\linewidth]{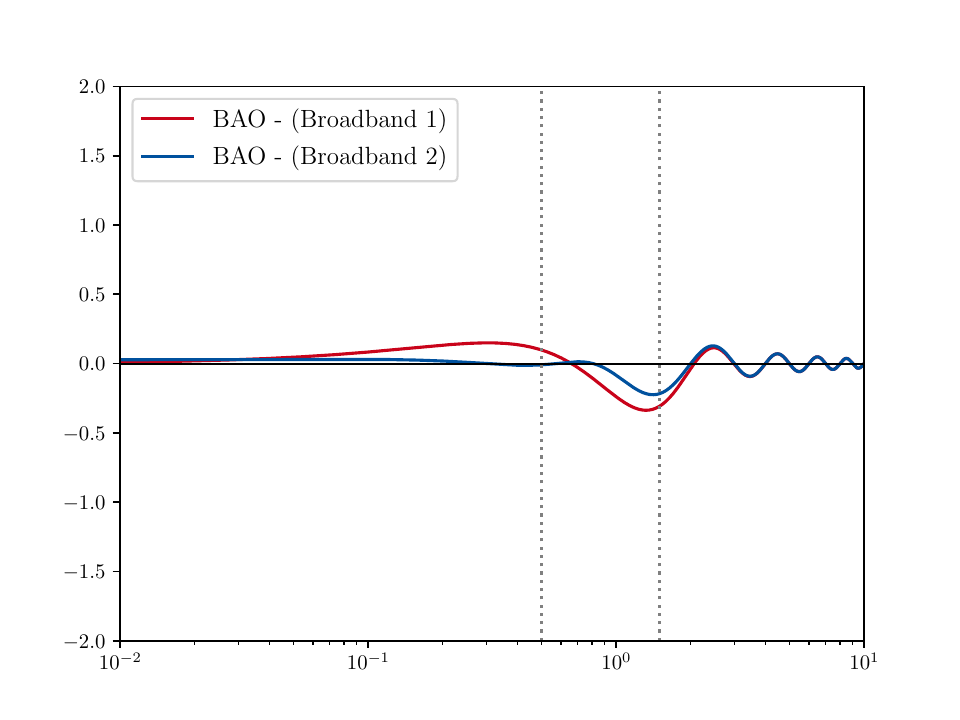}
    \caption{Simplified BAO model to show the ambiguity in the removal of the first peak/plateau. The function Broadband 1 is $1-\tanh(x)$ and Broadband 2 is $\frac{1}{2} - \frac{1}{2} \tanh[3 (x-0.6)]$. We also mark $x=1/2$ and $x=3/2$ by dotted vertical grey lines.}
    \label{fig:simple_bao}
\end{figure}

We can show this effect with a very simple toy example: Consider a simplified BAO of the form $f(x) = \sin(\pi x)/(\pi x)$, see also \cref{fig:simple_bao}. It is clear that for $x \gg 1$ the broadband should be close to zero and for $x \ll 1$ the broadband should be unity. However, the precise transition between these two regimes is not well defined. Either the broadband simply tracks the function until the first oscillation begins (essentially until the first zero-crossing), or the broadband already begins to smoothly interpolate at an earlier point. We show two such functions in \cref{fig:simple_bao}. The resulting difference (as a stand-in for the wiggle/no-wiggle decomposition) shows either a clear first peak at around $x \approx 1/2$ and $x \approx 3/2$ or no peak at $x\approx1/2$ and a reduced peak at $x\approx3/2$. We also observe that the slope at $x\approx1$ is strongly impacted as a consequence. Translating this back to $k = \pi x /r_d$\,, this gives the differences mentioned above.

Since neither assignment of a broadband can be considered more physical than another, this kind of ambiguity in the first two peaks and the slope at $x\approx 1$ or $k \approx \pi/r_d$ is unavoidable.
\section{Splines}

A spline is a simple numerical approach to interpolating data. The idea is that between any two data points ($x_i$, $y_i$) and ($x_{i+1}, y_{i+1}$) the interpolation uses a smooth function $f_i(x)$. What differentiates this approach from linear interpolation (for example) is that at each of the data positions (nodes~$x_i$) the function has to suffice conditions. The most obvious would for example be passing through the data points (and thus ensuring continuity), which requires that $f_i(x_i) = y_i$ and $f_i(x_{i+1})=y_{i+1}$. Given a sufficiently flexible interpolating function $f_i(x)$ more advanced conditions can be imposed, most commonly continuous differentiability, which requires that $f_i'(x_{i+1}) = f_{i+1}'(x_{i+1})$. We discuss specific approaches and implementations below.

\subsection{Cubic Spline}\label{app:ssec:cubic}
A cubic spline is simply a spline with a cubic function, i.e. a function of the form
\begin{equation}
    f_i(x) = a_i + b_i x + c_i x^2 + d_i x^3
\end{equation}
Obviously there are four degrees of freedom for such a function. Two are required to fix $f_i(x_i) = y_i$ and $f_i(x_{i+1})=y_{i+1}$. One more is used to require differentiability $f_i'(x_{i+1}) = f_{i+1}'(x_{i+1})$. In this case one further degree of freedom is open and is used to ensure that the second derivatives are also continuous $f_i''(x_{i+1}) = f_{i+1}''(x_{i+1})$. In total, there are $4N-2$ conditions (with $N$ being the number of data points). This leaves 2 final boundary conditions, which are typically requiring the derivatives at the endpoints to be those linearly estimated $f_{1}'(x_1) = (y_2-y_1)/(x_2-x_1)$ and $f_{N-1}'(x_{N}) = (y_{N}-y_{N-1})/(x_N-x_{N-1})$. In general such splines are very flexible without strong over-fitting or Runge-phenomena.

\subsection{Univariate spline}\label{app:ssec:univariate_spline}
The idea of a univariate spline is that instead of fixing the interpolating intervals to be the data positions ($x_i$, $x_{i+1}$), one instead introduces artificial points (so-called knots $t_j$) which lie somewhere between the data points. Depending on the approach there can be more or fewer knots than data points (but typically fewer). In most use cases, the final function $f(x)$ that is generated from joining each $f_j(x)$ defined between $t_j$ and $t_{j+1}$ to cover the full interval $x_1$ to $x_N$ has to obey certain conditions. First, the $f_j(x)$ are ensured to be continuous and as differentiable as their polynomial degree (which, unlike for Cubic splines, can be something other than $3$). Then, the number of knots is increased (starting from 2) until a certain condition is met. In our case this condition is always one of fitting the data points reasonably well, with 
\begin{equation}
    \sum (y_i - f(x_i))^2 < \mathfrak{s}
\end{equation}
For some smoothing strength $\mathfrak{s}$. Note that since the knots don't coincide with the data points $x_i$ and since the spline is of possibly higher order, there is often no analytical solution and the knot number and the individual polynomial coefficients of the $f_j(x)$ are found numerically. The choice of a smoothing strength is an important consideration, and we often optimized the values quoted here visually and \enquote*{by hand}.

The univariate splines are often represented in the B-spline basis, which is an alternative way of representing arbitrary splines. In the text we call a B-spline simply that for which $\mathfrak{s}=0$ (note that this still does not imply that $t_j = x_i$).

\section{Power spectrum in \texorpdfstring{$w_0$-$w_a$}{w0wa} cosmologies}\label{app:w0wa}
\begin{figure}
    \centering
    \includegraphics[width=0.7\linewidth]{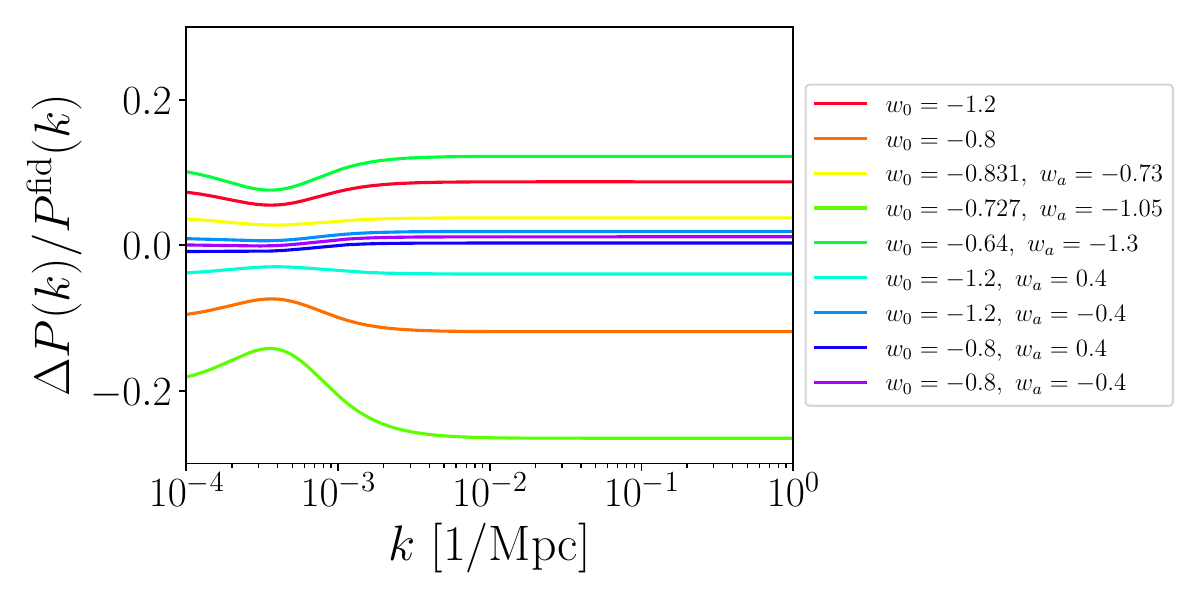} \\
    \includegraphics[width=0.7\linewidth]{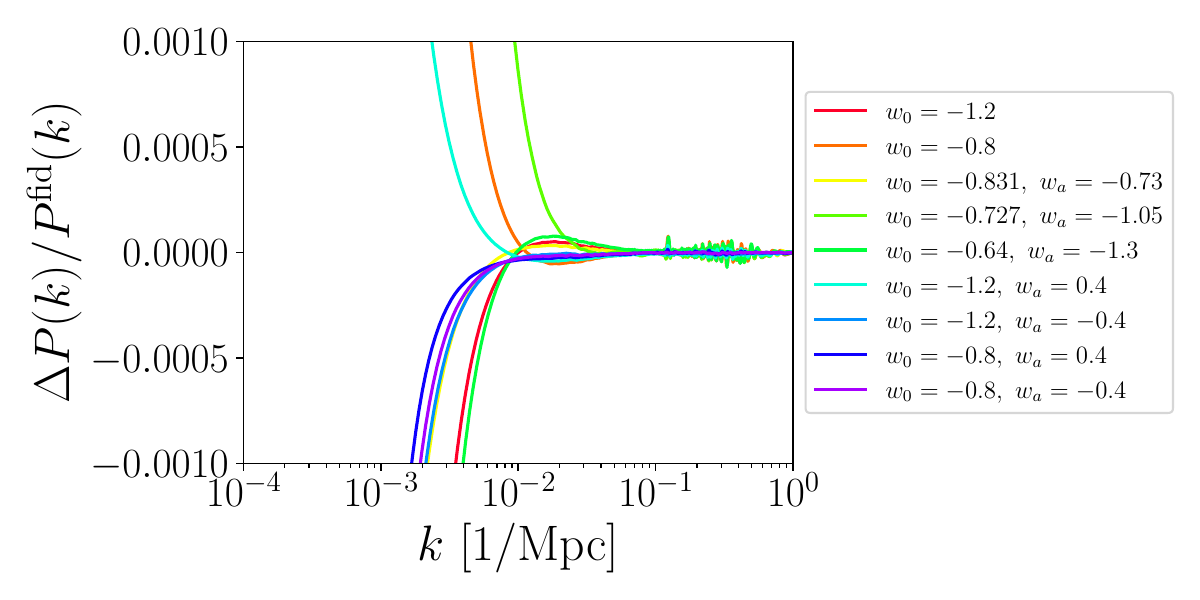}
    \caption{Relative difference in the power spectrum for a few selected $w_0$-$w_a$ cosmologies of \cref{fig:null_tests}. Top: Relative difference. Bottom: Relative difference, but re-normalized to coincident amplitude at smallest scales and zoomed into the $\pm 0.1\%$ range.}
    \label{app:fig:w0wa}
\end{figure}

We show in \cref{app:fig:w0wa} the impact of the CLP $w_0$-$w_a$ parameterized dark energy on the power spectrum, compared to the fiducial model. We observe in the right panel that there is a tiny leakage of power spectrum difference even to scales around $0.03h$/Mpc at the level of $<0.1\%$ which causes the minute differences in $m$ observed in \cref{fig:null_tests}.
\end{document}